\documentclass[12pt]{article}
\usepackage{amsthm, amsmath, amssymb}
\usepackage{latexsym,amsmath,amsfonts,amsthm,amssymb}

\newtheorem{theorem}{Theorem}[section]
\usepackage{tikz}
\usepackage[dvips]{epsfig}
\usepackage{enumerate}
\usepackage{bbm}
\usepackage{graphics}
\usepackage[font={footnotesize},margin=1cm]{caption}

\usepackage[breaklinks=false,pdfpagemode=None,pdfview=FitH,pdfstartview=FitH,citebordercolor={0 0 1},linkbordercolor={0 0 1},urlbordercolor={0 0 1},pagebordercolor={0 0 1},pdfborder={0 0 1}]{hyperref}

\usetikzlibrary{arrows,shapes}
\usetikzlibrary{decorations.pathmorphing}

\newcommand{\cd}{{\, \cdot \, }}

\newcommand{\calB}{{\mathcal{B}}}

\newcommand{\calD}{{\mathcal{D}}}
\newcommand{\calE}{{\mathcal{E}}}

\newcommand{\calH}{{\mathcal{H}}}

\newcommand{\calN}{{\mathcal{N}}}
\newcommand{\calO}{{\mathcal{O}}}

\newcommand{\calR}{{\mathcal{R}}}
\newcommand{\calS}{{\mathcal{S}}}

\newcommand{\res}{\mathrm{res}}

\newcommand{\prov}{\mathrm{prov}}
\newcommand{\tr}{\mathrm{tr}}

\newcommand{\lint}{\mathrm{lint}}

\newcommand{\veps}{\varepsilon}
\numberwithin{equation}{section}
\newtheorem*{assumptionlla}{Assumption LLA($\nu, C$)}
\newtheorem*{assumptiona1}{Assumption A1($c_b$)}
\newtheorem*{assumptiona2}{Assumption A2($\nu, \varepsilon_0$)}
\newtheorem{corollary}[theorem]{Corollary}
\newtheorem{proposition}[theorem]{Proposition}
\newtheorem{lemma}[theorem]{Lemma}
\title  {
        On Many-Body Localization for Quantum Spin Chains
                 }

\author{
John Z.\ Imbrie\footnote{
This research was conducted in part while
the author was visiting the Institute for Advanced Study in Princeton, supported by The Fund for Math and The Ellentuck Fund.
}
\\Department of Mathematics,
University of Virginia \\
Charlottesville, VA 22904-4137, USA
\\ {\tt imbrie@virginia.edu}
}
\date{}
\begin{document}
\maketitle
\begin{abstract}
For a one-dimensional spin chain with random local interactions,
we prove that many-body localization follows from a physically reasonable assumption that limits the amount of level attraction in the system. The construction uses a sequence of local unitary transformations to diagonalize the Hamiltonian and connect the exact many-body eigenfunctions to the original basis vectors.
\end{abstract}
\tableofcontents
\section{Introduction}\label{1}
\subsection{Background}\label{1.1}

The eigenfunctions of a single-particle Hamiltonian with a large random potential are localized: they decay exponentially with the distance from some center. Does the phenomenon of localization persist in a more realistic model with interacting particles? This question was raised in Anderson's original paper \cite{Anderson1958}, and subsequent work in the physics literature \cite{Fleishman1980, Giamarchi1987, Santos2004, Gornyi2005, Basko2006, Znidaric2008, Pal2010} supports the idea of many-body localization, on the basis of several theoretical perspectives and on numerical work.
	
	In this paper we focus on one of the simplest models where many-body localization should occur. We consider the many-body spin chain Hamiltonian on the lattice 
 $\Lambda = [-K, K']\cap \mathbb{Z}$ :
\begin{equation}
H = \sum\limits_{i=-K}^{K'} h_i S_i^\mathrm{z} + \sum\limits_{i=-K}^{K'} \gamma_i S_i^\mathrm{x} + \sum\limits_{i=-K-1}^{K'} J_i S_i^\mathrm{z} S^\mathrm{z}_{i + 1}.
\label{(1.1)}
\end{equation}
This operates on the Hilbert space $\calH = \bigotimes_{i\in\Lambda} \mathbb{C}^2$, with Pauli matrices 
\begin{equation}
S^\mathrm{x}_i =
\begin{pmatrix}
0 & 1\\ 1 & 0
\end{pmatrix}
,
 S^\mathrm{y}_i = \begin{pmatrix}
0 & -i\\ i & 0
\end{pmatrix}
,
 S^\mathrm{z}_i = \begin{pmatrix}
1 & 0\\ 0 & -1
\end{pmatrix}
\label{(1.2)}
\end{equation}
operating on the $i$\textsuperscript{th} variable. 
Variables outside of $\Lambda$ are frozen (projected out), \textit{i.e.} we set $S_i^\mathrm{z} = 1$ in (\ref{(1.1)}) for $i \notin \Lambda$ for $+$ boundary conditions.
Note that $H$ is diagonal in the basis used above, except for the second term involving $S_i^\mathrm{x}$. We write $\gamma_i = \gamma\Gamma_i$ with $\gamma$ small, and assume $h_i$, $\Gamma_i$, and $J_i$ are independent random variables, bounded by 1, with probability densities bounded by a fixed constant $\rho_0$. 

This model has random field, random transverse field, and random exchange interactions; it is a variant of the model studied in \cite{Pal2010}. It should have a transition from a many-body-localized phase for small $\gamma$ or $J$ to a thermalized phase if $\gamma$ and $J$ are large. (Note that a tensor product basis of eigenstates can easily be constructed for $H$ if either $\gamma$ or $J$ is zero.) We investigate properties of general eigenstates, not just those at low energies.

The notion of localization has to be adapted to the many-body context, for a couple of reasons. First, the configuration space includes the positions of all the particles (or the values of all the spins). Decay in this space is too much to ask for.  Second, whatever basis we choose for $\calH$, interactions connect a given state to nearby states everywhere in space. This means that a normalized eigenfunction will lose its amplitude exponentially with the volume. 

We will examine three signs of many-body localization. First, 
for the above Hamiltonian, the basis vectors are tensor products of $(1, 0)$ or $(0, 1)$ in each index. Thus the basis vectors are indexed by ``spin configurations'' $\sigma = \{\sigma_i\} \in \{-1,1\}^{|\Lambda|}$.
We have weak off-diagonal disorder, and one might expect that the eigenfunctions resemble basis vectors, which would imply that for most eigenstates, the expectation of $S^\mathrm{z}_i$ should be close to $+1$ or $-1$. This is a basic signal of many-body localization for $H$. The analogous statement for the one-body Anderson model is the fact that most eigenfunctions have the preponderance of amplitude near a particular site.

The second sign has to do with the product structure of $\calH$.  The Hilbert space is a tensor product of local vector spaces (instead of a direct sum, as in the single-body problem). In the absence of interactions, this structure carries over to the eigenstates. With weak interactions, one should see the tensor product structure emerge at long distances. For small $\gamma$, and for any eigenstate, correlations between local operators separated by a distance $r$ should decay like $\gamma^{\kappa r}$ for some $\kappa>0$. This is analogous to the exponential decay of eigenfunctions in the one-body problem, but it is a decay of entanglement, rather than amplitude. (Distant spins are very nearly in a product state.)

As in the one-body Anderson model, there should be a natural way to create a mapping between eigenstates and basis vectors, away from a dilute set of resonant regions. This is a third sign of many-body localization. 

The second term of (\ref{(1.1)}) is the Laplacian on the hypercube; it implements spin flips or hops between basis vectors differing at a single site in $\mathbb{Z}$.
But a key difference between $H$ and the one-body Anderson model is the random potential -- here it consists of the first and third terms of (\ref{(1.1)}). There is a lack of monotonicity, and in addition the number of random variables is only logarithmic in the dimension of $\calH$. This creates particular challenges for rigorous work. The term $\sum_i h_i S^\mathrm{z}_i$ is sufficient to break degeneracies associated with individual spin flips. However, we do not have full control over energy differences for configuration changes in extended regions, so an assumption about local eigenvalue statistics is a prerequisite for our results.
Specifically, we prove that if a physically reasonable assumption on the separation of eigenvalues is valid, then many-body localization holds (in the sense described above).

	Our methods will apply to more general models provided they have a few key properties in common with (\ref{(1.1)}). Specifically, there must be a tensor product basis in which the Hamiltonian is a diagonal matrix plus a local perturbation, with all terms having random coefficients. (Models with only some terms random, \textit{e.g.} (\ref{(1.1)}) with $\gamma_i$, $J_i$ fixed, could be considered as well, but a stronger assumption about the behavior of eigenvalue differences would be needed.) The dimension of the state space at a site should be finite. The diagonal part should have local interactions (\textit{e.g.} nearest-neighbor as above), and the random variables at each site $i$ should be able to move energy differences between pairs of basis vectors that differ only at $i$. Thus one may consider certain models of interacting particles in $\mathbb{Z}$ with hard-core conditions.
	
	There is a considerable literature of rigorous work on the phenomenon of single-particle Anderson localization, for example the proof of absence of diffusion in dimensions 2 or more \cite{Frohlich1983} and the proof of localization using the exponential decay of the average of a fractional moment of the Green's function \cite{Aizenman1993}. The latter work also applies to Hamiltonians on the Bethe lattice, which is relevant for models of many-body localization involving decay in Fock space along tree-like particle cascades \cite{Mirlin1991, Gornyi2005, Basko2006}. Like the spin chain, the Bethe lattice exhibits an exponential growth in the number of states as a function of the diameter of the system -- this is a key problem for rigorous work on many-body localization. There are also a number of results for a fixed number of interacting particles \cite{Aizenman2009,Chulaevsky2009,Chulaevsky2009a,Chulaevsky2011, Chulaevsky2011a, Klein2013, Fauser2014}. 
	
	Recent results on localization in many-body systems include a proof of dynaical localization for an isotropic random spin chain, using the Jordan-Wigner transformation to reduce the problem to an equivalent one-body Hamiltonian \cite{Hamza2012}.
Other results include a proof of an asymptotic form of localization (in a non-random system with frustration) \cite{DeRoeck2013} and a proof of localization when the disorder strength diverges rapidly with the volume \cite{Giscard2014}.

\subsection{Results}\label{1.2}

We will need to assume a property of limited level attraction for the Hamiltonian in boxes of varying size:
\begin{assumptionlla}
 (Limited level attraction) Consider the Hamiltonian $H$ in a box $\Lambda$ with $|\Lambda| = K + K' + 1 = n$. With the given probability distribution for $\{h_i, \Gamma_i, J_i\}$, its eigenvalues satisfy
\begin{equation}
P\left(\min_{\alpha \ne \beta} |E_{\alpha} - E_{\beta}| < \delta\right) \le \delta^{\nu} C^n, \label{(1.3)}
\end{equation}
for all $\delta>0$ and all n.
\end{assumptionlla}

We show that many-body localization holds (in a sense made precise below) for $\gamma$ small enough, provided \textbf{LLA($\nu, C$)} holds for some fixed $\nu, C$. Ideally, one would prove many-body localization without making such an assumption. However, at this point we lack the tools to adequately deal with such questions of level statistics. 
(For a step in this direction, see \cite{Imbrie2016} for a proof of a level-spacing condition for a block Anderson model.)
Nevertheless, \textbf{LLA($\nu, C$)} is a very mild assumption from the physical point of view, since random matrices normally have either neutral statistics ($\nu = 1$, \textit{e.g.} Poisson) or repulsive ones ($\nu > 1$, \textit{e.g.} GOE). Indeed, the thermalized phase should have significant level repulsion \cite{Oganesyan2007}. 
In fact, we only need (\ref{(1.3)}) for a particular value of $\delta$ ($\tilde{\varepsilon}^n$, where $\tilde{\varepsilon}$ is a small power of $\gamma$). 

For the purposes of this paper, many-body localization (MBL) consists of the following properties of the eigenvalues and eigenstates of $H$:
\begin{enumerate}[(i)]
  \item Existence of a labeling system for eigenstates by spin/metaspin configurations, with metaspins needed only on a dilute collection of resonant blocks. (As mentioned above, the spin variables used to label basis vectors can also be used to label the exact eigenstates, but the correspondence becomes somewhat arbitrary in resonant regions, so we use the term ``metaspin'' instead.)
  \item Faster-than-power-law decay of the probability of resonant blocks, which implies their diluteness. (This is critical to the whole concept of a labeling system -- without it the labeling system would lose its meaning.)
  \item Diagonalization of $H$ via a sequence of local rotations defined via convergent graphical expansions with exponential bounds. (Locality means that graphs depend only on the random variables in their immediate vicinity.)
  \item Bounds establishing closeness of expectations of local observables in any eigenstate to their na\"{i}ve $(\gamma = 0)$ values, when observables are not in resonant regions. (This makes precise the idea that eigenstates resemble the basis vectors.)
  \item Almost sure convergence of local energy differences and expectations of local observables as $\Lambda \rightarrow \mathbb{Z}$.
\item Exponential decay of connected correlations
$\langle \calO_i;\calO_j\rangle_\alpha \equiv \langle\calO_i\calO_j\rangle_\alpha-\langle\calO_i\rangle_\alpha \langle\calO_j\rangle_\alpha$ 
in each eigenstate, except on a set of rapidly decaying probability. (This shows the exponential loss of entanglement with distance for the subsystems associated with the observables.)
\item Faster-than-power-law decay of averaged connected correlations.
\end{enumerate}

The set of resonant regions will be constructed through an inductive procedure that generates local rotations to successively diagonalize the Hamiltonian. Further details and concrete bounds will be deferred to the main body of the paper. However, we state here a theorem that incorporates (iv), (vi), and (vii). It can be taken as a basic characterization of many-body localization. We will need a notion of state-averaging. 
Let $\alpha$ be a label for the eigenstates of $H$. Then let $\textrm{Av}_\alpha$ denote the average over the $2^n$ values of $\alpha$ (for a box of size $n$). The average can be with uniform weights (infinite temperature) or with any normalized energy-dependent weight function (\textit{e.g.} $\textrm{(const)}\exp{(-\beta E_\alpha)}$, which gives the usual ensemble for inverse temperature $\beta$).

\begin{theorem}\label{th1.1}
Let  $\nu, C$ be fixed.  There exists a $\kappa >0$ such that for $\gamma$ sufficiently small, \textbf{LLA($\nu, C$)} implies the following estimates. Let $\langle \cd \rangle_\alpha$ denote the expectation in the eigenstate $\alpha$. Then
\begin{equation}
\mathbb{E}\, \normalfont{\text{Av}}_\alpha \left| \langle S^\mathrm{z}_0 \rangle_\alpha\right| = 1 - O(\gamma^\kappa).
\label{(1.4)}
\end{equation}
Furthermore, for any $i$, $j$, 
\begin{equation}
\max_\alpha |\langle\calO_i;\calO_j\rangle_\alpha| \le \gamma^{|i-j|/3}\,\, \text{with probability}\,\, 1-(\gamma^\kappa)^{1+c_3(\log(|i-j|/8\vee1))^2},
\label{(1.5)}
\end{equation}
for some constant $c_3>0$. Here $\langle\calO_i;\calO_j\rangle_\alpha \equiv \langle\calO_i\calO_j\rangle_\alpha - \langle\calO_i\rangle_\alpha\langle\calO_j\rangle_\alpha$, with $\calO_i$ any operator formed from products of $S^\mathrm{x}_{i'}$, $S^\mathrm{y}_{i'}$  or $S^\mathrm{z}_{i'}$, for $i'$ in a fixed neighborhood of $i$. Finally,
\begin{equation}
\mathbb{E} \, \normalfont{\text{Av}}_\alpha |\langle\calO_i;\calO_j\rangle_\alpha| \le
(\gamma^\kappa)^{1+c_3(\log(|i-j|/8\vee1))^2}. \label{(1.6)}
\end{equation}
All bounds are uniform in $\Lambda$.
\end{theorem}

From (\ref{(1.4)}), we can see that with high probability, most states have the property that the expectation of $S_0^\mathrm{z}$ is close to $+1$ or $-1$, as is the case for the basis vectors. This would contrast with a thermalized phase, wherein states resemble thermal ensembles (a consequence of the eigenstate thermalization hypothesis -- see \cite{Deutsch1991, Srednicki1994, Rigol2008}). 
At infinite temperature, thermalization would imply that averages of eigenstate expectations of $S^\mathrm{z}_0$ go to zero as $\Lambda \rightarrow \infty$ \cite{Pal2010}.
Thus one sign of many-body localization is the violation of thermalization as in (\ref{(1.4)}). 

	Another sign of many-body localization would be the absence of transport. Although we have not looked at time-dependent quantities, essentially all of the eigenstates we have constructed have a distribution of energy that is nonuniform in space (\textit{i.e.} in $\mathbb{Z}$), and this necessarily persists for all time. So in a very basic sense, there is no transport in the system -- this is another feature of the lack of thermalization. 

	Our rigorous result on many-body localization is an important capstone to the physical arguments that have led to the idea of a many-body localized phase. Without full control of the approximations used, there remains the possibility that thermalization sets in at some very long length scale. Such a scenario would not show up in the numerics, and has been conjectured to occur in the nonrandom model of \cite{DeRoeck2013}.  
 
\subsection{Methods}\label{1.3}

We will perform a complete diagonalization of the Hamiltonian by successively eliminating low-order off-diagonal terms. The process runs on a sequence of length scales $L_k = (\frac{15}{8})^k$, and off-diagonal elements of order $\gamma^m$, $m \in [L_k, L_{k+1})$ will be eliminated in the $k^{\text{th}}$ step. The orthogonal rotations that accomplish this can be written as a convergent power series, provided nonresonance conditions are satisfied. Resonant regions are diagonalized as blocks in quasi-degenerate perturbation theory. The crux of the method is control of probabilities of resonances. It will be critical to maintain bounds exponential in the length scale of the resonance. Otherwise, the bounds will be overwhelmed by the exponential number of transitions that need to be tested for resonance. The method was developed in \cite{Imbrie2014a} for the single-body Anderson model.  This led to a new proof of the exponential localization result of \cite{Aizenman1993} via multiscale analysis, working directly with rotations, instead of resolvents. 
(We recommend \cite{Imbrie2014a} to serious readers of this article, as the key tools are developed in a much less complex setting.)
The key estimate that allows the procedure to work on all length scales is a uniform decay rate for a fractional moment for graphs with many independent energy denominators. This leads to exponential bounds on resonance probabilities, at least for graphs with mostly independent denominators. The method can be thought of as a KAM or block Jacobi \cite{Sleijpen2000} procedure. Each step is a similarity transformation implementing Rayleigh-Schr\"{o}dinger perturbation theory in a manner very close to that of \cite{Datta1996, Datta1999} (though in those works a single transformation was sufficient to break degeneracies in the Hamiltonian).
	
A number of authors have used related KAM constructions to prove localization for quasiperiodic and deterministic potentials \cite{Bellissard1983b, Bellissard1983a,Sinai1987, Chulaevsky1991, Chulaevsky1993, Eliasson1997, Eliasson2002}. More broadly, a number of different flows have been used to diagonalize matrices in various contexts \cite{Deift1983, Brockett1991, Grote2002, Wegner2006}. Related renormalization group ideas have appeared in  \cite{Schrieffer1966, White1992, Glazek1993, Bach1998}. 

The idea that quasi-local unitary transformations may be used to isolate local variables or conserved quantities in a many-body localized system appears in \cite{Huse2013,Serbyn2013,Bauer2013,Ros2014}. Here, we implement the rotations in a constructive manner, providing explicit expansions for the rotations along with bounds that quantify the notion of locality. Such expansions are new, even in the single-body context \cite{Imbrie2014a}.

\section{First Step}\label{2}

The basic features of our method are easy to understand in the first step. Here we focus on single spin flips. We know that perturbation theory will be under control in regions where no resonant transitions occur. 
So in Section \ref{2.1}, we identify resonant regions and prove that they form a dilute subset of $\mathbb{Z}$. Away from these regions, energy denominators cannot get too small, so first-order perturbation theory is under control. 
In Section \ref{2.2}, we we use this to define a rotation (change of basis) that diagonalizes the Hamiltonian up to terms that are second order in $\gamma$, in the nonresonant region. We give a graphical expansion for the rotated Hamiltonian exhibiting quasi-locality (\textit{i.e.} a term with range $r$ is exponentially small in $r$). In Section \ref{2.3}, we deal with the resonant regions by performing rotations that diagonalize the Hamiltonian there. Although these rotations are not under perturbative control, their effects are limited due to the diluteness of resonant regions and to the smallness of connections to nonresonant regions. 
In Section \ref{2.4}, we show that the effect of the rotations on observables is small.

\subsection{Resonant Blocks}\label{2.1}

Resonances occur when transitions induced by off-diagonal matrix elements produce an energy change that is smaller than some cutoff $\varepsilon$. In the spin chain, a transition is a spin flip at a site $i$. If we start from spin configuration $\sigma$, let the flipped spin configuration be $\sigma^{(i)}$:
\begin{equation}
\sigma^{(i)}_j =
\begin{cases}
-\sigma_j, & j = i;\\
\sigma_j, & j \ne i.
\end{cases}
 \label{(2.1)}
\end{equation}
Let
\begin{equation}
E(\sigma) = \sum_{i = -K}^{K'} h_i \sigma_i + \sum_{i=-K-1}^{K'} J_i \sigma_i \sigma_{i + 1} \label{(2.2)}
\end{equation}
denote the diagonal entry of $H$ corresponding to $\sigma$. (We take $\sigma_i = 1$ for $i\notin \Lambda$.) Then
\begin{equation}
E(\sigma) - E (\sigma^{(i)}) = 2 \sigma_i (h_i + J_i \sigma_{i + 1} + J_{i - 1} \sigma_{i - 1}). \label{(2.3)}
\end{equation}
We say that the site $i$ is resonant if $|E(\sigma) - E(\sigma^{(i)})| < \varepsilon$ for at least one choice of $\sigma_{i - 1}, \sigma_{i + 1}$. The probability that $i$ is resonant is bounded by $4 \rho_0 \varepsilon$. We take $\varepsilon$  to be a small power of the coupling constant for spin flips: $\varepsilon \equiv \gamma^{1/20} << 1$.

Let $\calS_1 = \{i \in \Lambda : i$ is resonant$\}$. Then we may decompose $\calS_1$ into resonant blocks $B^{(1)}$ using nearest-neighbor connections. The probability that two sites $i, j$ lie in the same block is bounded by $(4 \rho_0 \varepsilon)^{|i-j|+1}$.
(Conditioning on $\{\Gamma_i,J_i\}$, we obtain a product of independent probabilities for each site $m$ with $i \le m \le j$; $\rho_0$ is the bound assumed above on the probability densities and the factor of 4 accounts for choices of neighbor spins $\sigma_{m-1}$ and $\sigma_{m+1}$.)

\subsection{Effective Hamiltonian}\label{2.2}

Let us group the off-diagonal terms of $H$ as follows:

\begin{align}
J^{(0)} = \sum\limits_{i\in\Lambda} \gamma_i S^\mathrm{x}_i = \sum\limits_{i\in\Lambda} J^{(0)} (i) & = J^{(0)\text{per}} + J^{(0)\text{res}} \notag \\ 
& = \sum\limits_{i\notin \calS_1} J^{(0)\text{per}} (i) + \sum\limits_{i\in \calS_1} J^{(0)\text{res}} (i),
\label{(2.4)}
\end{align}
where $J^{(0)\text{res}}$ contains terms with $i \in \calS_1$ (``resonant terms''), and $J^{(0)\text{per}}$ contains terms in the nonresonant region (``perturbative terms''). Then define the antisymmetric basis-change generator
\begin{equation}
A = \sum \limits_{i \notin \calS_1} A (i), \label{(2.5)}
\end{equation}
with the local operator $A(i)$ given by its matrix elements:
\begin{equation}
A(i)_{\sigma \sigma^{(i)}} = \frac{J^{(0)\text{per}}(i)_{\sigma \sigma^{(i)}}}{E(\sigma) - E(\sigma^{(i)})} = \frac{\gamma_i}{E(\sigma) - E(\sigma^{(i)})}. \label{(2.6)}
\end{equation}
All other matrix elements of $A(i)$ are zero; $A(i)$ only connects spin configurations differing by a single flip at $i$. Nonresonance conditions ensure that all matrix elements of $A(i)$ are bounded by $\gamma/\varepsilon = \gamma^{19/20}$. In fact, $\|A(i)\| \le \gamma/\varepsilon$, since each row/column has a single term with that bound. Also, $\|J^{(0)}(i)\| \le \gamma$.

Next, we define the basis change $\Omega = e^{-A}$. Let $H_0$ be the diagonal part of $H$. Note that $[A, H_0] = -J^{(0)\text{per}}$, so that $[A, H] = -J^{(0)\text{per}} + [A, J^{(0)}]$. This leads to a cancellation of the term $J^{(0)\text{per}}$ in $H$:
\begin{align}
 H^{(1)} &= e^A H e^{-A} = \sum\limits_{n = 0}^{\infty} \frac{(\text{ad}\,A)^n}{n!} H \notag \\
& = H_0 + J^{(0)\text{res}} + J^{(0)\text{per}}  + \sum\limits_{n = 1}^{\infty}
\frac{ (\text{ad}\,A)^{n-1} (-J^{(0)\text{per}}) + (\text{ad}\,A)^n (J^{(0)\text{per}} + J^{(0)\text{res}})}{n!} \notag \\
& = H_0 + J^{(0)\text{res}} + \sum\limits_{n = 1}^{\infty} \frac{n}{(n + 1)!} (\text{ad}\,A)^n J^{(0)\text{per}} + \sum\limits_{n = 1}^{\infty} \frac{(\text{ad}\,A)^n}{n!} J^{(0)\text{res}} \notag \\
& = H_0 + J^{(0)\text{res}} + J^{(1)}. \label{(2.7)}
\end{align}

Note that $A$ and $J^{(0)}$ are given by a sum of local operators, so the commutators in $J^{(1)}$ likewise will be given as a sum of local operators (although the range of the operator grows as $n$, the order of the commutator). However, even though $A(i)$ and $J(i)$ only act on the spin at site $i$, the matrix elements of $A(i)$ depend on the spins at $i - 1$ and $i + 1$. Therefore, $[A(i), A(j)]$ and $[A(i), J(j)]$ do not in general vanish when $|i - j| = 1$. They do vanish when $|i - j|\ge 2$.

We may give graphical expansions for $J^{(1)}$ by expanding each $J$, $A$ in $(\text{ad}\,A)^n J^{(0)\text{per}}$ and $(\text{ad}\,A)^n J^{(0)\text{res}}$ as a sum of operators localized at individual sites.

\noindent Thus
\begin{equation}
(\text{ad}\,A)^n J^{(0)\text{per}} = \sum \limits_{i_0, i_1,\ldots,i_n} (\text{ad}\,A(i_n)) \cdots (\text{ad}\,A(i_1)) J^{(0)} (i_0). \label{(2.8)}
\end{equation}
We must have dist$(i_p, \{i_0,\ldots,i_{p - 1}\})\le 1$; otherwise the commutator with $A(i_p)$ vanishes. Note that the number of choices for $i_p$, given $i_0,\ldots,i_{p - 1}$, is no greater than $p + 2$. Thus we have, in effect, a combinatoric factor $(n + 2)!/2!$ which is controlled by the prefactors $n/(n + 1)!$ or $1/n!$, leaving only a factor $(n + 1)(n + 2)$. There are $2^n$ terms from writing out an $n^{\text{th}}$ order commutator, so the series is geometrically convergent. We may write
\begin{equation}
J^{(1)}_{\sigma\tilde{\sigma}} = \sum \limits_{g_1: \sigma \rightarrow \tilde{\sigma}} J^{(1)}_{\sigma\tilde{\sigma}} (g_1), \label{(2.9)}
\end{equation}
where $g_1$ represents a walk in spin configuration space that is connected in the sense described above. More precisely, $g_1$ prescribes an ordered product of operators $A(i_p)$ or $J^{(0)}(i_0)$ arising from an admissible sequence $i_0, i_1,\ldots,i_n$ after expanding the commutators (admissible means each $i_p$ is with a distance 1 of $\{i_0, i_1,\ldots,i_{p-1}\}$, for a nonvanishing commutator). Each operator implements a spin flip, hence we obtain a walk on the hamming cube $\{-1,1\}^{n+1}$. The interaction term $J^{(1)}_{\sigma\tilde{\sigma}} (g_1)$ is the product of the specified matrix elements and a factor $n/(n+1)!$ from (\ref{(2.7)}), along with a sign and a binomial coefficient from expanding the commutators and gathering like terms. See \cite{Imbrie2014a}, eq. 2.18.
A similar graphical expansion can be derived when the similarity transformation is applied to any local operator, \textit{e.g.} $S_0^\mathrm{x}$, $S_0^\mathrm{y}$, or $S_0^\mathrm{z}$. Nonresonance conditions imply that
\begin{equation}
|J^{(1)}_{\sigma \tilde{\sigma}} (g_1)| \le \frac{\gamma(\gamma / \varepsilon)^{|g_1| - 1}}{(|g_1| - 1)!}, \label{(2.10)}
\end{equation}
where $|g_1| = n + 1$ is the number of operators $(A \, \text{or}\, J)$ in $g_1$. As discussed above, the sum over $g_1$ involving a particular $J^{(0)}(i_0)$ converges geometrically as $\gamma(c \gamma/ \varepsilon)^n$.

\subsection{Small Block Diagonalization}\label{2.3}

We need to decompose the resonant region $\calS_1$ into a collection of subsets of $\mathbb{Z}$ that we will term ``blocks.'' These are essentially connected components of $\calS_1$, although it is necessary to complicate the definitions by adding collar neighborhoods to the blocks and to distinguish between ``small'' and ``large'' blocks. In the $k^{\textrm{th}}$ step, we consider interaction terms with range less than $L_k = (\frac{15}{8})^k$. By adding a collar of width $L_k -1$ around blocks, we ensure that interactions connecting across the collar are of order $L_k$ or higher. At the same time, we require the diameter of small blocks to be  $< L_k$. In this way, the minimum range (which translates to the order of perturbation theory in $\gamma$) matches up well with the size of small blocks, and so for them, factors $\gamma^{L_k}$ or $\varepsilon^{L_k}$ will beat the sum over states in the block.
(We will diagonalize the Hamiltonian within small blocks, and then an $n$-site resonant block has $2^n$ eigenstates.)

We need to go further by requiring small blocks to be isolated, in the sense that a small block with $n$ sites in step $k$ is separated from other blocks on that scale (and later scales) by a distance $> d_m \equiv \exp(L^{1/2}_{m + m_0})$ when $n \in [L_{m - 1}, L_m)$. Here $m_0>0$ is a fixed integer to be chosen later -- see the discussion following (\ref{(4.8a)}).  

All these issues arose in the treatment of the one-body problem in \cite{Imbrie2014a}, but there the number of states in a block of size $n$ is only $n$, so we were able to use collars logarithmic in $n$.
Here, we have to take collars linear in $n$, which pushes us into a regime with extended separation conditions. The additional separation $\exp(O(n^{1/2}))$ ensures that blocks do not clump together and ruin the exponential decay. In both cases the goal is to ensure that the combinatorics of graphical sums behave well in the multiscale analysis. The distance condition should be familiar to readers of \cite{Frohlich1983}. As in that work, the construction ensures that uniform exponential decay is preserved away from resonant regions. But this benefit comes at the cost of working with loosely connected resonant blocks with weak probability decay. 

For step one, small blocks are those with diameter 0 or 1, since $L_1 = \tfrac{15}{8}$, and they have a separation distance $d_1$ from other blocks. They will be denoted $b^{(1)}$. We add a 1-step collar neighborhood (the set of sites at distance 1 from $b^{(1)}$, and denote the result $\bar{b}^{(1)}$. The rest of $\calS_1$ is given a 1-step collar to form $\overline{\calS}_{1'}$. Its components are denoted $\bar{B}^{(1')}$. We may also refer to $B^{(1')} = \bar{B}^{(1')} \cap \calS_1$ and $\calS_{1'} = \overline{\calS}_{1'} \cap \calS_1$. This enables us to identify the ``core'' resonant set that produced a large block $\bar{B}^{1'}$.

Let us separate terms that are internal to the collared blocks $\bar{b}^{(1)}$ and $\bar{B}^{(1)}$ from the rest. The former are the troublesome ones (resonance probabilities are hard to control, which leads to lack of control of the ad expansion), so they need to be ``diagonalized away.'' (Compare with \cite{Imbrie2014a}, eq. 2.20.) Put
\begin{equation}
\begin{aligned}
 J^{(1)\text{int}}_{\sigma \tilde{\sigma}} &= J^{(0)\text{res}}_{\sigma \tilde{\sigma}} + \sum\limits_{g_1 : \sigma \rightarrow \tilde{\sigma}, \,g_1 \cap \calS_1 \ne \varnothing, \,g_1 \subset \overline{\calS}_1} J^{(1)}_{\sigma \tilde{\sigma}} (g_1) = J^{(1)\text{sint}}_{\sigma \tilde{\sigma}} + J^{(1)\text{lint}}_{\sigma \tilde{\sigma}},  \\
 J^{(1)\text{ext}}_{\sigma \tilde{\sigma}} &= \sum\limits_{g_1 : \sigma \rightarrow \tilde{\sigma}\,\textrm{such that} \,g_1 \cap \calS_1 = \varnothing \, \textrm{or} \, g_1 \not\subset \overline{\calS}_1}  J^{(1)}_{\sigma \tilde{\sigma}} (g_1), 
 \end{aligned}
 \label{(2.11)}
\end{equation}
where $J^{(1)\text{sint}}$ contains terms of $J^{(1)\text{int}}$ whose graph is contained in a small block $\bar{b}^{(1)}$, and $J^{(1)\text{lint}}$ contains terms whose graph is contained in a large block $\bar{B}^{(1)}$. Then
\begin{equation}
H^{(1)} = H_0 + J^{(1)\text{ext}} + J^{(1)\text{sint}} + J^{(1)\text{lint}}. \label{(2.12)}
\end{equation}

The next step is to diagonalize $H_0 + J^{(1)\text{sint}}$ within small blocks $\bar{b}^{(1)}$. Let $O$ be the matrix that accomplishes this. It is a tensor product of matrices acting on the spin space for each small block (and identity matrices for spins elsewhere). Each block rotation affects only the spin variables internal to the block $\bar{b}^{(1)}$. Let $\bar{\bar{b}}^{(1)}$ be a one-step neighborhood of $\bar{b}^{(1)}$. The rotation depends on the spins in $\bar{\bar{b}}^{(1)} \setminus \bar{b}^{(1)}$ and on the random variables in $\bar{\bar{b}}^{(1)}$ only. The procedure here may seem overly complicated (after all, single site rotations could have been performed at the outset, simplifying the first step considerably). But we prefer to use a standard procedure so that the first step serves as a guide to later steps. The rotation produces a new effective Hamiltonian
\begin{equation}
H^{(1')} = O^{\text{tr}} H^{(1)} O = H_0^{(1')} + J^{(1')} + J^{(1)\text{lint}}, \label{(2.13)}
\end{equation}
where
\begin{equation}
H_0^{(1')} = O^{\text{tr}} (H_0 + J^{(1)\text{sint}}) O \label{(2.14)}
\end{equation}
is diagonal, and
\begin{equation}
J^{(1')} = O^{\text{tr}} J^{(1)\text{ext}} O = \sum_{g_1} O^{\text{tr}} J^{(1)} (g_1) O. \label{(2.15)}
\end{equation}
Note that $J^{(1)\text{lint}}$ is unaffected by the rotation. However, terms in $J^{(1)\text{ext}}$ that connect to small blocks are rotated. This necessitates an extension of the graph $g_1$ since transitions within a small block are produced. In effect, all the states in a small block can be thought of as a ``metaspin" taking $2^{|\bar{b}^{(1)}|} = 8$ values (the same as the number of spin configurations in $\bar{b}^{(1)}$). Because of the rotation, there is no canonical way of associating states with spin variables, so we will often use generic labels $\alpha, \beta$ for block states. Let $g_{1'}$ label the set of terms obtained from the matrix product $O^{\text{tr}} J^{(1)} (g_1) O$. Thus $g_{1'}$ specifies $\sigma, g_1, \tilde{\sigma}$ and we may write
\begin{equation}
J^{(1')}_{\alpha\beta} = \sum_{g_{1'}: \alpha \rightarrow \beta} J^{(1')}_{\alpha \beta} (g_{1'}), \label{(2.16)}
\end{equation}
where
\begin{equation}
J^{(1')}_{\alpha \beta} (g_{1'}) =  O^{\text{tr}}_{\alpha \sigma} J^{(1)\text{ext}}_{\sigma \tilde{\sigma}}(g_{1'}) O_{\tilde{\sigma}\beta}. \label{(2.16a)}
\end{equation}
Since the matrix elements of $O$ for any block are bounded by 1, we maintain the bound $J^{(1')}_{\alpha \beta} (g_{1'}) \le \gamma (\gamma / \varepsilon)^{|g_{1'}| - 1}/(|g_{1'}| - 1)!$, where $|g_{1'}| = |g_1|$, the size of the graph ignoring the rotation steps. In the first step, the rotation matrices are small (8 $\times$ 8), so the spin sums implicit in (\ref{(2.16)}) are controlled by the smallness of the couplings. As we proceed to later steps, we need to be sure that coupling terms are small enough to control sums over $\sigma, \tilde{\sigma}$ for larger rotation matrices.

\subsection{Expectations of Observables}\label{2.4}

\indent We are trying to prove estimates on expectations of local observables in eigenstates of $H$. For example, we would like to compute the expectation $\langle S^\mathrm{z}_0 \rangle_\alpha$ in any eigenstate $\alpha$. 
We are trying to prove (\ref{(1.4)}), which can be written as
\begin{equation}
\mathbb{E} \; \text{Av}_{\alpha}\; |\langle S^\mathrm{z}_0 \rangle_{\alpha}| = 1 - O(\varepsilon), \label{(2.17)}
\end{equation}
(recall that $\varepsilon = \gamma^{1/20}$). However, at this stage of our analysis, we only have approximate eigenfunctions given by the columns of $\Omega O$. As a warm-up exercise, then, let us prove the following result:
\begin{proposition}\label{prop:2.1}
Let $\varepsilon=\gamma^{1/20}$ be sufficiently small. Then  
\begin{equation}
\mathbb{E}\; \mathrm{Av}_{\alpha}\bigg| \sum_{\sigma, \tilde{\sigma}} (O^{\mathrm{tr}} \Omega^{\mathrm{tr}})_{\alpha\sigma} (S^\mathrm{z}_0)_{\sigma \tilde{\sigma}} (\Omega O)_{ \tilde{\sigma}\alpha}\bigg| = 1 - O(\varepsilon). \label{(2.18)}
\end{equation}
\end{proposition}
As we proceed to better and better approximate eigenfunctions, this bound will become (\ref{(2.17)}). 

\textit{Proof.} Of course, $S^\mathrm{z}_0$ is diagonalized in the $\sigma$-basis, so $(S^\mathrm{z}_0)_{\sigma \tilde{\sigma}} = \sigma_0 \delta_{\sigma_0 \tilde{\sigma}_0}$, and (\ref{(2.18)}) becomes
\begin{equation}
\mathbb{E} \; \text{Av}_{\alpha}\bigg|\Big| \sum_{\sigma} (O^{\text{tr}} \Omega^{\text{tr}})_{\alpha\sigma} \sigma_0 (\Omega O)_{\sigma\alpha}\Big|-1\bigg| \le O(\varepsilon). \label{(2.19)}
\end{equation}

Our construction depends on the collection of resonant blocks
$\bar{b}^{(1)}, \bar{B}^{(1)}$; let us call it $\calB$. Thus
(\ref{(2.19)}) is best understood by inserting a partition of unity $\sum_{\mathcal{B}} \chi_{\mathcal{B}}$ 
under the expectation, where $\chi_\calB$ is an indicator for the event that the set of resonant blocks is $\calB$.
Once this is done, we have two cases to consider: either 0 is in a resonant block, or it is not. If 0 is in a large block, there is actually no contribution because there is no rotation, so $\sigma_0 = \alpha_0 = \pm 1$. If 0 is in a small block, we can expect substantial mixing, leading to an expectation for $S^\mathrm{z}_0$ anywhere between -1 and 1, for any $\alpha$. Thus for an upper bound, we can replace the integrand of (\ref{(2.19)}) with an indicator $\mathbbm{1}_0(\calB)$ for the event that 0 lies in a small block. We have established that the probability that a site is resonant is bounded by $4\rho_0\varepsilon$. Due to the collar, 0 need not be a resonant site, but if it is not, then one of its two neighbors is; thus we get a bound of $12\rho_0\varepsilon$.

In the case where 0 is not in a resonant block, we rotate $S^\mathrm{z}_0$ :
\begin{equation}
\Omega^{\text{tr}} S^\mathrm{z}_0 \Omega = \sum_{n = 0}^{\infty} \frac{(\text{ad}\,A)^n}{n!} S^\mathrm{z}_0 \equiv \sum_{g_1} S^\mathrm{z}_0 (g_1). \label{(2.20)}
\end{equation}
This expansion is very much like the one derived for $J^{(1)}$; in particular it represents the expectation as a sum of local graphs. Note that the empty graph with $n = 0$ does not contribute, because it has modulus 1 and so disappears in (\ref{(2.19)}). As we have seen, the sum over $g_1$ converges geometrically, so the norm of the matrix $M = \sum_{|g_1| \ge 1} S^\mathrm{z}_0 (g_1)$ is bounded by $O(\gamma / \varepsilon)$. The rotation by $O$ can affect terms with $g_1$ reaching to a block, but the norm is preserved, so $\text{Av}_\alpha|(O^{\text{tr}}MO)_{\alpha\alpha}|$ is likewise bounded by $O(\gamma / \varepsilon)$. Thus the contribution from this case to (\ref{(2.19)}) is $O(\gamma / \varepsilon)$, uniformly in $\mathcal{B}$ (provided the set of blocks $\mathcal{B}$ does not contain 0). This completes the proof of (\ref{(2.19)}), and hence also (\ref{(2.18)}). \qed

It should be clear that a similar analysis can be performed to give an expansion for the approximate expectation of any local operator, such as products of spin operators $S^\mathrm{x}_i$, $S^\mathrm{y}_i$, or $S^\mathrm{z}_i$ at collections of sites $i$. If we consider the first-step connected correlation $\langle \mathcal{O}_i ; \mathcal{O}_j \rangle^{(1)}_\alpha$ for operators localized at or near $i, j,$ there will be a cancellation of terms except for graphs extending from $i$ to $j$. If we insert an indicator for the event that no more than half the ground between $i$ and $j$ is covered by resonant blocks, we obtain exponential decay. The probability of half coverage of $[i, j]$ by resonant blocks likewise decays exponentially (we will have to settle for weaker probability decay in later steps). Thus we see that
\begin{equation}
|\langle \mathcal{O}_i ; \mathcal{O}_j \rangle^{(1)}_\alpha| \le (c \gamma / \varepsilon)^{|i - j|/2}, \;\; \text{with probability}\;\; 1 - (c \varepsilon)^{|i -j|/2}. \label{(2.21)}
\end{equation}

\section{The Second Step}\label{3}

It will be helpful to illustrate our constructions in a simpler context before proceeding to the general inductive step. The second step is based on the same operations described above for the first step. However, complications ensue because multi-denominator graphs appear starting with second order perturbation theory; for example (\ref{(3.1)}) has an explicit denominator as well as denominator(s) in ${J}^{(1')}_{\sigma\tilde{\sigma}}(g_{1'})$. We deal with this by using a graph-based notion of resonance -- see (\ref{(3.2)}) below. The probability that a graph is resonant is controlled by means of a Markov inequality -- see (\ref{(3.5)}), (\ref{(3.6)}) below. The other new feature that appears in the second step is the distinction between ``short'' and ``long'' graphs and the resummation of long graphs -- see (\ref{(3.11)}) below. Both of these ideas will play a key role in maintaining uniformity of exponential decay rates in the general step.

\subsection{Resonant Blocks}\label{3.1}

Recall that we use a sequence of length scales $L_k = (\frac{15}{8})^k$, with graphs sized in the range $[L_{k - 1}, L_k)$ considered in the $k$\textsuperscript{th} step. So we will allow graphs of size 2 or 3 in the perturbation in the second step. Graphs that intersect small resonant blocks have been rotated, so they now produce transitions in the ``metaspin" space of the block. Still, it would be cumbersome to maintain a notational distinction between ordinary spins and metaspins, so we will use $\sigma, \tilde{\sigma}$ to label spin/metaspin configurations. Labeling of states in blocks is arbitrary, but we may choose a one-to-one correspondence between ordinary spin configurations in a block $\bar{b}^{(1)}$ and metaspins/states in $\bar{b}^{(1)}$.

Each graph $g_{1'}$ induces a change in spin/metaspin configuration in the sites/blocks of $g_{1'}$. For each $g_{1'}$ corresponding to a term of $J^{(1')}$, with $2 \le |g_{1'}| \le 3$, 
$\sigma \ne \tilde{\sigma}$, $g_{1'} \cap \calS_{1'} = \varnothing$, we define
\begin{equation}
A^{(2)\prov}_{\sigma \tilde{\sigma}} (g_{1'}) = \left| \frac{{J}^{(1')}_{\sigma\tilde{\sigma}}(g_{1'})}{E^{(1')}_{\sigma} - E^{(1')}_{\tilde{\sigma}}}\right|. \label{(3.1)}
\end{equation}
Here $E^{(1')}_{\sigma}$ denotes a diagonal entry of $H_0^{(1')}$. The graph $g_{1'}$ changes the spin/metaspin locally in 1, 2, or 3 sites/blocks; hence the energy difference in (\ref{(3.1)}) is local as well.These are ``provisional" $A^{(2)}$ terms because not all of them will be small enough to include in $A^{(2)}$. Note that intra-block terms with $g_{1'} \subset \overline{\calS}_1$ are in $J^{(1)\text{int}}$, so are not part of $J^{(1')}$ -- \textit{c.f.} (\ref{(2.11)}). But in contrast to \cite{Imbrie2014a}, we allow intra-block terms with $g_{1'} \cap \overline{\calS}_1^{\text{c}}\ne\varnothing$ to occur in (\ref{(3.1)}) -- this is made possible by the level-spacing assumption. Nevertheless, we only consider off-diagonal terms here; in general, diagonal terms will renormalize energies, but will not induce rotations directly. Note that energies $E^{(1')}_\sigma$ are given by unperturbed values $\sum h_i \sigma_i + \sum J_i \sigma_i \sigma_{i + 1}$ away from blocks, because corrections are second or higher order in $\gamma \, (|g_1| \ge 2)$, so no change in $H_0$ is implemented in (\ref{(2.12)}). Nontrivial changes in metaspin energies $E^{(1')}_\sigma$ arise only in small blocks $\bar{b}^{(1)}$, where the rotation (\ref{(2.14)}) generates a new diagonal matrix $H_0^{(1')}$.

We say that $g_{1'}$ from $\sigma$ to $\tilde{\sigma}$ is resonant in step 2 if $|g_{1'}|$ is 2 or 3, and if either of the following conditions hold:
\begin{equation}\label{(3.2)}
\begin{aligned}
\textrm{I}&. \quad |E^{(1')}_\sigma - E^{(1')}_{\tilde{\sigma}}| < \varepsilon^{|g_{1'}|}, \\
\textrm{II}&.  \quad A^{(2)\text{prov}}_{\sigma \tilde{\sigma}}(g_{1'}) > \frac{(\gamma / \varepsilon)^{|g_{1'}|}}{ (|g_{1'}| - 1)!^{2/9}} \; \text{with} \; |I(g_{1'})| \ge \tfrac{7}{8} |g_{1'}|. 
\end{aligned}
\end{equation}
Here, $I(g_{1'})$ is the smallest interval in $\mathbb{Z}$ covering all the sites or blocks $\bar{\bar{b}}^{(1)}$ that contain flips of $g_{1'}$, and $|I(g_{1'})|$ is the number of sites or blocks $\bar{\bar{b}}^{(1)}$ in $I(g_{1'})$
(\textit{i.e.} the number of blocks $\bar{\bar{b}}^{(1)}$ plus the number of sites not in such blocks).
Condition II graphs have few duplicated sites, which means most energies are independent -- this leads to good Markov inequality estimates. Graphs with $|I(g_{1'})| < \frac{7}{8} |g_{1'}|$ do not reach as far, so less decay is needed, and inductive estimates will be adequate. The combination will help us prove uniform bounds on the probability that $A^{(k)}$ fails to decay exponentially. 

We define the scale 2 resonant blocks. Examine the set of all step 2 resonant graphs $g_{1'}$. Note that we include in $g_{1'}$ information about the starting spin configuration $\sigma$, since $g_{1'}: \sigma \rightarrow \tilde{\sigma}$. We need to specify $\sigma$ on $I(g_{1'})$ plus one neighbor on each side of any flip of $g_{1'}$, because energies depend on $\sigma$ one step away from sites/blocks where flips occur. These graphs involve sites and small blocks $\bar{b}^{(1)}$. They do not touch large blocks $B^{(1')}$, because of the above restriction to graphs such that $g_{1'} \cap \calS_{1'} = \varnothing$. The set of sites/blocks that belong to resonant graphs $g_{1'}$ are decomposed into connected components. The result is defined to be the step 2 resonant blocks $B^{(2)}$. They do not touch large blocks $B^{(1')}$. Small blocks $\bar{b}^{(1)}$ can be linked to form a step 2 block, but unlinked small blocks are not held over as scale 2 blocks.

We need to reorganize the resonant blocks produced so far, to take into account the presence of new resonant blocks and to define new small blocks $b^{(2)}$. The result is a collection of small blocks $b^{(i)}$ for $i = 1, 2$ and a leftover region $\calS_{2'}$; they must satisfy the following diameter and separation conditions for $i, j \le 2$:
\begin{align}\label{(3.3)}
\text{diam}(b^{(i)}) &< L_i; \nonumber \\
\text{dist}(b^{(i)}_1, b^{(j)}_2) &> d_m \equiv \exp(L^{1/2}_{m + m_0}), \;\; \text{if} \;\;\text{min}\lbrace |b_1^{(i)}|, |b_2^{(j)}|\rbrace \in [L_{m - 1}, L_m);\nonumber \\
\text{dist}(b^{(i)}, \calS_{2'}) &> d_m\;\; \text{if} \;\;|b^{(i)}| \in [L_{m - 1}, L_m).
\end{align}
Here we define $|b^{(2)}|$ to be the number of sites or blocks $\bar{b}^{(1)}$ in $b^{(2)}$. 
However, any block $b^{(2)}$ with fewer than two sites/blocks is considered to have size 2. (We link this to  the minimum graph size for step 2 because that affects probability bounds, which in turn determine workable separation distances.) 
It is easy to see that there is a unique way to decompose the complete resonant region into a maximal set of small blocks satisfying (\ref{(3.3)}), plus the leftover region
$\calS_{2'}$. We have already instituted proximity connections on scale $d_1$; now we introduce new connections on scale $d_2$ if required by (\ref{(3.3)}).
If one of the resulting blocks fails to satisfy the diameter condition, it is transferred to $\calS_{2'}$.  Since the $B^{(2)}$ blocks were not introduced until the second step, this process may force some of the step 1 small blocks $b^{(1)}$ into $\calS_{2'}$ or into some $b^{(2)}$.  
 
Let $\calS_2$ denote $\calS_{2'}$ plus the small blocks $b^{(2)}$.
We add a 3-step collar to $\calS_2$. 
(As in step 1, collars serve to contain the ``troublesome'' graphs and define regions for block diagonalization.)
Then $\overline{\calS}_2$ is the collared version of $\calS_2$, and its components are the collared small blocks $\bar{b}^{(2)}$ and large blocks $\bar{B}^{(2')}$. The union of the $\bar{B}^{(2')}$ is denoted $\overline{\calS}_{2'}$, and then each $B^{(2')} \equiv \calS_2 \cap \bar{B}^{(2')}$.

The blocks defined above can be thought of as connected clusters for a generalized percolation problem. The following proposition provides control on the decay of the associated connectivity function.
\begin{proposition}\label{prop:3.1}
Let $P_{ij}^{(2)}$ denote the probability that $i$, $j$ lie in the same block $B^{(2)}$. 
For a given $\nu, C$, let $\varepsilon = \gamma^{1/20}$ be sufficiently small, and
assume  \textbf{LLA($\nu, C$)}. Then
\begin{equation}
P^{(2)}_{ij} \le (c \rho_1 \varepsilon^s)^{(|i - j|^{(1)}\vee 2)/2}. \label{(3.7)}
\end{equation}
Here $s=\frac{2}{7}$, and $|i - j|^{(1)}$ is a notation for the distance from $i$ to $j$ with blocks $\bar{\bar{b}}^{(1)}$ contracted to points.
\end{proposition}

\textit{Proof.} There must be a collection of resonant graphs $g_{1'}$ connecting $i$ to $j$. However, because of dependence, we cannot simply take the product of the probabilities for each graph. As in \cite{Imbrie2014a}, we find a sequence of non-overlapping graphs which combine to cover at least half the distance from $i$ to $j$. Here distance is measured in the metric $|i - j|^{(1)}$, in which small blocks $\bar{\bar{b}}^{(1)}$ are contracted to points. Let $g_{1',1}$ be the graph covering the site $i$ and extending farthest to the right. Then let $g_{1',2}$ be the graph that extends farthest to the right from $g_{1',1}$ (without leaving a gap). Continue until the site $j$ is covered. It should be clear that the odd graphs do not overlap one another; likewise the even graphs are non-overlapping. (Any overlap would mean the in-between graph could have been dropped.) We may bound the probability of the whole collection of graphs by the geometric mean of the probabilities of the even and odd subsequences of $\{g_{1',k}\}$. As the complete sequence extends continuously from $i$ to $j$, we will obtain exponential decay in the distance from $i$ to $j$ (but losing a factor of 2 in the rate due to the geometric mean).

The above construction reduces the problem of bounding $P_{ij}^{(2)}$ to the estimation of resonance probabilities for cases I and II in (\ref{(3.2)}). With $|g_{1'}| = 2$, there is no case I since $\sigma = \tilde{\sigma}$ if the two flips are at the same site (if that site is in a $\bar{b}^{(1)}$, the term is internal to $\bar{b}^{(1)}$ and so is in $J^{(1)\text{sint}}$, not $J^{(1')}$). With $|g_{1'}| = 3$, $|I(g_{1'})| = 1, 2$, or 3. If $|I(g_{1'})| = 1$ or 2, we are in case I. Let $i$ be the site where $\sigma \ne \tilde{\sigma}$. If $i$ is not in a block $\bar{b}^{(1)}$, then as explained above, the energies are given by their unperturbed values, so the energy difference from the flip at $i$ is $\pm 2h_i + \text{const}$. The probability can be bounded by $\rho_1 \varepsilon^{s|g_{1'}|}$, for some constant $\rho_1$ depending only on $\rho_0$, the bound on the probability densities.
Alternatively, a bounded probability density implies a bound on the $-s = -\frac{2}{7}$ moment of $h_i$:
\begin{equation}
\sup\limits_{a \in \mathbb{R}}\; \mathbb{E}\; |2h_i - a|^{-s} \le \rho_1, \label{(3.3a)}
\end{equation}
which leads to the same estimate via a Markov inequality.
If $\sigma$ differs from $\tilde{\sigma}$ only in a block, then the energy difference $E^{(1')}_\sigma - E^{(1')}_{\tilde{\sigma}}$ is a difference of block energies. Here we need to make a similar assumption
\begin{equation}
\sup\limits_{a \in \mathbb{R}} \; \mathbb{E} \; |E^{(1')}_\sigma - E^{(1')}_{\tilde{\sigma}} - a|^{-s} \le \rho_1. \label{(3.4)}
\end{equation}
In general, we need to assume there is a constant $\rho_1$ such that the energy differences in resonant blocks $\bar{b}^{(k)}$ have $-s$ moments bounded as in (\ref{(3.4)}), with a bound like $\rho_1^{L_k}$. This is equivalent to a statement about H\"{o}lder continuity of block energy differences, with bounds exponential in the volume of the block. 
This follows from our level-spacing assumption \textbf{LLA($\nu, C$)} -- more details will be given in the general step.
Given (\ref{(3.3a)}), (\ref{(3.4)}) we can say that the case I probabilities are all bounded by $(\rho_1 \varepsilon^s)^{|g_{1'}|}$.

For case II, we have graphs with 2 or 3 flips, all at different sites/blocks. In general, if a graph has $k$ flips at $k$ different sites/blocks, this gives rise to a tree graph of energy denominators on $k + 1$ ``vertices," \textit{i.e.} the $k + 1$ spin configuration energies linked by $k$ denominators $E^{(1')}_\sigma - E^{(1')}_{\tilde{\sigma}}$. Each link to a new site introduces a new random variable into the energy, so each denominator is independent. As a result, a Markov inequality with (\ref{(3.3a)}), (\ref{(3.4)}) can be used to bound the probability. For example, consider a particular $g_{1'}$ with $|g_{1'}| = 2$ and no blocks. We estimate as follows:
\begin{equation}
P \left(A^{(2)\text{prov}}_{\sigma\tilde{\sigma}} > (\gamma/\varepsilon)^2\right)  \le \mathbb{E}\; \frac{(A^{(2)\text{prov}}_{\sigma\tilde{\sigma}})^s }{ (\gamma / \varepsilon)^{2s}} \label{(3.5)}
 \le \varepsilon^{2s}\; \mathbb{E}\; \frac{1}{|2h_i + a|^s |2h_i + 2h_{i + 1} + b|^s}.
\end{equation}
Here $A^{(2)\text{prov}}_{\sigma \tilde{\sigma}}$ is given by (\ref{(3.1)}) with $J^{(1)}_{\sigma\tilde{\sigma}}$ having the structure $A(i) J(i + 1)$, and $a$, $b$ are $h$-independent constants determined by the exchange interactions with neighboring spins. We may integrate over $h_{i + 1}$ with $h_i$ fixed, using (\ref{(3.3a)}); then a second application of (\ref{(3.3a)}) bounds the right-hand side of (\ref{(3.5)}). In general, we find that
\begin{equation}
P \left(A^{(2)\text{prov}}_{\sigma\tilde{\sigma}} > (\gamma/\varepsilon)^{|g_{1'}|}\right) \le (\rho_1 \varepsilon^s)^{|g_{1'}|}. \label{(3.6)}
\end{equation}

It is worth noting that under nonresonance conditions from this step (the negation of (\ref{(3.2)})) and nonresonance conditions inherited from the first step, all $A^{(2)\text{prov}}_{\sigma \tilde{\sigma}}(g_{1'})$ have good bounds, not just the ``straight" graph of condition II. For example, the three flip graph with one repeated site has two denominators $\ge \varepsilon$ and one $\ge \varepsilon^3$. The overall bound is $\gamma^3/\varepsilon^5 = \gamma^{11/4}$, which is adequate since we
are looking for decay like $\gamma^{|I(g_{1'})|}$ and $|I(g_{1'})|=2$.
Similar estimates will work for the ``crooked" non-condition II graphs for the general step.

As explained above, we may combine the estimates $(\rho_1 \varepsilon^s)^{|g_{1'}|}$ on the probabilities of case I and II graphs to obtain the bound (\ref{(3.7)}); note that $|g_{1'}| \ge 2$. This completes the proof. \qed

Note that as long as $i, j$ are not in the same block, $|i - j|^{(1)} \ge |i - j|/5$ due to the separation conditions. (For large enough $m_0$, blocks $\bar{\bar{b}}^{(1)}$ are much farther apart than their diameters. So the worst case for this inequality is for $i$ adjacent to the block containing $j$.) In later steps, more stringent separation conditions will ensure that $|i - j|^{(k)}$ remains comparable to $|i - j|$. This is important because when we sum over collections of non-overlapping resonant graphs covering half the distance from $i$ to $j$, we have combinatoric factors $c^{|i - j|}$, and the decay in $|i - j|^{(1)}$ is adequate to control them. The combinatoric factors come from sums over $g_{1'}$, but these include sums over initial and final spin configurations in the blocks $\bar{b}^{(1)}$ touched by $g_{1'}$. Thus the ``combinatoric volume" is the full $|i -j|$. Note that as discussed earlier, there are factorials in $|g_{1'}|$ to consider, but since $|g_{1'}| \le 3$ this is not an issue we need to worry about here.

Let us define $Q^{(2)}_{ij}$ to be the probability that $i$, $j$ lie in the same small block $\bar{b}^{(2)}$. In the $P^{(2)}_{ij}$ bound, we considered only resonant graphs new to the second step. Here we allow new resonances (for which $2 \le |g_{1'}| \le 3$),
as well as old resonances. But keep in mind that isolated $b^{(1)}$ are no longer present in $B^{(2)}$. Hence if there is no $g_{1'}$, there must be at least two $b^{(1)}$ blocks. Either way, the probability is bounded by $(c \rho_1 \varepsilon^s)^2$. Recall that we have imposed the condition that the diameter of $b^{(2)}$ is $< L_2$, so we have a maximum diameter of 3. Thus
\begin{equation}
Q^{(2)}_{ij} \le (c \rho_1 \varepsilon^s)^2 \mathbbm{1}_{|i - j| \le 3}. \label{(3.8)}
\end{equation}
As we proceed to later steps, $P^{(k)}_{ij}$ will maintain uniform exponential decay, but $Q^{(k)}_{ij}$, being more loosely connected, will decay more slowly, like $\varepsilon^{O(k^2)}$ with $|i - j| \le 4L_k$. Still, the decay is faster than any power of $|i - j|$, and it is sufficient to ensure that small blocks are unlikely. (Note that when $k \rightarrow \infty$, all blocks will be small.)

\subsection{Perturbation in the Nonresonant Couplings}\label{3.2}

We group terms in $J^{(1')}$ into ``perturbative" and ``resonant" categories and write
\begin{equation}
J^{(1')} = J^{(1')\text{per}} + J^{(1')\text{res}}, \label{(3.9)}
\end{equation}
where $J^{(1')\text{per}}$ contains terms $g_{1'}: \sigma \rightarrow \tilde{\sigma}$ with $2 \le |g_{1'}| \le 3$, $\sigma \ne \tilde{\sigma}$, and $g_{1'} \cap \calS_2 = \varnothing$ (meaning all the sites/blocks in $g_{1'}$ are in $\calS_2^{\text{c}}$). 
Note that unlike \cite{Imbrie2014a}, we allow intra-block terms in $J^{(1')\text{per}}$; using  \textbf{LLA}$(\nu, C)$, they are manageable in (\ref{(3.1)}) and hence also here.
Graphs connected to the resonant region $\calS_2$, large graphs $(|g_{1'}| \ge 4)$ and diagonal terms $(\sigma = \tilde{\sigma})$ form $J^{(1')\text{res}}$. We put
\begin{equation}
A^{(2)}_{\sigma\tilde{\sigma}} = \sum_{g_{1'}: \sigma \rightarrow \tilde{\sigma}} A^{(2)}_{\sigma\tilde{\sigma}} (g_{1'}) = \sum_{g_{1'}: \sigma \rightarrow \tilde{\sigma}} \frac{J^{(1')\text{per}}_{\sigma\tilde{\sigma}}(g_{1'})}{E^{(1')}_\sigma - E^{(1')}_{\tilde{\sigma}}}. \label{(3.10)}
\end{equation}

\textit{Long and short graphs; jump transitions.} 
We say a graph $g_{1'}$ is \textit{long} if $|g_{1'}| > \frac{8}{7} |I(g_{1'})|$. Otherwise, it is \textit{short}.
We will need to resum terms with long graphs, for given initial and final spin configurations $\sigma, \tilde{\sigma}$ and a given interval $I = I(g_{1'})$.
The data $\{\sigma, \tilde{\sigma}, I\}$ determine a \textit{jump transition}. 
Long graphs are extra small -- for example, see the discussion following (\ref{(3.6)}) -- so for probability estimates we do not need to keep track of individual graphs, and we can take the supremum over the randomness. Let $g_{1''}$ denote either a short graph from $\sigma$ to $\tilde{\sigma}$ or a  jump transition taking $\sigma$ to $\tilde{\sigma}$ on an interval $I$. The length of $g_{1''}$ is defined to be $|g_{1''}|=|I|\vee \tfrac{7}{8}L_1$. The jump transition represents the collection of all long graphs from $\sigma$ to $\tilde{\sigma}$ for which $I(g_{1'})=I$. Thus we define
\begin{equation}
A^{(2)}_{\sigma \tilde{\sigma}} (g_{1''}) = 
\begin{cases}
A^{(2)}_{\sigma \tilde{\sigma}} (g_{1'}), \text{if} \, g_{1''} = g_{1'}, \text{a short graph};\\
\sum\limits_{\text{long}\, g_{1'}: \sigma \rightarrow \tilde{\sigma}} A^{(2)}_{\sigma \tilde{\sigma}} (g_{1'}), \text{if} \, g_{1''} \, \text{is long.}
\end{cases}
\label{(3.11)}
\end{equation}

We may now define the basis-change operator $\Omega^{(2)} = \exp (-A^{(2)})$ and the new effective Hamiltonian
\begin{equation}
H^{(2)} = \Omega^{(2)\text{tr}} H^{(1')} \Omega^{(2)}. \label{(3.12)}
\end{equation}
Recalling that $H^{(1')} = H_0^{(1')} + J^{(1')} + J^{(1)\text{lint}} \; \text{with} \; H^{(1')}_{0, \sigma \tilde{\sigma}} = E^{(1')}_{\sigma} \delta_{\sigma \tilde{\sigma}}$, we obtain
\begin{align}
H^{(2)} & = H^{(1')}_0 + J^{(1')\text{res}} + J^{(1)\text{lint}} + \sum\limits_{n = 1}^{\infty} \frac{n}{(n + 1)!} (\text{ad}\,A^{(2)})^n J^{(1')\text{per}} +  \sum\limits_{n = 1}^{\infty} \frac{(\text{ad}\,A^{(2)})^n}{n!}  J^{(1')\text{res}} \notag \\
& = H^{(1')}_0 + J^{(1')\text{res}} + J^{(1)\text{lint}} + J^{(2)}. \label{(3.13)}
\end{align}
Since $J^{(1')}$ is second or third order in $\gamma$, all terms of $J^{(2)}$ are fourth order or higher.

The local structure of $J^{(2)}$ arises as before because $A^{(2)}, J^{(1')}$ are both sums of local operators. In particular, $[A^{(2)}(g_{1'}), J^{(1')} (\tilde{g}_{1'})] = 0$ if dist$(g_{1'}, \tilde{g}_{1'}) > 1$. (In later steps, the energies will receive new terms manifesting couplings over greater distances, and then a greater distance will be required for commutativity.) Suppressing spin indices, we have, for example,
\begin{equation}
(\text{ad}\,A^{(2)})^n J^{(1')\text{per}} = \sum\limits_{g_{1', 0}, \ldots, g_{1', n}} \quad \big(\text{ad}\,A^{(2)}(g_{1', n})\big) \cdots \big(\text{ad}\,A^{(2)} (g_{1', 1})\big) J^{(1')} (g_{1', 0}). \label{(3.14)}
\end{equation}
When summing over $g_{1', p}$ with $\text{dist}(g_{1', p}, \{g_{1', 0}, \ldots, g_{1', p - 1}\}) \le 1$, there are no more than $3p + 4$ choices for the starting site/block for $g_{1', p}$.
(The maximum number of sites/blocks in $\{g_{1', 0}, \ldots, g_{1', p - 1}\}$ is $3p$, and there
are up to three additional choices on the left and one on the right that can lead to a nonvanishing commutator.)
Hence the sums over the initial points for the walks $g_{1', 0}, \ldots, g_{1', p}$ lead to a combinatoric factor no greater than $n!c^{|g_2|}$, for some constant $c$. Here $g_2$ is the walk in spin configuration space giving the sequence $g_{1', 0}, \ldots, g_{1', n}$, and $|g_2|$ is the sum of the lengths of the sub-walks. The length of a graph is the number of transitions (or steps, if we think of a graph as a walk in spin configuration space). Blocks $\bar{b}^{(1)}$ do not affect graph lengths, but they do affect the counting of graphs, because the number of possible transitions in a block grows exponentially in the size of the block. 
For example, a block $\bar{b}^{(1)}$ of three sites counts as one unit of graph length, but there are $2^3(2^3-1)$ possible transitions in the block.
Nevertheless, separation conditions ensure that the length of the region covered by $g_2$ is no greater than a fixed multiple of $|g_2|$. Altogether, the sum over $g_2$ (including its subgraphs $g_{1', 0}, \ldots, g_{1', n}$ and its initial spin configuration) is controlled by a combinatoric factor $n! c^{|g_2|}$. This is acceptable since we have factors of $1/n!$ in (\ref{(3.13)}), and bounds on $A^{(2)}, J^{(1')}$ which decay exponentially in each $|g_{1'}|$. (Recall that $|A^{(2)}_{\sigma\tilde{\sigma}} (g_{1'})| \le (\gamma / \varepsilon)^{|g_{1'}|}/ (g_{1'} - 1)!^{2/9}$ from nonresonance conditions -- the negation of (\ref{(3.2)}) -- and the discussion following (\ref{(3.6)}); $J^{(1')}_{\sigma\tilde{\sigma}} (g_{1'})$ is bounded in (\ref{(2.10)}).)

We give a graphical representation for the new interaction 
\begin{equation}
J^{(2)}_{\sigma\tilde{\sigma}} = \sum\limits_{g_2 : \sigma \rightarrow \tilde{\sigma}} J^{(2)}_{\sigma\tilde{\sigma}} (g_{2}), \label{(3.15)}
\end{equation}
and from the abovementioned bounds,
\begin{equation}
|J^{(2)}_{\sigma \tilde{\sigma}} (g_2)| \le \gamma (\gamma / \varepsilon)^{|g_2| - 1}/ (g_2!)^{2/9}. \label{(3.16)}
\end{equation}
Here we introduce a notation $g_2!$ for $n!$ times the product of $(|g_{1', p}| - 1)!$ over the subgraphs of $g_2$. (We did not need to be concerned about such factors for $A$'s, for which $|g_{1'}| \le 3$, but long graphs can occur in $J^{(1')\text{res}}$.) Terms in $J^{(2)}$ are short one denominator (compared to $|g_2|$, the number of transitions), and this accounts for the form of the bound (\ref{(3.16)}). Note that the graph $g_2$ transitions from $\sigma$ to $\tilde{\sigma}$, so it is actually specifying a particular entry of the matrix $J^{(2)}_{\sigma \tilde{\sigma}} (g_2)$, with the others equal to zero. 

\subsection{Small Block Diagonalization}\label{3.3}

In the last section we defined the small blocks that will be diagonalized here. They have core diameter $< L_2$ (3 or less) and a 3 step collar. The core can be formed from two 1-site blocks $b^{(1)}$, or from a 2- or 3-site block $b^{(1)}$ or $b^{(2)}$. These were the cases considered in the bound (\ref{(3.8)}) on the probability of a block $\bar{b}^{(2)}$ containing $i, j$. Let us reorganize the interaction terms in (\ref{(2.9)}) as follows:
\begin{align}
J^{(1')\text{res}} + J^{(1)\text{lint}} + J^{(2)} & = J^{(2)\text{ext}} + J^{(2)\text{int}} \notag \\
& = J^{(2)\text{ext}} + J^{(2)\text{sint}} + J^{(2)\text{lint}}. \label{(3.17)}
\end{align}
Here $J^{(2)\text{int}}$ contains terms whose graph intersects $\calS_2$ and is contained in $\overline{\calS}_2$. Then $J^{(2)\text{lint}}$ includes terms of $J^{(2)\text{int}}$ that are contained in large blocks $\bar{B}^{(2)}$, and $J^{(2)\text{sint}}$ includes terms of $J^{(2)\text{int}}$ that are contained in small blocks $\bar{b}^{(2)}$, as well as second-order diagonal terms for sites in $\calS^\text{c}_2$. All remaining terms of $J^{(2)}$ and $J^{(1')\text{res}}$ are included in $J^{(2)\text{ext}}$. This includes terms fourth order and higher in $J^{(1')\text{res}}$ that did not participate in the step 2 rotation (\ref{(3.9)}), (\ref{(3.10)}).

Now we let $O^{(2)}$ be the matrix that diagonalizes $H_0^{(1')} + J^{(2)\text{sint}}$. By construction, $J^{(2)\text{sint}}$ acts locally within small blocks, so $O^{(2)}$ is a tensor product of small-block rotations. Then define
\begin{align}
H^{(2')} & = O^{(2)\text{tr}} H^{(2)} O^{(2)} \notag \\
& = O^{(2)\text{tr}} (H^{(1')}_0 + J^{(2)\text{ext}} + J^{(2)\text{sint}} + J^{(2)\text{lint}}) O^{(2)}  \notag\\
& = H^{(2')}_0 + J^{(2')} + J^{(2)\text{lint}}, \label{(3.18)}
\end{align}
where
\begin{equation}
H^{(2')}_0 = O^{(2)\text{tr}} (H^{(1')}_0 + J^{(2)\text{sint}}) O^{(2)} \label{(3.19)}
\end{equation}
is diagonal, and
\begin{equation}
J^{(2')} = O^{(2)\text{tr}} J^{(2)\text{ext}} O^{(2)}. \label{(3.20)}
\end{equation}
The diagonal elements of $H_0^{(2')}$ define the energies $E_\sigma^{(2')}$; by construction they incorporate block energies and second-order energies for sites in $\calS_2^{\text{c}}$.

The new interaction has an expansion analogous to (\ref{(2.16)}):
\begin{equation}
J^{(2')}_{\alpha \beta} = \sum\limits_{g_{2'} : \alpha \rightarrow \beta} J ^{(2')}_{\alpha \beta} (g_{2'}), \label{(3.21)}
\end{equation}
where $g_{2'}$ specifies rotation matrix elements $O^{(2)\text{tr}}_{\alpha \sigma}, O^{(2)}_{\tilde{\sigma} \beta}$, and a graph $g_2$ or $g_{1'}$ transitioning from $\sigma$ to $\tilde{\sigma}$. We specify that the length $|g_{2'}|$ is the same as that of the pre-rotation graph, even though part of the graph may be covered by a block $\bar{b}^{(2)}$.

Let us define cumulative rotations
\begin{equation}
\begin{aligned}
R^{(1')} &= \Omega^{(1')} O^{(1)} , \\
R^{(2')} &= R^{(1')} \Omega^{(2)} O^{(2)} .
\end{aligned}
\label{(3.22)}
\end{equation}
Then it should be clear that the arguments used to prove Proposition \ref{prop:2.1}  will allow us to obtain a similar result for the eigenfunctions approximated in step 2:
\begin{proposition}\label{prop:3.2}
For a given $\nu, C$ let $\varepsilon = \gamma^{1/20}$ be sufficiently small, and assume  \textbf{LLA}$(\nu, C)$. Then
\begin{equation}
\mathbb{E} \; \mathrm{Av}_{\alpha}  \bigg|\sum\limits_{\sigma, \tilde{\sigma}} R^{(2')\mathrm{tr}}_{\alpha \sigma} S^\mathrm{z}_0 R^{(2')}_{\tilde{\sigma}\alpha}\bigg| = 1 - O(\varepsilon). \label{(3.23)}
\end{equation}
\end{proposition}
\textit{Proof.} The graphical expansions produced by commutators with $S^\mathrm{z}_0$ are very much like the ones we have been working with. We need to control the probability that 0 is in a small block $\bar{b}^{(1)}$ or $\bar{b}^{(2)}$, that is, $Q^{(1)}_{00} + Q^{(2)}_{00}$. From step 1, we know $Q^{(1)}_{00}$ is $O(\rho_1 \varepsilon)$, and from (\ref{(3.8)}) we have that $Q^{(2)}_{00}$ is $O(\rho_1 \varepsilon^s)^2$. Thus we obtain (\ref{(3.23)}). Likewise we can prove a bound analogous to (\ref{(2.21)}) for connected correlations; further details on this will be left to the general step.  \qed

\section{The General Step}\label{4}

\begin{figure}[h]
\centering
\includegraphics[width=.8\textwidth]{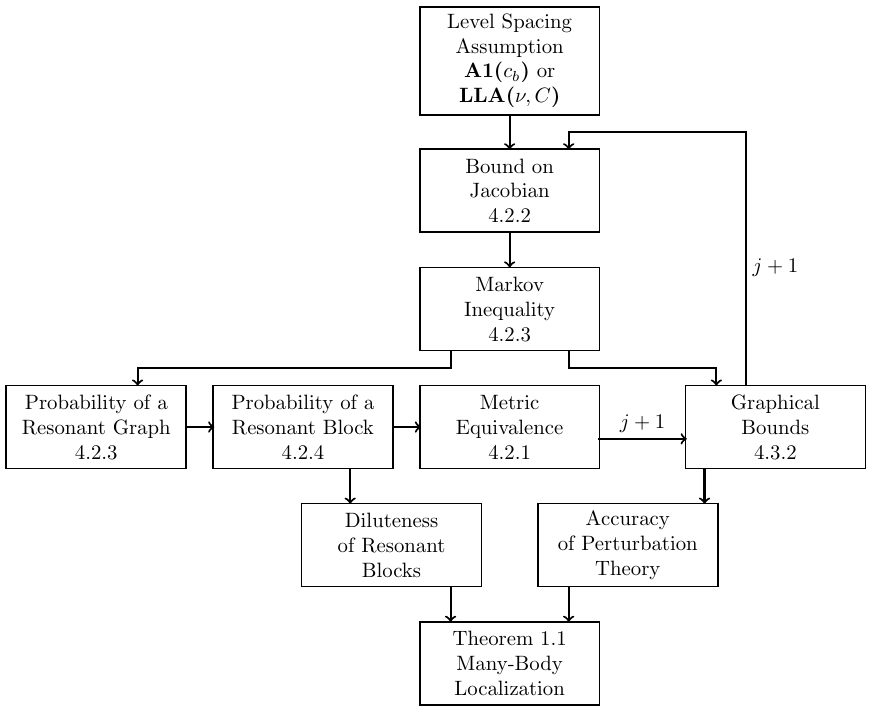}
\caption{An example of an inductive construction. 
An arrow with the notation ``$j+1$'' indicates that a result from step $j$ is assumed when making the argument in step $k = j+1$.\label{flow}}
\end{figure}

Our presentation of the $k^{\text{th}}$ step follows the same plan as the first two steps. In Section \ref{4.1}, we lay out the inductive bounds that need to be proven. In Section \ref{4.2}, resonant blocks are defined, and probability estimates are made to ensure their diluteness. Graphical estimates leading to inductive control over interaction terms are presented in Section \ref{4.3}. There are dependencies between these two sections, insofar as estimates and constructions from previous steps are used when needed -- see Figure \ref{flow}. 
Control over the counting of graphs is used throughout; it is presented in Subsection \ref{4.2.1}. We use a stronger form of the level-spacing assumption in Chapter \ref{4} -- it will be relaxed to \textbf{LLA}($\nu,C$) in Chapter \ref{5}. It leads to control over the Jacobian for the change of variables between uncorrected and corrected versions of energy differences. A Markov inequality can then be used to show that each graph obeys an exponential bound with a high probability; the probability of failure of the bound is also exponentially small. These probabilities feed into  generalized percolation estimates on the connectivity functions for resonant blocks. 
Percolation aspects feed into necessary results on metric equivalence: the lack of graphical decay across resonant blocks means that decay needs to be measured in a metric where blocks are contracted to points.  The equivalence of this metric with the usual one is demonstrated in Subsection \ref{4.2.1}.

\subsection{Starting Point for the $k^{\text{th}}$ Step}\label{4.1}

Let $j = k - 1$. We describe the situation as it stands after $j$ steps; thus this section
describes an inductive hypothesis that has been checked up through $j=2$. 

We have small blocks $b^{(j)}$ with diameter $<L_j$ that are well-separated from other parts of $\calS_j$ as per (\ref{(3.3)}). Couplings in $J^{(j')}$ are $O(\gamma^{L_j})$, which will be sufficient to control sums over states in $\bar{b}^{(j)}$, since their number is no more than exponential in $|\bar{b}^{(j)}| $. Rotations have been performed in each $\bar{b}^{(j)}$ to diagonalize the Hamiltonian there up to terms of order $L_j$. Collar neighborhoods of width $L_j-1$ ensure that none of the couplings expanded in the $j$\textsuperscript{th} step reach into the large blocks $B^{(j')}$. At each stage we prove a bound
\begin{equation}
P^{(j)}_{xy} \le (c\rho_1 \varepsilon^s)^{(|x - y|^{(j-1)}\vee L_{j-1})/8} \label{(4.1)}
\end{equation}
on the probability that $x, y$ lie in a block $B^{(j)}$. Here $|x - y|^{(j)}$ refers to the distance from $x$ to $y$ with blocks $\bar{\bar{b}}^{(m)}$ contacted to points for $m \le j$.

Graphs $g_{j'}$ are \textit{multigraphs} (or multiscale graphs), each of which is based on a sequence of subgraphs $g_{(j-1)'}$ from the previous scale. Each of the subgraphs
$g_{(j-1)'}$ is based in turn on a sequence of further subgraphs $g_{(j-2)'}$. Continuing in this fashion, we obtain for each $i < j$ a sequence of \textit{level i subgraphs} $g_{i'}$.
When unwrapped down to the starting level, we obtain a sequence of spin flips; thus one may think of $g_{j'}$ as a walk in the space of spin configurations. (The collection of graphs $g_{j'}$ has to be enlarged to take into account non-local terms in the energies $E^{(j')}_{\sigma}$ -- specifically the way such terms mediate commutator connections between graphs. This will be discussed in detail in Subsection \ref{4.3.2}.) When unwrapped to the first scale, we obtain spatial graphs $g^{\text{s}}_{j'}$ 
of spin/metaspin flips and associated
denominator graphs $g^\text{d}_{j'}$. Resummed sections appear as jump steps with no denominators. Rotation matrix elements introduce intra-block ``flips" of the metaspin variable labeling all the states in the block.

We inherit bounds from the $j$\textsuperscript{th} step:
\begin{equation}
|A^{(j)}_{\sigma \tilde{\sigma}} (g_{(j - 1)''})| \le
\begin{cases}
(\gamma / \varepsilon)^{|g_{(j - 1)''}|} / (g_{(j - 1)''}!)^{2/9}, \;\text{in general}; \\
\gamma^{|g_{(j - 1)''}|}, \;\text{if} \; g_{(j - 1)''} \; \text{is a jump step}. 
\end{cases}
\label{(4.2)}
\end{equation}
Here we make use of an inductive formula for the factorials that appear in our procedure:
\begin{equation}\label{(4.3)}
g_{(j - 1)''}! \equiv 
\begin{cases}
1, \text{ if } g_{(j - 1)''} \text{ is a jump step};\\
n! \prod_{p = 0}^n g_{(j - 2)'', p}!, \text{ otherwise}.
\end{cases}
\end{equation}
Here $g_{(j - 2)'', 0} \ldots, g_{(j - 2)'', n}$ are the subgraphs of $g_{(j - 1)''}$ -- see (\ref{(3.14)}). 
Thus the factorial of a graph at a given level is defined recursively in terms of the factorials of its subgraphs.
(In the first step, there are no subgraphs or jump steps, so $g_{1'}! \equiv n!$, corresponding to $(\text{ad}\,A(i_n)) \cdots (\text{ad}\,A(i_1)) J^{(0)} (i_0)$, \textit{c.f.} (\ref{(2.8)})).
As one unwraps the graph, factorials from earlier scales accumulate -- but the process stops whenever one reaches a jump step, for which $g_{i''}! \equiv 1$. The idea is that the ad expansion generates a factor of $1/n!$, which is available to help control graphical sums. Jump steps correspond to sums of graphs; the factorials have already been ``used up'' in controlling those sums, so they do not appear anymore in bounds such as (\ref{(4.2)}).

In a similar fashion, the length $|g_{(j - 1)''}|$ is defined to be the sum of the lengths of its subgraphs, if $g_{(j - 1)''}$ is not a jump step. The length of a jump step on an interval $I$ is defined for any $i$ to be
\begin{equation}
|g_{i''}| = |I| \vee \tfrac{7}{8}L_i, \label{(4.4)}
\end{equation}
where $|I|$ is the length of $I$ in the metric $|x - y|^{(i)}$ in which blocks $\bar{\bar{b}}^{(\tilde{i})}$ on scale $\tilde{i} \le i$ are contracted to points. 

At each level we have the ``core'' small blocks $b^{(i)}$, where resonant graphs occur. Adding a collar of width $L_{i}-1$, we obtain $\bar{b}^{(i)}$, where rotations are performed. Adding a second collar of width $\frac{15}{14}L_{i-1}$ about $\bar{b}^{(i)}$, we obtain $\bar{\bar{b}}^{(i)}$, which is the region of dependence of the energies of $\bar{b}^{(i)}$ after the rotations. All these distances are measured in the metric $|x-y|^{(i-1)}$.

Interaction terms $J^{(j')}$ and $J^{(j)\text{lint}}$ have graphical expansions as in (\ref{(3.21)}):
\begin{equation}
J^{(j')}_{\sigma \tilde{\sigma}} = \sum_{g_{j'}: \sigma \rightarrow \tilde{\sigma}} J^{(j')}_{\sigma \tilde{\sigma}} (g_{j'}), \label{(4.5)}
\end{equation}
with bounds
\begin{equation}
|J^{(j')}_{\sigma \tilde{\sigma}} (g_{j'})| \le \gamma(\gamma/ \varepsilon)^{|g_{j'}| - 1} / (g_{j'} !)^{2/9}. \label{(4.6)}
\end{equation}

\subsection{Resonant Blocks}\label{4.2}

Following the constructions from the second step, consider a graph $g_{j'}$ that labels a term of $J^{(j')}$ (so $g_{j'}$ does not intersect $\calS_j$). We define a reduced graph $\bar{g}_{j'}$ that will be used for indexing event sums. It is defined from $g_{j'}$ by forgetting all substructure inside jump steps. In addition, there is a set of subgraphs of $g_{j'}$ that are called ``erased" -- these will be defined in Subsection \ref{4.2.3}. For erased subgraphs, we forget the order of further subgraphs, putting them into a standard left-to-right order
in $\mathbb{Z}$. 
(There is no issue of commutativity of operators for erased subgraphs when considering event sums, because erased subgraphs are replaced by their upper bounds, and then the order is irrelevant. See the next paragraph for a more extensive explanation of the principles behind this construction.)
The length $|\bar{g}_{j'}|$ is the same as $|g_{j'}|$, and $\bar{g}_{j'}!$ is the same as $g_{j'}!$ (which has no factorials on jump steps). If $L_{j} \le |g_{j'}| < L_{j + 1}$, we define
\begin{equation}
A^{(k)\prov}_{\sigma \tilde{\sigma}} (\bar{g}_{j'}) = \left|\frac{\tilde{J}^{j'}_{\sigma \tilde{\sigma}}(\bar{g}_{j'})}{E^{(j')}_{\sigma} - E^{(j')}_{\tilde{\sigma}}}\right|, \label{(4.7)}
\end{equation}
where $\tilde{J}^{(j')}_{\sigma \tilde{\sigma}} (\bar{g}_{j'})$ is the same as $J^{(j')}_{\sigma \tilde{\sigma}} (g_{j'})$ except: (1) jump steps $g_{i''}$ that are subgraphs of $g_{j'}$ are replaced with their upper bound $\gamma^{|g_{i''}|}$ from (\ref{(4.2)}) and (2) erased subgraphs $g_{i''}$ are replaced with their upper bound $(\gamma / \varepsilon)^{|g_{i''}|} / (g_{i''} !)^{2/9}$ from (\ref{(4.2)}). We say that $\bar{g}_{j'}$ from $\sigma$ to $\tilde{\sigma}$ is resonant if either of the following conditions hold:
\begin{equation}
\begin{aligned}
\textrm{I}&. \quad  |E^{(j')}_{\sigma} - E^{(j')}_{\tilde{\sigma}}| < \varepsilon^{|\bar{g}_{j'}|},  \\
\textrm{II}&. \quad  A^{(k)\text{prov}}_{\sigma \tilde{\sigma}} (\bar{g}_{j'}) > (\gamma / \varepsilon)^{|\bar{g}_{j'}|} / (\bar{g}_{j'} !)^{2/9} \;\text{with}\; |I(\bar{g}_{j'})| \ge \tfrac{7}{8} |\bar{g}_{j'}|. 
\end{aligned}
\label{(4.8)}
\end{equation}
Generalizing the step 2 definition, let $I(g_{j'})$ be the smallest interval in $\mathbb{Z}$ covering all the sites or blocks $\bar{\bar{b}}^{(\tilde{j})}$ with $\tilde{j} \le j$ that contain flips of $g_{j'}$, and let $|I(g_{j'})|$ be the number of sites or blocks $\bar{\bar{b}}^{(\tilde{j})}$ in $I(g_{j'})$. The same definitions hold for $I(\bar{g}_{j'})$.
It is important to understand that $\bar{g}_{j'}$ is a graph implementing a transition from some $\sigma$ to some $\tilde{\sigma}$. Thus it specifies $\sigma$ and $\tilde{\sigma}$ as well as the transitions that walk from $\sigma$ to $\tilde{\sigma}$. Jump steps of $\bar{g}_{j'}$ specify a single transition of this walk, altering the spin configuration on the interval $I$ of the jump step. The transition energies depend on $\sigma$ out to a distance $\frac{15}{14}L_j$ from the flips of $\bar{g}_{j'}$, in the $|\cd|^{(j)}$ metric (this will be verified inductively -- see Subsection \ref{4.3.2}). So $\bar{g}_{j'}$ has to specify $\sigma$ out to that distance. This leads to a proliferation of possibilities for $\bar{g}_{j'}$, but it is harmless because it is only exponential in $L_j$ (or in $|\bar{g}_{j'}|$).

Let us take a moment to explain the key ideas behind this construction, as they are critical to the design of a procedure that yields bounds uniform in $k$. A resonant graph can be thought of as an event with a small probability. In order for a collection of graphs to be rare, we need to be able to sum the probabilities. In the ideal situation, where there are no repeated sites/blocks in the graph, the probability is exponentially small, so it can easily be summed. However, when graphs return to previously visited sites, dependence between denominators develops, and then the Markov inequality that is used to estimate probabilities begins to break down. Subgraphs in a neighborhood of sites with multiple visits need to be ``erased," meaning that inductive bounds are used, and they do not participate in the Markov inequality. (This means we use $P(AC > B\overline{C}) \le \mathbb{E}(A\overline{C})/(B \overline{C}) = \mathbb{E}(A)/B$ when $\overline{C}$ is bound for $C$ -- so the variation of $C$ is not helping the bound.) When there are a lot of return visits, a graph's interval $I(g_{j'})$ is shortened by at least a factor $\frac{7}{8}$, and it goes into a jump step, where again we use inductive bounds. In this case, we have more factors of $\gamma$, and hence a more rapid decay in $|I(g_{j'})|$, and this provides the needed boost to preserve the uniformity of decay in the induction. (Fractional moments of denominators are finite, no matter the scale, which provides uniformity for ``straight" graphs with few returns.) The net result is uniform probability decay, provided we do not sum over unnecessary structure, \textit{i.e.} the substructure of jump steps and the order of subgraphs for erased graphs. 
Note that jump
steps represent sums of long graphs, so when taking absolute values it is best to do it
term by term. This is why in (\ref{(4.7)}) we replaced jump steps with their upper bound.
(The jump step bound (\ref{(4.2)}) is also a bound on the sum of the absolute values of the
contributing graphs. Thus we may  manipulate the sum at the level of individual graphs, when needed, for example when taking derivatives in Subsection \ref{4.2.2}.)
A single bound on 
$A^{(k)\prov}_{\sigma \tilde{\sigma}} (\bar{g}_{j'})$ will imply corresponding bounds for $A^{(k)}_{\sigma \tilde{\sigma}} (g_{j'})$ for any $g_{j'}$ that reduces to $\bar{g}_{j'}$. One just needs to combine inductive bounds on the erased sections and jump steps with the probabilistic bound on $A^{(k)\prov}_{\sigma \tilde{\sigma}} (\bar{g}_{j'})$. One may think of $A^{(k)\prov}_{\sigma \tilde{\sigma}} (\bar{g}_{j'})$ as a sort of universal socket into which any graph $g_{j'}$ can be plugged, provided its subgraphs obey the required inductive bounds. There is no point in attempting to sum over events labeled by $g_{j'}$ because (1) it would involve summing the same event many times over and (2) probability bounds are not good enough to allow such an uneconomical procedure.

We define the scale $k$ resonant blocks. Examine the set of all resonant graphs $\bar{g}_{j'}$. The set of sites/blocks that belong to resonant graphs $\bar{g}_{j'}$ are decomposed into connected components. These are the step $k$ resonant blocks $B^{(k)}$. They do not touch
any of the large blocks $B^{(j')}$ from the previous step.
Small blocks $\bar{b}^{(1)}, \ldots, \bar{b}^{(j)}$ can be absorbed into blocks $B^{(k)}$, but only if they are part of a resonant graph $g_{j'}$. 

As in step 2, we reorganize the resonant blocks produced so far, to take into account the presence of new resonant blocks and to define new small blocks $b^{(k)}$. The result is a collection of small blocks $b^{(i)}$ for $i \le k$ and a leftover region $\calS_{k'}$; they must satisfy the following diameter and  separation conditions for $i, \tilde{i} \le k$:
\begin{align}\label{dist}
\text{diam}(b^{(i)})&< L_i;\nonumber \\
\text{dist}(b^{(i)}_1, b^{(\tilde{i})}_2) &> d_m \equiv \exp(L^{1/2}_{m + m_0}),\;\; \text{if}\;\; \min\{|b^{(i)}_1|, |b^{(\tilde{i})}_2|\} \in [L_{m - 1}, L_m);\nonumber \\
\text{dist}(b^{(i)}, \calS_{k'}) &> d_m, \;\; \text{if}\;\; |b^{(i)}| \in [L_{m-1},L_m).
\end{align}

Here $|b^{(i)}|$ is the ``core" volume, $\textit{i.e.}$, the number of sites or blocks $\bar{b}^{(\tilde{i})}, \tilde{i} < i$ in $b^{(i)}$. But we establish the following convention: any block $B^{(i)}$ with fewer than $L_{i - 1}$ sites/blocks is considered to have size $L_{i - 1}$ when calculating volumes. This is because $L_{k - 1}$ is the minimum graph size considered in step $k$, and resonance probabilities are correspondingly small. This convention carries over to small blocks formed out of $B^{(i)}$ at stage $\tilde{i} \ge i$. Note that the rules (\ref{dist}) apply to small blocks on all scales up through $k$. This means that a $b^{(i)}$ with $i < k$ can be absorbed into the new resonant region if it is close enough to a $B^{(k)}$. It is easy to see that there is a unique way to decompose the complete resonant region (including blocks $B^{(i')}, B^{(k)}$ and $b^{(i)}$ with $i < k$) into a maximal set of small blocks on scales up through $k$ satisfying (\ref{dist}), plus a leftover ``large block'' region $\calS_{k'}$. One may proceed by forming proximity connections on successive length scales $d_m$. At each stage, connected components satisfying diameter and distance rules can be extracted as small blocks, and eliminated when constructing connected components on the next scale. We do not produce any new blocks $b^{(i)}$ with $i < k$, but previous ones can be absorbed into $\calS_{k'}$ or into a $b^{(k)}$.

Let $\calS_k$ denote $\calS_{k'}$ plus the small blocks $b^{(k)}$.
We add to $\calS_k$ a collar of width $L_k-1$ in the metric $|\cd|^{(j)}$.  
(As in previous steps, collars ensure a minimum graph size for connections to the outside, and define regions for block diagonalization.)
Then $\overline{\calS}_k$ is the collared version of $\calS_k$, and its components are the collared small blocks $\bar{b}^{(k)}$ and large blocks $\bar{B}^{(k')}$. The union of the $\bar{B}^{(k')}$ is denoted $\overline{\calS}_{k'}$, and then each $B^{(k')} \equiv \calS_k \cap \bar{B}^{(k')}$. We also define $\bar{\bar{b}}^{(k)}$ by adding a second collar of width $\frac{15}{14}L_j$ in the metric $|\cd|^{(j)}$; this constitutes the extent of dependence on the spin configuration for quantities associated with $\bar{b}^{(k)}$.


The ``geometric mean" construction of Section \ref{3.1} shows that if $x, y$ belong to the same resonant block $B^{(k)}$, then there must be a sequence of resonant graphs connecting $x$ to $y$ with the property that the even and odd subsequences consist of non-overlapping graphs. Thus we may focus on probability estimates for individual resonant graphs.

\subsubsection{Graphical Sums}\label{4.2.1}

It will be helpful to use this subsection to describe how we control sums over multiscale graphs $g_{j'}$ or $\bar{g}_{j'}$. The goal is to replace any sum of graphs with a corresponding supremum, multiplied by a \textit{combinatoric factor}. Any sum $\sum_g |f(g)|$ can be bounded by $\text{sup}_g|f(g)|c (g)$ provided $\sum_g c(g)^{-1} \le 1$. Then $c(g)$ is the combinatoric factor. From (\ref{(4.8)}), we see that $A$'s will obey a bound exponentially small in $|g_{j'}|$, times $(g_{j'}!)^{-2/9}$. We will obtain similar bound on resonance probabilities. Thus we need to be sure that combinatoric factors $c^{|g_{j'}|}(g_{j'}!)^{2/9}$ will be sufficient to control sums over $g_{j'}$.

As we shall explain in Subsection \ref{4.3.2}, any time there is a gap between a subgraph $g_{(i - 1)', p}$ and the collection $\{g_{(i - 1)', 0}, \ldots, g_{(i - 1)', p - 1}\}$, it will be ``filled in" by bridging gaps between graphs on earlier scales. These gap graphs result from expanding the difference between two denominators that arise from a commutator $\text{ad}\,A^{(i)}(g_{(i - 1)', p})$ applied to an operator associated with $g_{(i - 1)', 0}, \ldots, g_{(i - 1)', p - 1}$. Gap graphs are terms in expansions for the energies ($\textit{i.e.}$ diagonal entries of the Hamiltonian), and have the same structure as off-diagonal graphs.

Let us consider first the situation where $g_{j'}$ does not move through any blocks. We need to consider the combinatoric factors needed to control the sum over the structure of the level $i$ subgraphs, $g_{i'}$, of $g_{j'}$.
(Let us assume for the moment that $g_{j'}$ is not a jump step.)
Each subgraph $g_{i'}$ has subgraphs $g_{(i - 1)', 0}, \ldots, g_{(i - 1)', n}$. We need to sum over the positions of the starting point (first flip) of each $g_{(i - 1)', p}$. There can be gaps of size $\le \frac{15}{14} L_{i - 1}$ between each $g_{(i - 1)', p}$ and the ones that came before. Na{\"{i}}vely, the sum over $g_{(i - 1)',p}$ could produce a factor $O(L_i)p$, or $O(L^n_i)n!$ in total. But we use this bound only when summing over long graphs. (As in step 2, we say a graph $g_{j'}$ is \textit{long} if $|g_{j'}| > \frac{8}{7} |I(g_{j'})|$. Otherwise, it is \textit{short}.) For long graphs, we sum directly the series $(\text{ad}\,A)^n/n!$, so a combinatoric factor $n!$ is admissible. But then the $1/n!$ is gone from the estimate. This is the reason $g_{j'}!$ was defined with no contribution from jump steps, which represent sums of long graphs. 

Now let us consider the situation where $g_{i'}$ is short. There can be very little overlap between the $g_{(i - 1)', p}$ -- any overlap shortens $|I(g_{i'})|$, which must be at least $\frac{7}{8}|g_{i'}|$. We claim that no more than $2n/9$ graphs $g_{(i - 1)', p}$ can fail to break new ground. (Let us call them floating graphs.) The others are pinned to the left or right side of the growing graph, and do not produce factors of $p$ -- we call these pinned graphs. This means that short graphs can be controlled with a combinatoric factor $O(L^n_i)n^{2n/9} = O(L^n_i)(n!)^{2/9}$. To check this claim, consider first the case with no gaps. The ratio $\ell_{\text{p}}:\ell_{\text{f}}$ between the total length of the pinned graphs and the total length of the floating graphs must be at least 7:1. The lengths of graphs vary by no more than a factor of $\tfrac{15}{8} <2$. Hence the ratio $n_{\text{p}}:n_{\text{f}}$ between the number of pinned graphs and the number of floating graphs must be at least 7:2, which verifies in the claim in this case. If gaps are present, then (as explained in Section \ref{4.3}) the graphs bridging the gaps double back. This means that half the length of the bridging graphs is ``wasted,'' \textit{i.e.} it does not extend $I(g_{i'})$. Consequently, the graph length available for floating graphs decreases as gaps increase, and the number of floating graphs is even less than $2n/9$. In detail, let us suppose that the length of the pinned graphs plus the length of the gap graphs is $\ell_{\text{p}}(1+\delta)$.
Then the condition for short graphs implies that the ratio $\ell_{\text{p}}(1+\delta):\ell_{\text{p}}\delta/2 + \ell_{\text{f}}$ must be at least 7:1. Hence $\ell_{\text{f}} \le \ell_{\text{p}}(1-5\delta/2)/7$. Allowing as before a factor $\le 2$ between the sizes of pinned and floating graphs, we see that $n_{\text{f}} \le 2n_{\text{p}}(1-5\delta/2)/7$. Hence the ratio $n_{\text{f}}/n$ is no greater than $(2-5\delta)/(9-5\delta)\le \frac{2}{9}$, which completes the proof of the claim.

If we have a jump step $g_{i''}$, then the sum over substructure has already been taken care of in the inductive bound (\ref{(4.2)}). The sum over the jump itself is controlled with a combinatoric factor $c^{|g_{i''}|}$, which bounds the number of initial and final configurations for the jump.

To complete the bound, we need to take the product over all the subgraphs $g_{i'}$. Since the $g_{i'}$ each have size $\ge L_i$, there are no more than $|g_{j'}|/ L_i$ factors of $L_i$. So we obtain a bound $\exp(O(1)|g_{j'}|L^{-1}_i \log L_i)$ times a product of (2/9)$^\text{th}$-power factorials. In view of the geometric increase, $L_i = (\frac{15}{8})^i$, the product over $i < j$ gives a bound $c^{|g_{j'}|}(g_{j'}!)^{2/9}$, which is just what is required. One way to think of this estimate is to compute the ``combinatoric factor per site" $L_i^{1/L_i}$ by dividing each factor $L_i$ amongst $L_i$ steps. As the product of $L_i^{1/L_i}$ is bounded, the combinatorics are under control. Note that super-linear growth of $L_i$ with $i$ is required. We may control in a similar manner various other combinatoric factors bounded by $c^n$ when the subgraph $g_{i'}$ has $n + 1$ subgraphs. For example, the number of terms in $(\text{ad}\,A)^nJ$, the choice of jump step or regular step, and the sum on $n$.
Similarly, one needs to choose the denominators that will be differenced when forming gap graphs: the sums can be controlled by combinatoric factors $c^n$ for the subgraphs $g_{i'}$ of $g_{j'}$.

\medskip
\noindent{\textbf{Metric Equivalence}}

We need to establish some facts about comparability of the metric $|x - y|^{(j)}$ with $|x - y|$. The issue is that graphs exhibit decay in $|x - y|^{(j)}$ (blocks $\bar{\bar{b}}^{(i)}, i \le j$ contracted to points) but there are counting factors exponential in the size of blocks -- specifically sums over states or metaspins in blocks $\bar{b}^{(i)}$ and background spin configurations in $\bar{\bar{b}}^{(i)} \setminus \bar{b}^{(i)}$. Recall that blocks $b^{(j)}$ have size $< L_j$; $\bar{b}^{(j)}$ includes a collar of width $L_j-1$ in the metric $|x - y|^{(j - 1)}$; and $\bar{\bar{b}}^{(j)}$ includes a second collar of width $\frac{15}{14} L_{j-1}$. Comparability of the metrics will ensure that the size of $b^{(j)}$ increases by no more than a fixed factor, \textit{e.g.} 8, in forming $\bar{\bar{b}}^{(j)}$. Then the state-counting factor $2^{|\bar{b}^{(j)}|}$ and the background-spin counting factor $2^{|\bar{\bar{b}}^{(j)}|-|\bar{b}^{(j)}|}$ can be controlled by the smallness of the graph, $(\gamma / \varepsilon)^{L_j}$, or its probability, $\varepsilon^{L_j}$. This is a crucial element of our method, because the maintenance of uniform exponential decay is essential for controlling state sums. Separation distances that grow rapidly with block size ensure that the fraction of distance lost to blocks is summable, and so $|x - y|^{(j)}$ is always at least a positive fraction of $|x - y|$. The construction is similar in spirit to that of \cite{Frohlich1983}.

The separation rule is that blocks $b^{(j)}$ have diameter $< L_j$, and that pairs of blocks satisfy
\begin{equation}
\text{dist}(b^{(i)}_1, b^{(j)}_2) > d_m \equiv \exp(L^{1/2}_{m + m_0}), \;\; \text{if} \;\; \min\{|b^{(i)}_1|, |b^{(j)}_2|\} \in  [L_{m - 1}, L_m). \label{(4.8a)} 
\end{equation}
This type of rule can generate blocks with large, hierarchically organized gaps. But there is a limit to how spread-out blocks can be. Consider a block $b^{(j)}$ with $|b^{(j)}| \in \lfloor L_{m - 1}, L_m)$. If any gap in $b^{(j)}$ is greater than $d_{m - 1}$, then it would divide $b^{(j)}$ into two parts. At least one of the parts would have volume $< L_{m - 1}$ while being separated by a distance $> d_{m - 1}$. This is impossible, because it would mean that part would have become a separate small block at some step $i \le j$. (The diameter of the part is obviously smaller than $L_j$, since it is a subset of $b^{(j)}$ and diam$(b^{(j)}) < L_j$.) 
There are no more than $L_m-2$ gaps, so
the limit on gap size implies that
\begin{equation}
\text{diam} (b^{(j)}) \le (L_m - 2) d_{m - 1} \le q^m d_{m-1}, \label{(4.9)}
\end{equation}
where $q \equiv \frac{15}{8}$, $L_m = q^m$.
As these blocks are separated from each other and from larger blocks by a distance $d_m$, we see that the fraction of the distance between larger blocks that is occupied by blocks with $|b^{(i)}| \in [L_{m-1},L_m)$ is bounded by $r_m \equiv q^m d_{m-1}/d_m$.
Proceeding from larger to smaller values of $m$, we find that the fraction of distance that is free of blocks of any size is at least
\begin{equation}
\prod_{m=1}^\infty (1-r_m) \ge \exp\left(-2\sum_{m=1}^\infty r_m \right), \label{(4.9aa)}
\end{equation}
provided $r_m \le \frac{1}{2}$.
The super-exponential growth of $d_m$ with $m$ ($d_m =d_{m-1}^{\sqrt{q}}\approx d_{m - 1}^{1.37}$) implies that $\sum_{m=1}^\infty r_m$ is small, for large enough $m_0$.
In detail, we may put $\zeta = 1-q^{-1/2} \approx .27$ and write
\begin{align}\label{(4.9ab)}
r_m &= q^m \exp\left(L_{m-1+m_0}^{1/2} - L_{m+m_0}^{1/2}\right) = q^m e^{-\zeta q^{(m+m_0)/2} }
\nonumber
\\
&\le \frac{4!}{\zeta^4}\left(\frac{\zeta^4}{4!}q^m e^{-\zeta q^{(m+m_0)/4}}\right)e^{-\zeta q^{(m+m_0)/4}} \le \frac{4!}{\zeta^4}e^{-\zeta q^{(m+m_0)/4}},
\end{align}
and then it is clear that $\sum_{m=1}^\infty r_m$ can be made arbitrarily small by choosing $m_0$ sufficiently large. By (\ref{(4.9aa)}), we obtain the desired smallness of the fraction of distance occupied by blocks satisfying our separation conditions. This forms the basis for the following two lemmas, one controlling the expansion of blocks due to collars, and one on metric equivalence.

\begin{lemma}\label{lem:4.1}
Let $m_0$ be sufficiently large. Then the following bound holds for any $i$:
\begin{equation}\label{(4.9a)}
\text{diam}(\bar{\bar{b}}^{(i)}) \le 8 \, \text{diam}(b^{(i)}).
\end{equation}
\end{lemma}
\textit{Proof.}
Note that a block $b^{(j)}$ has diameter in $[L_{j - 1}, L_j)$, because if it were smaller than $L_{j - 1}$, it would have been a block $b^{(i)}$ with $i < j$. Let us assume (\ref{(4.9a)}) inductively  for $i < j$. Any blocks $b^{(i)}$ that might appear in the collar must obey the separation condition with respect to $b^{(j)}$. From the discussion in the paragraph above,  we can choose $m_0$ so that blocks $\bar{\bar{b}}^{(i)}$ with $i < j$ take up a small fraction of the width of the collar. Since $b^{(j)}$ has a minimum diameter $L_{j - 1}$, we see that adding the two collars (the first of width $L_j$ and the second of width $\frac{15}{14}L_{j-1})$ expands its size in $|\cd|^{(j - 1)}$ by no more than a factor $(2(L_j + \frac{15}{14}L_{j-1}) + L_{j - 1})/ L _{j - 1} \le 7$. Allowing for blocks $\bar{\bar{b}}^{(i)}$ with $i < j$, we increase the factor to 8 and recover the inductive assumption. \qed

Although we cannot expect comparability of metrics for points very close to blocks, the next lemma gives comparability when the distance involved is at least as large as the current length scale.
\begin{lemma}\label{lem:4.2}
Let $m_0$ be sufficiently large. Then the following bound holds for any $x$, $y$, $j$ with $|x - y|^{(j)} \ge L_j$:
\begin{equation}\label{(4.9b)}
|x - y| \le 6 |x - y|^{(j)}.
\end{equation}
\end{lemma}
\textit{Proof.}
The worst case for this estimate will be when  $|x - y|^{(j)} = L_j$, for example if $x$ is in a block $\bar{\bar{b}}^{(j)}$ and $y$ is a distance $L_j$ away. The block $\bar{\bar{b}}^{(j)}$ has maximum size $3L_j + 2\cdot\frac{15}{14}L_{j-1} \le 5L_j$ in $|\cd|^{(j - 1)}$. Allowing for a small amount of expansion from blocks on other scales, we find that $|x - y| \le 6 |x - y|^{(j)}$. \qed

This result on metric equivalence is used in a number of places to handle situations where graphs of size $L_j$ touch a block $\bar{\bar{b}}^{(j)}$ and/or blocks $\bar{\bar{b}}^{(i)}$ with $i < j$. We have an exponential factor like $\varepsilon^{L_j}$ to work with, but the decay has to be spread out over the blocks as well as the graph. But with metric equivalence, we get at least $1/6$ of the original decay rate. This also allows us to handle the associated state sums as well, while preserving exponential decay. Similar issues arise in the single-body analysis of \cite{Imbrie2014a}, but they were easier to handle because state sums were linear (rather than exponential) in the volume, so collars could be chosen logarithmic (rather than linear) in the volume of a block.

\subsubsection{The Jacobian}\label{4.2.2}

In order to estimate probabilities of a resonant graph, we use a Markov inequality -- see (\ref{(4.17)}) below.
This provides a bound in terms of the expectation of the graph to the $s = \frac{2}{7}$ power. The graph has a number of energy denominators, so it is important that we are able to demonstrate the finiteness of the $-s$ moment of the product of the denominators of the graph. Each denominator corresponds to the energy change from flipping a set of spins (or metaspins, in the case of block energies). To leading order, the energy change from flippping $\sigma_i$ is $\pm h_i$, plus something independent of $h_i$. As the random variables $h_i$ have a bounded density, the $-s$ moments are bounded. A similar bound applies to block energy differences, provided we assume noncriticality of their dependence on the random variables. The goal of this subsection is (1) to describe when it is possible to obtain good bounds on $-s$ moments of a product of denominators and (2) to estimate the Jacobian that results from using the energy denominators as integration variables, instead of the parameters $h_i$, etc. that appear in the Hamiltonian (\ref{(1.1)}). All these issues apear in a simpler context in \cite{Imbrie2014a}, where the same ``change of variable method'' is used to bound $-s$ moments of products of energy denominators.

The main result of this subsection is the Jacobian bound, Proposition \ref{prop:4.1}. Using a convenient normalization, the Jacobian determinant is $\pm 1$ to leading order for a ``good'' set of denominators (see below). Allowing for perturbative corrections, derivatives of energy differences are close to their leading behavior, and hence the Jacobian determinant can be bounded above and below by an exponential in the number of denominators.

Let us begin the analysis with a discussion of the structure of the graphs that arise from our construction.
Each graph has a number of energy denominators produced at various scales. We need to work with a hierarchically organized structure of denominators. Each denominator can be visualized as an arch over the collection of sites/blocks flipped in the associated subgraph. Denominators from later steps arch over earlier ones. Arches are either strictly contained in one another or else completely disjoint. This structure follows from the way denominators are introduced -- see (\ref{(3.10)}), (\ref{(4.7)}). However, if two subgraphs have a site/block in common, the underlying random variables are identified, creating unwanted dependence, which shows up graphically as overlapping arches. 
Therefore, we will need to require that as one proceeds up the hierarchy, there are no repeat visits, which means each new denominator introduces a new independent variable.
As long as this is the case, there is no linear relation amongst the denominators, and they can be used as integration variables, provided we can bound the Jacobian. 
(If there are repeat visits, then we will need to throw out some denominators -- \textit{i.e.} replace them with uniform bounds -- in order to find a set that has the hierarchical property. The details of how this is done will be deferred to the next subsection.)

To leading order, each energy difference in the graph is a sum of energy differences of the sites/blocks of the graph. So we think of each site/block energy difference as an independent variable (but no more than one variable for each site/block). For example, a flip at site 1 produces an energy difference $2h_1 + a_1$. A flip at site 2 will produce an energy difference $2h_2 + a_2$ if compared to the previous configuration; it produces $2h_1 + 2h_2 + a_1 + a_2$ if compared to the starting configuration. Either way, the two denominators are independent. (Here $a_1, a_2$ are $h$-independent constants coming from the exchange interaction $J_iS^\mathrm{z}_i S^\mathrm{z}_{i + 1}$.) If a third flip occurs at either site 1 or 2, the new denominator will not be independent, because it can be written as a linear combination of the first two denominators. In general, any flip at a new site will introduce a denominator independent of the ones that come before, simply because it introduced a new independent variable: the energy difference at the new site. Blocks will be treated as fat sites with metaspin variables. We allow for only one energy difference in the block to be considered independent (otherwise we would need to assume probabilistic properties about multiple energy differences in blocks). So only steps to new sites/blocks introduce new variables that generate independent denominators. Our assumed bounds on the distribution of energy differences allow us to integrate the $-s$ power of each denominator in turn, keeping the unintegrated variables fixed. We are free to choose any set of independent denominators, even an incomplete (non-spanning) set. Any remaining denominators, including all non-independent denominators, are bounded in sup norm. 

The above discussion applies to the leading order approximation for energies as a simple sum of contributions from sites/blocks. Terms like $J_iS^\mathrm{z}_iS^\mathrm{z}_{i + 1}$ do not depend on the random variables in the sites/blocks they connect, so they merely introduce constant shifts as in the example above. However, as we saw in the second step, energies $E^{(i')}_\sigma$ receive perturbative contributions with $\textbf{h}$-dependent energy denominators.  In the $k^{\text{th}}$ step, energies $E^{(j')}_{\sigma}$ are updated to $E^{(k')}_{\sigma}$ through perturbative terms and from the eigenvalues of newly formed small blocks. Denominators in a graph for $E^{(k')}_{\sigma}$ depend on $E^{(i')}_{\sigma}$ for $ i < k$. We want to check the dependence on the underlying random variables by differentiating with respect to $h_i$ (or with respect to other random variables at our disposal). We apply the chain rule repeatedly, down to the first scale if possible. But when a derivative hits a block energy at the scale of its formation, we stop and apply our basic assumption on non-criticality of those energies. The graphical expansions are well controlled, so it should not be a surprise that they can be differentiated, and the Jacobian connecting a set of independent denominators of a graph to the underlying random variables is close to the noninteracting case. 

\textit{Behavior of block energies.} Block energy differences $E^{(j')}_{\alpha} - E^{(j')}_{\beta}$ depend on random variables out to a distance $\frac{15}{14}L_{j}$ from a block $\bar{b}^{(j)}$ in $|\cd|^{(j - 1)}$. This is from dependence of graphs on energies $E^{(i')}, i \le j - 1$ -- see (\ref{range}) below. Recall that our parameter space is $\textbf{h} = (h_i, J_i, \Gamma_i)$ with $\gamma_i = \gamma \Gamma_i$. The random variables $h_i, J_i, \Gamma_i$ are independent, each having a distribution supported on [-1,1], with a density bounded uniformly by a constant $\rho_0$. Thus if the block has size $n$ there are no more than $\eta n$ parameters, for some fixed $\eta$. All eigenvalues have bounded derivatives with respect to these variables, since the operators $S^\mathrm{z}_i, S^\mathrm{x}_i$ are bounded by 1 in norm. We need to assume that all energy differences are \textit{noncritical}, \textit{i.e.} at least one direction in the space of random variables produces a nonzero directional derivative. In practice, we will use a coordinate system in which energy differences move to first order with respect to one of the coordinates. Let us focus on the case of spherical coordinates since (as we will show in Chapter \ref{5}) radial derivatives can be bounded from below if there is a minimum level splitting. The angular coordinates will be treated as ``spectator" coordinates, that is, fractional moments of energy denominators are bounded by integrating over the radial variable with bounds independent of the angles. Let $r = |\textbf{h}|$ be the Euclidean norm of $\textbf{h}$, and use it as the radial coordinate. We make the following assumption:
\begin{assumptiona1}
(Non-criticality of energy differences) For all $\mathbf{h}$ and all pairs of eigenstates $\alpha, \beta$ for the matrix $H_0^{(j - 1)'} + J^{{(j)}\text{sint}}$ in a small block $\bar{b}^{(j)}$ of size $n$, the difference of eigenvalues satisfies
\begin{equation}
\left|\frac{\partial}{\partial r} (E^{(j')}_{\alpha} - E^{(j')}_{\beta})\right|^{-1} \le c^n_b, \label{(4.10)}
\end{equation}
\end{assumptiona1}
\noindent
We will relax this assumption in Chapter \ref{5} by allowing failure on a set of small probability. But for simplicity, we work here with this strong non-criticality assumption.

For ease of exposition, let us introduce variables $\tilde{h}_x$ for each site/block $x$. For sites, $\tilde{h}_x = 2h_x $ so that bare energy differences move at unit speed. For each block in the graph of independent denominators, we let $\tilde{h}_x$ be the coordinate in the radial direction, rescaled so that $E^{(j')}_{\alpha} (\tilde{h}_x) - E^{(j')}_{\beta} (\tilde{h}_x)$ moves at unit speed. The eigenvalues are distinct, except on a curve of measure $0$, given by the vanishing of the discriminant. (The discriminant cannot vanish identically. Take any magnetic fields $h_1, \ldots, h_n$ in the block with no relation $\pm h_1 \pm h_2 \ldots \pm h_n = 0$. As they are scaled to infinity together, they dominate all other terms in the Hamiltonian, and the eigenvalues are separated.) Therefore the eigenvalues are continuously differentiable. We illustrate with a simple one-site example, 
$H = (\begin{smallmatrix} h & \gamma \\ \gamma & -h \end{smallmatrix})$, with both $h, \gamma$ random. The eigenvalue difference is $2 \sqrt{h^2 + \gamma^2} = 2r = \tilde{h}$ in polar coordinates, and the radial derivative of the eigenvalue difference is 2. Of course, if $\gamma$ is not allowed to vary, we would have a critical point at $h = 0$, where assumption \textbf{A1}($c_b$) would fail.

We should point out that there is some arbitrariness in the selection of $\alpha, \beta$ in the blocks of the denominator graph. The denominators form a 
tree graph of energy denominators linking distinct spin configuration energies,
since by construction each new denominator goes to a configuration with a new site/block flipped. The choice of how to order the denominators can affect the choice of $\alpha, \beta$ in a block. Intervening flips may cause a change in state in the block, so that a second denominator may reflect a different state change at a block. However, once a block variable is linked into the expanding denominator graph, it is treated as fixed, along with any other energy difference in the block (we do not have enough information to treat other differences as independent.) Subsequent denominators depend on new variables outside the block, so they are independent of previous ones. There is also some arbitrariness in the selection of variables when there are fewer denominators than sites/blocks being flipped. In such cases extra variables may be treated as fixed, all analysis done uniformly with respect to the values of the extra variables.

It will be helpful to order the denominators in the graph $g_{j'}$ as follows. First run through the denominators from the first step (single spin flips). Then proceed to denominators introduced in the second step (graphs $g_{1'}$ with two or three steps -- the new denominator coming from an $A^{(2)}_{\sigma \tilde{\sigma}}$ with $\sigma \tilde{\sigma}$ differing by two or three flips. Continue through all length scales up to $j$. Each denominator in the sequence introduces a new independent variable $\tilde{h}_y$, and it is convenient to number the $\tilde{h}_y$'s in the same order as the denominators. Let us introduce the notation
\begin{equation}
D^{(j')}_{\sigma \tilde{\sigma}} = E^{(j')}_{\sigma} - E^{(j')}_{\tilde{\sigma}} \label{(4.10a)}
\end{equation}
for the denominator connecting $\sigma$ to $\tilde{\sigma}$. The dependence on $\tilde{h}_y$ is given by the leading term from scale 0 (or from scale $i$ for energies of blocks $\bar{b}^{(i)}$) plus corrections given by graphical expansions. Thus
\begin{equation}
\frac{\partial D^{(j')}_{\sigma \tilde{\sigma}}}{\partial \tilde{h}_y} = \pm (1 - \delta_{\sigma(y)\tilde{\sigma}(y)}) + \frac{\partial}{\partial \tilde{h}_y} \sum_{i = 1}^{j - 1} \sum_{g_{i'}} \left(J^{(i')}_{\sigma \sigma} (g_{i'}) - J^{(i')}_{\tilde{\sigma} \tilde{\sigma}}(g_{i'})\right), \label{(4.11)}
\end{equation}
where the Kronecker $\delta$ makes the leading term $\pm 1$ if and only if $\sigma (y) \ne \tilde{\sigma}(y)$. The corrections come from diagonal entries of $J^{(i')}$ that were absorbed into $H_0^{(j')}$, 
and also from differences $E^{(i')}_{\sigma} - E^{(i')}_{\tilde{\sigma}}$ of energies of blocks
 $\bar{b}^{(i)}$ after diagonalization; see  (\ref{(3.17)})-(\ref{(3.21)}) and the analogous equations in the general step, (\ref{(4.37)})-(\ref{(4.39)}). 
To keep the notation in (\ref{(4.11)}) simple, let us allow $J^{(i')}_{\sigma \tilde{\sigma}}$ to refer either to the initial block energies or to the perturbative corrections at subsequent scales. We defer for a moment the discussion of block energy differences, and focus now on the perturbative terms. By the inductive hypothesis,
they obey the bound (\ref{(4.6)}). Note that our ordering convention implies that the matrix $\pm (1 - \delta_{\sigma(y)\sigma(\tilde{y})})$ is lower triangular, with $\pm 1$'s on the diagonal, and $\pm 1$ or $0$ below the diagonal. (Here the pair $\sigma \tilde{\sigma}$ runs over $n$ choices coming from the $n$ denominators, and $y$ runs over the $n$ independent variables.) Thus to leading order, the eigenvalues of the Jacobian are $\pm 1$. Our challenge now is to show the corrections are small.

If we apply the chain rule to the $\tilde{h}_y$-derivative on the right-hand side of (\ref{(4.11)}), the derivatives flow to the denominator energies in each term $J^{(i')}_{\sigma \sigma} (g_{i'})$ or $J^{(i')}_{\tilde{\sigma} \tilde{\sigma}} (g_{i'})$. We need to be cognizant of the fact that jump steps in $g_{i'}$ are actually sums of (long) graphs, and these graphs have denominators which will depend on $\tilde{h}_y$. So for the purposes of this discussion, we expand out every jump step at level $i < j$ into the sum of all its constituent graphs, see (\ref{(3.11)}) or (\ref{(4.28)}). The grouping of graphs into jump steps is a convenient way to keep track of the way estimates from smaller scales merge with Markov inequality bounds to produce bounds on longer scales. But when needed, we may go back to the underlying sum of graphs. Another point to mention is the fact that the energy graphs $g_{i'}$ do not have the ``independent denominator" property that we assume for $g_{j'}$ -- it is not needed because we have inductive bounds on those graphs. 
By the Leibniz rule, the $\tilde{h}_y$-derivative produces
a sum of terms with one of the denominators in $g_{i'}$ duplicated and the corresponding denominator differentiated in the numerator. The extra denominator can be bounded from below as in (\ref{(4.8)}I), and the result is an extra factor of $\veps^{-|g_{i'}|}$. The original graph is bounded as in (\ref{(4.6)}). Altogether, we obtain a bound
\begin{equation}
\bigg|\frac{\partial}{\partial \tilde{h}_y} J^{(i')}_{\sigma \sigma} (g_{i'})\bigg| \le \frac{(c \gamma / \varepsilon^2)^{|g_{i'}|}}{(g_{i'}!)^{2/9}} \begin{matrix}\text{sup} \\ \tau \tilde{\tau}\end{matrix} \bigg|\frac{\partial D^{(i')}_{\tau \tilde{\tau}}}{\partial \tilde{h}_y}\bigg|, \label{(4.12)}
\end{equation}
where the constant $c$ is inserted to account for the sum of denominators $\tau \tilde{\tau}$ in $g_{i'}$. 

Let $\tau \Delta \tilde{\tau}$ denote the set of sites/blocks where $\tau \ne \tilde{\tau}$. 
The denominator $D^{(i')}_{\tau \tilde{\tau}}$ is part of a graph that covers $\tau \Delta \tilde{\tau}$ and it extends to a place where $\sigma \ne \tilde{\sigma}$, because otherwise the difference in the second term of (\ref{(4.11)}) would vanish.

We may repeat the process, inserting (\ref{(4.11)}) on scales $i < j$. There is a leading term $\pm 1$, plus further graphical expansions. The $\pm 1$ terms and the expansion terms are localized near the places where $\tau \ne \tilde{\tau}$, so there needs to be a sum over such sites -- but there are no more than $|g_{i'}|$ of them, so the sum can be handled with an increase in the constant $c$ in (\ref{(4.12)}). The process stops at scale 1 (if $\tilde{h}_y$ is $2h_y$) or at scale $i$ (if $\tilde{h}_y$ is the radial variable for a block $\bar{b}^{(i)}$); at that point the derivative produces a factor $\pm 1$. 
Throughout the process, a graphical connection is maintained to $\sigma \Delta \tilde{\sigma}$;
hence when it concludes, there is a connection from $y$ to $\sigma \Delta \tilde{\sigma}$. In fact, there is a double connection, because the graph must flip each site/block not in $\sigma \Delta \tilde{\sigma}$ at least two times, so as to return it to its starting value.
Repeatedly applying the bounds from the last section on graphical sums, we find that
\begin{equation}
\frac{\partial D^{(j')}_{\sigma \tilde{\sigma}}}{\partial \tilde{h}_y} = \pm (1 - \delta_{\sigma (y) \tilde{\sigma} (y)}) + O(\gamma / \varepsilon^2)
^{1 + \text{dist}(y, \sigma \Delta \tilde{\sigma})}, \label{(4.13)}
\end{equation}
where $\text{dist}(y, \sigma \Delta \tilde{\sigma})$ is the number of steps from $y$ to $\sigma \Delta \tilde{\sigma}$.

Let us return to the case of derivatives of block energy differences. When a derivative hits a block energy, it flows to the Hamiltonian of the block and we have to take the expectation of the resulting operator. The Hamiltonian is given by a leading term plus a graphical expansion with $\tilde{h}$-dependence in the denominators. We are only concerned with terms that depend on $\tilde{h}_y$ for $y$ outside the block. All the terms are bounded in norm as in (\ref{(4.6)}), so we can proceed to apply the chain rule as discussed above, and the resulting bounds serve to control the derivatives of block energies. As for the perturbative terms, we end up with decay from the block to $y$ for the derivative of a block energy with respect to $\tilde{h}_y$, and (\ref{(4.13)}) remains valid.

The denominator $D^{(j')}_{\sigma \tilde{\sigma}}$ is part of a graph that specifies a sequence of flips taking $\sigma$ to $\tilde{\sigma}$. We write $D^{(j')}_{\sigma \tilde{\sigma}}$ as a telescoping sum of energy differences for each spin flip:
\begin{equation}
D^{(j')}_{\sigma \tilde{\sigma}} = \sum_x d_x^{(j')}, \label{(4.14)}
\end{equation}
where $d_x^{(j')}$ is an energy difference as in (\ref{(4.10a)}), arising from changing the spin at $x$ from its value in $\sigma$ to its value in $\tilde{\sigma}$.
Note that $d_x^{(j')}$ depends on the spin configuration out to a distance $O(L_j)$, but we do not make the dependence explicit in the notation. Of course, each $d_x^{(j')}$ is just a local difference of energies, so as a special case of (\ref{(4.13)}) we have an estimate:
\begin{equation}
\mathcal{J}_{xy} \equiv \frac{\partial d_x^{(j')}}{\partial \tilde{h}_y} = \pm \delta_{x y} + O(\gamma/ \varepsilon^2)^{1 + \text{dist} (x, y)}. \label{(4.15)}
\end{equation}
Now letting $x, y$ run over the $n$ sites/blocks associated with the independent variables $\tilde{h}_y$, we may write the Jacobian in matrix notation: $\mathcal{J} = \tilde{I} + \Delta$. Here $\tilde{I}$ is a modified identity matrix, with signs allowed. The matrix $\Delta$ has absolute row and column sums bounded by
$O(\gamma/\varepsilon^2)$, from the decay in (\ref{(4.15)}). By a standard result on matrix norms, the same bound applies to $\|\Delta\|$ (see for example Proposition 10.6 of \cite{Aizenman2015}). Hence $\| \Delta\| = O(\gamma/ \varepsilon^2)$. By Weyl's inequality, $\Delta$ cannot move the eigenvalues of $\tilde{I}$ by more than that amount. Therefore,
\begin{equation}
\big|\log \left|\det \mathcal{J}\right|\big| \le O (\gamma / \varepsilon^2)n. \label{(4.16)}
\end{equation}
Let $L$ be the matrix $1 - \delta_{\sigma (y) \tilde{\sigma}(y)}$, which, as previously noted, is lower triangular. (But now, with the signs removed, it has $1$'s on the diagonal.) The matrix $L$ expresses the relation (\ref{(4.14)}) between the denominators and the single flip energies $d_x^{(j')}$. Thus the full Jacobian $\partial D^{(j')}_{\sigma \tilde{\sigma}}/ \partial \tilde{h}_y$ is $L\mathcal{J}$, and its determinant is the same as $\text{det} \, \mathcal{J}$, with the same bound (\ref{(4.16)}). Thus we obtain the main result of this subsection:

\begin{proposition}\label{prop:4.1}
For a given $c_b$, let $\varepsilon = \gamma^{1/20}$ be sufficiently small. Assume \textbf{A1}($c_b$) and inductive bounds (\ref{(4.6)}) and (\ref{(4.8)}I). Then any system of $n$ hierarchically organized denominators obeys the following Jacobian bound:
\begin{equation}
\left| \log \bigg| \det \frac{\partial D ^{(j')}_{\sigma \tilde{\sigma}}}{\partial \tilde{h}_y} \bigg|\right| \le O(\gamma / \varepsilon^2)n. \label{(4.16a)}
\end{equation}
\end{proposition}
Note that the choice of normalization for the variables $\tilde{h}_y$ obscures the size of the lower bound (\ref{(4.10)}) on the rate of variation of eigenvalue differences with the original variables. When estimating fractional moments of energy denominators, we will need to include factors $c_b^{n_y}$ when the block at $y$ has size $n_y$. 

It turns out that the multi-denominator estimates of the next subsection can be organized so as to avoid working with block energy variables. Those estimates are mainly about getting uniform exponential decay, and due to the diluteness of resonant blocks, the decay can be extracted from the spaces between blocks. Nevertheless, the case $n = 1$ of (\ref{(4.16a)}) is indispensable for controlling single-denominator resonances, in particular for block energies.

\subsubsection{Resonant Graphs}\label{4.2.3}

The Jacobian bounds allow us to estimate probabilities of resonant graphs. 
The arguments are based on similar estimates in the corresponding subsection of \cite{Imbrie2014a}.
First, let us consider case I of the resonant condition (\ref{(4.8)}). This is a single denominator estimate, so by (\ref{(4.16a)}) with $n = 1$, we see that the energy difference moves at close to unit speed with the variation of any one of the variables corresponding to flips of spins/metaspins in the transition $\sigma \rightarrow \tilde{\sigma}$. So the resonance probability is proportional to $\varepsilon^{|\bar{g}_{j'}|}$. In the case of a block variable for some $\bar{b}^{(i)}$, there is also a factor of the area of the sphere of radius $r$ in $\mathbb{R}^{\eta L_i}$, since there are up to $\eta L_i$ random variables in $\bar{\bar{b}}^{(i)}$. Putting $m = \eta L_i$, we have that $r \le \sqrt{m}$, and then the area is bounded by 
$r^m\cdot 2\pi^{m/2}/\Gamma(m/2) \le c_{\text{area}}^{L_i}$ 
for some $c_{\text{area}}$. Also, the rescaling $r \rightarrow \tilde{h}_y$ leads to a factor $c_b^{|\bar{\bar{b}}^{(i)}|} \le c^{L_i}$ from (\ref{(4.10)}). As discussed at the end of Subsection \ref{4.2.1}, our constructions ensure that $|\bar{g}_{j'}|$ is always large enough so that exponentials in block sizes can be absorbed into the exponential decay in $|\bar{g}_{j'}|$. Thus the probability for condition (\ref{(4.8)}I) is bounded by $(\rho_1 \varepsilon)^{|\bar{g}_{j'}|}$, for some constant $\rho_1$.

Next, let us consider probability bounds for case II of (\ref{(4.8)}), which applies to short graphs with $I(\bar{g}_{j'}) \ge \frac{7}{8}|\bar{g}_{j'}|$.
Consider the simplest case in which all the denominators in $g_{j'}$ are independent. This will happen if there are no multiple visits to sites/blocks in $g_{j'}$, which means that each flip introduces a new independent variable. This means there are no erased subgraphs -- those are introduced for the general case (below) to deal with repeat  visits. (Note that a variable for a block $\bar{b}^{(i)}$ must be part of a graph that extends into $\bar{\bar{b}}^{(i)}$, which we count as a repeat visit. So we do not have block variables in this example.) 

We claim that
\begin{equation}\label{(4.16b)}
A^{(k)\prov}_{\sigma \tilde{\sigma}} (\bar{g}_{j'}) \le \frac{\gamma^{|\bar{g}_{j'}|}}{\bar{g}_{j'}!}\prod_{\tau \tilde{\tau}\in G^\text{d}_k} \left|E^{(i')}_{\tau} - E^{(i')}_{\tilde{\tau}}\right|^{-1}. 
\end{equation}
Here $G^\text{d}_k$ is the denominator graph for $A^{(k)\text{prov}}_{\sigma \tilde{\sigma}} (\bar{g}_{j'})$, which is the same as the denominator graph of $J^{(j')}_{\sigma \tilde{\sigma}} (g_{j'})$, plus the denominator for $\sigma \tilde{\sigma}$ -- recall the definition (\ref{(4.7)}). The bound (\ref{(4.16b)}) arises from unwrapping the definitions. In particular, as explained after (\ref{(4.7)}), 
jump steps on scales $i<j$ are replaced by their upper bound $\gamma^{|g_{i''}|}$ -- this means that jump steps contribute their share of factors of $\gamma$. (These factors represent the improved bound 
$|A^{(i')}_{\sigma \tilde{\sigma}}(g_{(i - 1)''})| \le \gamma^{|g_{(i - 1)''}|}$, which is used in place of jump steps in $A^{(k)\text{prov}}_{\sigma \tilde{\sigma}}(\bar{g}_{j'})$, see (\ref{(4.2)})). 
But this is the only use of inductive bounds -- the rest of the graph has all its subgraphs expanded out to the level of elementary flips. Thus $J^{(j')}_{\sigma \tilde{\sigma}} (g_{j'})$ is obtained from the ad expansion (\ref{(4.30)}) and the rotation (\ref{(4.39)}) applied to $J^{(j-1)'}$. Continuing down to the first level, each spin flip comes with a factor of $\gamma$, as required in (\ref{(4.16b)}). The factorials $1/n!$ and $n/(n+1)! \le 1/n!$ from the ad expansion accumulate exactly as per the definition (\ref{(4.3)}) -- including the lack of carry-forward at jump steps, which arises from the lack of factorials in the jump step bound $\gamma^{|g_{i''}|}$. 

Applying the Markov inequality to (\ref{(4.16b)}), we obtain 
\begin{align}
P\left(A^{(k)\prov}_{\sigma \tilde{\sigma}} (\bar{g}_{j'}) > \frac{(\gamma / \varepsilon)^{|\bar{g}_{j'}|}}{(\bar{g}_{j'}!)^{2/9}}\right) &\le \mathbb{E} \; \frac{(A^{(k)\text{prov}}_{\sigma \tilde{\sigma}}(\bar{g}_{j'}))^s (\bar{g}_{j'}!)^{2s/9} }{ (\gamma / \varepsilon)^{s|\bar{g}_{j'}|}} \notag\\
& \le \frac{\varepsilon^{s|\bar{g}_{j'}|}}{(\bar{g}_{j'}!)^{2/9}} \; \mathbb{E} \prod_{\tau \tilde{\tau}\in G^\text{d}_k} \left|E^{(i')}_{\tau} - E^{(i')}_{\tilde{\tau}}\right|^{-s}. \label{(4.17)}
\end{align}
As in Proposition \ref{prop:3.1}, we take $s = \frac{2}{7}$. 
In the second inequality, we have used the available factor $(\bar{g}_{j'}!)^{-s}$ from (\ref{(4.16b)}); after a partial cancellation it becomes $(\bar{g}_{j'}!)^{7s/9}=(\bar{g}_{j'}!)^{2/9}$ in the denominator.
We have chosen constants so as to equalize the factorials between both sides of the Markov inequality. What remains are the denominators: the fact that their $s$-moments are bounded allows us to extract good decay of the probability with $|\bar{g}_{j'}|$. The Jacobian estimate (\ref{(4.16a)}) from the last section shows that we can effectively use the denominators as integration variables. 

In detail, it is convenient to insert a partition of unity for each denominator, according to whether 
$D_{\tau\tilde{\tau}}^{(j')} \equiv E_\tau^{(i')} - E_{\tilde{\tau}}^{(i')}$ is bounded by 1 or not. The choice of which ones are smaller than 1 entails a factor 2 per denominator, in total no more than an exponential in $|\bar{g}_{j'}|$. For the large denominators, we have $|D_{\tau\tilde{\tau}}^{(j')}|^{-s} \le 1$. For the small ones, we use (\ref{(4.16a)}) to estimate
\begin{equation}
\label{(4.17a)}
\mathbb{E}\,\prod_{\tau \tilde{\tau}}|D_{\tau\tilde{\tau}}^{(j')}|^{-s}
\le \int \prod_{\tau \tilde{\tau}}\left[|D_{\tau\tilde{\tau}}^{(j')}|^{-s}\,dD_{\tau\tilde{\tau}}^{(j')}\right]\bigg| \det \frac{\partial D ^{(j')}_{\tau \tilde{\tau}}}{\partial \tilde{h}_y} \bigg|\prod_yC(y) 
\le c^n\prod_yC(y).
\end{equation}
Here $n$ is the number of small denominators; $C(y) = \frac{1}{2}\rho_0$ for sites (with the rescaling $\tilde{h}_y = 2h_y$, its density is half that of $h_y$); $C(y) = c_b^{|\bar{\bar{b}}^{(i)}|}c_{\text{area}}^{L_i}$ for blocks (from rescaling and polar coordinates, as in the (\ref{(4.8)}I) bound above).
As explained in the first paragraph of this subsection, the factors for blocks are always controlled by an exponential in $|\bar{g}_{j'}|$. Thus there is a constant $\rho_1$ such that
\begin{equation}
P\left(A^{(k)\prov}_{\sigma \tilde{\sigma}} (\bar{g}_{j'}) > (\gamma / \varepsilon)^{|\bar{g}_{j'}|} / (\bar{g}_{j'}!)^{2/9}\right) \le (\rho_1 \varepsilon^s)^{|\bar{g}_{j'}|} / (\bar{g}_{j'}!)^{2/9}. \label{(4.18)}
\end{equation}
We are using the same constant $\rho_1$ for probability bounds for both conditions, (\ref{(4.8)}I) and (\ref{(4.8)}II).
Note that if $c_b$ becomes large, then $\rho_1$ grows as well, but no faster than a fixed power of $c_b$.
(This is determined by the maximum ratio of block sizes to graph size, and is under control because of the comparability of metrics, see the last paragraph of Subsection \ref{4.2.1}.)

Next we consider the general argument for bounding the probability of (\ref{(4.8)}II). The idea is that sections of the graph where repeated sites/blocks occur have to be short -- otherwise the graph cannot reach far enough in $\mathbb{Z}$ to satisfy the condition $|I(\bar{g}_{j'})| \ge \frac{7}{8}|\bar{g}_{j'}|$. Inductive bounds will be applied for sections of $\bar{g}_{j'}$ that overlap with each other. For such sections, there is no gain in the Markov inequality, and no decay as in (\ref{(4.18)}) in the size of the subgraph. But we retain decay on a substantial fraction of $\bar{g}_{j'}$, and so a bound similar to (\ref{(4.18)}) can still be proven.

Let us review some basic facts about the structure of $\bar{g}_{j'}$. It consists of a number of subgraphs $\bar{g}_{(j - 1)'}$, and each $\bar{g}_{(j - 1)'}$ consists of subgraphs $\bar{g}_{(j - 2)'}$, and so on. (We can also have subgraphs two or more levels down.) 
Jump steps are replaced with their bounds $\gamma^{|g_{i''}|}$, as discussed after (\ref{(4.7)}), so they do not appear in these lists of subgraphs.
Thus we have a hierarchical, nested structure. A subgraph corresponding to an $A$ factor has an overall denominator; subgraphs corresponding to $J$ factors have no overall denominators. Any further $A$ subgraphs come with their own overall denominators. Thus the denominators respect the hierarchical organization of $\bar{g}_{j'}$. Each denominator depends to leading order only on the variables within its associated subgraph. When there are no repeated sites, the hierarchical organization of denominators translates to the spatial structure of denominators (and leads to the lower triangularity of the matrix $L$ of the previous subsection). For each subgraph, recall that $I(\bar{g}_{i'})$ denotes the smallest interval in $\mathbb{Z}$ covering all the sites or blocks $\bar{\bar{b}}^{(\tilde{i})}$ with $\tilde{i} \le i$ that contain flips of $\bar{g}_{i'}$. Clearly, if $I(g) \cap I(\tilde{g}) = \varnothing$, then there is no dependence between the variables in $I(g)$ and $I(\tilde{g})$. Any overlap between $I(g)$ and $I(\tilde{g})$ will necessarily shorten the distance $|I(\bar{g}_{j'})|$ that $\bar{g}_{j'}$ can span.

If we look at the entire interval $I(\bar{g}_{j'})$, there will be a set of disjoint segments in $\mathbb{Z}$ where sites/blocks are covered more than once due to repeated flips of $\bar{g}_{j'}$. We call these segments ``looping segments," because the graph is looping back to previously visited sites. We can assume that any two looping segments $T_{\alpha}, T_{\beta}$ have at least one site or block $\bar{\bar{b}}^{(i)}$ between that is not in a looping segment. Let $|T_{\alpha}|$ denote the size of a segment, where each site/block with $n$ visits is weighted by a factor $n - 1$. Visits are counted by looking at the individual flips of $\bar{g}_{j'}$; any jump step counts as a visit to all the sites/blocks covered by the jump. Note that some flips of $\bar{g}_{j'}$ can occur at places that are subsequently subsumed into a block -- nevertheless all the flips inside such a block count as separate visits to the block. Every time $\bar{g}_{j'}$ returns to a previously visited site/block, it fails to extend $I(\bar{g}_{j'})$. Hence $|\bar{g}_{j'}| - |I(\bar{g}_{j'})|$ is at least as large as the sum of the lengths $|T_{\alpha}|$ of the looping segments. 
Therefore, by the condition for short graphs (\ref{(4.8)}II),
\begin{equation}
\sum_\alpha |T_\alpha| \le |\bar{g}_{j'}| - |I(\bar{g}_{j'})| \le \tfrac{1}{8}|\bar{g}_{j'}|.
\label{(4.18a)}
\end{equation}

Let us consider the denominator graph prior to the identification of variables in the looping segments; each flip of $\bar{g}_{j'}$ is associated with an independent variable. All denominators are independent. This can be seen by observing inductively that the property holds for $A$ subgraphs; $J$ subgraphs always have a free variable, which makes the overall denominator on the next scale independent. Of course, a repeat visit to a looping segment forces us to identify variables of the flips with earlier variables, and independence is lost. However, denominators from a sufficiently long length scale do retain their independence. Let $\ell =2\max_{\alpha}|T_{\alpha}|$, and let $i$ be such that $\ell \in [L_{i - 1}, L_i)$. Consider the denominator subgraph $\mathcal{D}_i$ formed by links introduced at step $i$ and afterwards, with graph length in the range $[L_i, L_{i + 1})$. As a subgraph of a graph with independent denominators, the denominators of $\mathcal{D}_i$ are independent. Furthermore, even after the identification of variables, the denominators in $\mathcal{D}_i$ are independent. This is because each denominator covers at least $L_i$ variables, and since identifications occur within disjoint $T_{\alpha}$ with $2|T_{\alpha}| \le \ell < L_i$, there will always be a free variable for each denominator. If there are $n$ visits to a site/block, then $n$ variables are identified there, and the site/block contributes $n - 1$ to $|T_{\alpha}|$. Hence no more than $2|T_{\alpha}|$ variables are lost in $T_{\alpha}$.
(The worst case is when there are two visits per site/block of $T_{\alpha}$, which leads to a loss of two variables for each unit of $|T_{\alpha}|$.)
Once a denominator extends outside of $T_{\alpha}$, it spans an independent variable adjacent to $T_{\alpha}$ -- by construction, there are gaps of size $\ge 1$ between the $T_{\alpha}$, and they contain only singly-visited sites/blocks.) Note that jump steps have no denominators, as they have been replaced with their upper bound. So jump steps do not produce any dependence between variables.

\begin{figure}[h]
\centering
\includegraphics[width=.8\textwidth]{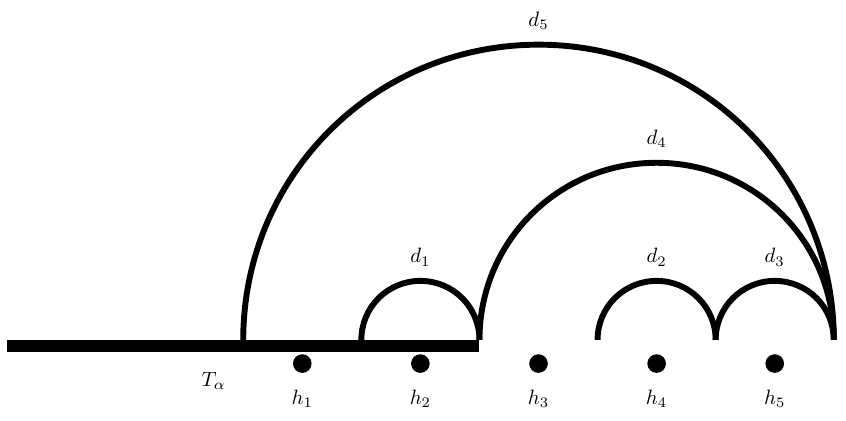}
\caption{Denominators to the right of $T_{\alpha}$. In this example, denominators $d_2, d_3, d_4, d_5$ are no longer independent once $h_1$ and $h_2$ are frozen. Therefore, $d_4$ is dropped, along with subsidiary denominators $d_2, d_3$. This frees up $h_3$. Then $d_5$ has an independent variable. \label{erase}}
\end{figure}

We now describe the ``erasure'' procedure; in particular we give an algorithm for determining
the set of erased subgraphs.
As we add denominators from the $(i - 1)^{\text{st}}$ step and below, some will be internal to one of the looping segments (\textit{i.e.}, all the variables on which the denominator depends are in the looping segment). We will have to replace the corresponding $A$'s by uniform bounds $(\gamma / \varepsilon)^{|g|}/ (g!)^{2/9}$ from (\ref{(4.2)}). The denominator is effectively erased from the denominator graph, along with all denominators nested inside. There may be denominators on scale $i - 1$ or above that connect $T_{\alpha}$ to its complement. (Again, we may visualize a denominator as an arch that encompasses all of its variables.) In order to keep those denominators independent, we may need to drop (\textit{i.e.} erase) a denominator on scale $i - 1$ or below on either side of $T_{\alpha}$. This is so that a variable on either side of $T_{\alpha}$ is freed up ($\textit{i.e.}$ it is no longer used to integrate short denominators, so it is available for integrating long denominators.) Thus, when necessary, we choose a denominator that extends at least one step away from $T_{\alpha}$. It may start inside of $T_{\alpha}$ or at the first site outside of $T_{\alpha}$. In order to free up the denominator on scale $i$, we need to erase a denominator that is directly subsidiary to it, $\textit{i.e.}$ up to scale $i - 1$. See Figure \ref{erase}.

Through this construction, we obtain a denominator graph $\calD_{i - 1}$, consisting of non-erased denominators on scales $\ge i - 1$. All of the denominators are independent, even with the variables in $T_{\alpha}$'s with $|T_{\alpha}| \in [L_{i - 1}, L_i)$ frozen. We continue through shorter length scales, erasing denominators as needed to preserve independence after the freezing of variables in the $T_{\alpha}$. Each looping segment with $|T_{\alpha}| \in [L_{i - 1}, L_i)$ may have a collar of erased sections of width $< L_i$ on each side. The looping segment ``spoils" an interval of size no longer than $2|T_{\alpha}| + 2L_i \le (2 + 2 \cdot \frac{15}{8}) |T_{\alpha}| < 6 |T_{\alpha}|$. (One could do better with a more detailed analysis.) Recall that the total length of all the looping segments is no greater than $\frac{1}{8}|\bar{g}_{j'}|$. Thus the total length of the ``spoiled" intervals where non-probabilistic bounds are employed is no greater than $\frac{3}{4}|\bar{g}_{j'}|$. This leaves at least $\frac{1}{4}$ of $|\bar{g}_{j'}|$ free of dependence issues, \textit{i.e.} with either independent
denominators or jump steps. Therefore, we will get a comparable number of factors of $\varepsilon^s$ from the Markov inequality.

Note that block variables are automatically eliminated with this procedure, because graphs traversing $\bar{\bar{b}}^{(i)} \setminus \bar{b}^{(i)}$ are treated as multiple visits to $\bar{\bar{b}}^{(i)}$, which leads to erasures near the block. They cannot be eliminated when estimating probabilities for case I of (\ref{(4.8)}). However, that is a single-denominator estimate, so the issue of correlation between denominators does not arise.

We may now return to the probability bound as in (\ref{(4.18)}), only now any graph with $|I(\bar{g}_{j'})| \ge \frac{7}{8} |\bar{g}_{j'}|$ is allowed. We claim that
\begin{equation}
P \left(A^{(k)\prov}_{\sigma \tilde{\sigma}} (\bar{g}_{j'}) > \frac{(\gamma / \varepsilon)^{|\bar{g}_{j'}|}}{(\bar{g}_{j'}!)^{2/9}}\right) \le (\rho_1 \varepsilon^s)^{|\bar{g}_{j'}|/4} / (\bar{g}^\text{e}_{j'}!)^{2/9}. \label{(4.19)}
\end{equation}
Here $\bar{g}^\text{e}_{j'}!$ is a modified factorial,
which contains only the factorials in non-erased sections.
As described above, the erased sections of $g_{j'}$ contribute factors of $(\gamma / \varepsilon)^{|g|}/(g!)^{2/9}$ instead of $\gamma^{|g|}/g!$ in the expectation. Thus they match up with corresponding factors on the other side of the inequality $A^{(k)\prov}_{\sigma \tilde{\sigma}} (\bar{g}_{j'}) > (\gamma / \varepsilon)^{|\bar{g}_{j'}|}/(\bar{g}_{j'}!)^{2/9}$. Hence they contribute no smallness to the bound on the probability, and we are forced to make due with the modified factorial $\bar{g}^\text{e}_{j'}!$ in (\ref{(4.19)}). But at least $1/4$ of $\bar{g}_{j'}$ is clear of dependence issues, so we are able to glean $|\bar{g}_{j'}|/4$ factors of $\varepsilon^s$ in the Markov inequality. 
Recall that the graph $\bar{g}_{j'}$ is defined from $g_{j'}$ by forgetting the order of subgraphs in erased sections. The order only affects denominators internal to erased sections; the rest of the graph is unaffected, as it only involves the energies at the start and finish of erased sections. So in $\bar{g}_{j'}$ we sum over the usual graphical structures, but some subgraphs are specified as erased, which means any further collections of subgraphs are ordered from left to right in $\mathbb{Z}$. We do not need to worry about whether these graphs' initial and final states match up properly. The point is to organize the event sum so that the erased factorials are not needed. The single event $\{A^{(k)\text{prov}}_{\sigma \tilde{\sigma}}(\bar{g}_{j'}) \le (\gamma / \varepsilon)^{|\bar{g}_{j'}|}/ (\bar{g}_{j'}!)^{2/9}\}$ 
ensures that every $A^{(k)}_{\sigma \tilde{\sigma}}(g_{j'})$ is similarly bounded, for
any $g_{j'}$ that reduces to $\bar{g}_{j'}$ after the erasure procedure. So there is no point in defining separate events for each $\bar{g}_{j'}$. Note that sums over subgraphs in erased sections are organized from left to right in $\mathbb{Z}$, so no factorials are needed to control them. (The leftmost point of any subgraph starts at a point within or immediately adjacent to the previous subgraph, so there is no factor that depends on the number of subgraphs in a collection.) This only works for event sums; graphs such as $A^{(k)}_{\sigma \tilde{\sigma}}(g_{j'})$ depend on the ordering of all their subgraph collections, so in graphical expansions we retain that structure and the factorials that go with it.

The Markov inequality produces decay from integrating denominators only in non-erased sections of a graph. One might ask what is the purpose of carrying along the erased sections, whose contributions cancel out in (\ref{(4.19)}). The answer is that in order to control event sums, a graphical connection is needed between initial and final states of erased sections. Then the complete sum over resonant events can be controlled -- using only the decay from non-erased sections, which constitute at least $\frac{1}{4}$ of the length of the graph. Note that the order of subgraphs in erased sections is not needed to construct a connection between initial and final state -- any path will do.

\subsubsection{Block Probabilities}\label{4.2.4}

The following proposition gives the core result on the exponential decay of the connectivity function for the resonant blocks $B^{(k)}$.
\begin{proposition}\label{prop:4.2}
Let $P^{(k)}_{xy}$ denote the probability that $x, y$ lie in the same block $B^{(k)}$. For a given $c_b$, 
 let $\varepsilon = \gamma^{1/20}$ be sufficiently small, and assume \textbf{A1}($c_b$). Then
\begin{equation}
P^{(k)}_{xy} \le (c \rho_1 \varepsilon^s)^{(|x - y|^{(j)} \vee L_j)/8}. \label{(4.20)}
\end{equation}
\end{proposition}
Recall that $j = k - 1$ and $|x - y|^{(j)}$ is the metric in which blocks $\bar{\bar{b}}^{(1)}, \ldots, \bar{\bar{b}}^{(j)}$ are contracted to points. Uniform comparability of $|x - y|^{(j)}$ with $|x - y|$ was established in Subsection \ref{4.2.1}. 

\textit{Proof.} The bound (\ref{(4.19)}) ensures that there is a positive density of factors of $\varepsilon^s$ as we estimate $P^{(k)}_{xy}$. As in the proof of (\ref{(3.7)}), we lose a factor of two in the decay from the geometric mean construction. Other aspects of the sum over graphs were described in Subsection \ref{4.2.1}, in particular the fact that only (2/9)$^{\text{th}}$-power factorials are needed to control sums over collections of subgraphs. As we just discussed, those factorials are not needed for erased sections, since we are now summing over coarser partitions of events based on $\bar{g}_{j'}$, but still satisfying the bound (\ref{(4.19)}). Resonant graphs in the $k$\textsuperscript{th} step have $L_j \le |g_{j'}| < L_{j + 1}$, which leads to the minimum in the exponent above.
Thus we obtain (\ref{(4.20)}). \qed

Next we consider the small block connectivity functions.
\begin{proposition}\label{prop:4.3}
Let $c_b$ be given.
Let  $Q^{(k)}_{xy}$ denote the probability that $x,y$ lie in the same small block $\bar{b}^{(k)}$.  There exists a constant $c_2 > 0$ such that for $\varepsilon = \gamma^{1/20}$ sufficiently small, \textbf{A1}($c_b$) implies 
\begin{equation}
Q^{(k)}_{xy} \le (c \rho_1 \varepsilon^s)^{1 + c_2 k^2} \mathbbm{1}_{\{|x - y| \le 4 L_k\}}. \label{(4.23)}
\end{equation}
Let $R^{(k)}_{xy}$ denote the probability that $x,y$ lie in the same small block $\bar{b}^{(i)}$ on any scale $i \le k$. There exists a constant $c_3 > 0$ such that for $\varepsilon = \gamma^{1/20}$ sufficiently small, \textbf{A1}($c_b$) implies
\begin{equation}
R^{(k)}_{xy} \le (c \rho_1 \varepsilon^s)^{1 + c_{3} (\log (|x - y| / 4 \vee 1))^2} \mathbbm{1}_{\{|x - y| \le 4 L_k\}}. \label{(4.24)}
\end{equation}
\end{proposition}
\textit{Proof.} A block $\bar{b}^{(k)}$ can arise from a fairly spaced-out collection of resonant blocks from earlier scales. However, the factors of $\varepsilon^s$ in (\ref{(4.20)}) control the sum over core sets $b^{(k)}$ consistent with the event $Q^{(k)}_{xy}$. Let us break up $b^{(k)}$ into components $\mathcal{C}^{(m)}$ by connecting any pair of sites with separation distance $< d_{m + 1}$. Each component $\mathcal{C}^{(m)}$ has 1 or more subcomponents $\mathcal{C}^{(m - 1)}$, and so on. If there is more than one subcomponent, the sum over each separation distance produces a combinatoric factor $d_{m + 1}$.
Our separation rules state that if a subset has volume in $[ L_m, L_{m + 1})$ and is separated by a distance $d_m$, then it would form a small block $b^{(j)}$ on an earlier scale,
in which case it would not be part of $b^{(k)}$. 
Hence any component $\mathcal{C}^{(m - 1)}$ of $b^{(k)}$ (separated as it is from other such components by at least a distance $d_m$) has a minimum volume $L_{m+1}$.
Thus the combinatoric factor per site of $\mathcal{C}^{(m - 1)}$ is bounded by $d_{m + 1}^{1 / L_{m + 1}}$. If we combine the combinatoric factors per site produced by subcomponent sums on all scales we obtain
\begin{equation}
\prod_m (d_{m + 1})^{1 / L_{m + 1}} = \exp\bigg(\sum_m L_{m + m_0 + 1}^{1/2} L^{-1}_{m + 1}\bigg) \le c_0. \label{(4.20a)}
\end{equation}
Since (\ref{(4.20)}) provides a factor $\varepsilon^s$ for each site of $b^{(k)}$, it should be clear that the sum over $b^{(k)}$ is under control.

We established in Subsection \ref{4.2.1} that the separation conditions imply that any small block with core volume $|b^{(k)}| \in [L_{m - 1}, L_m)$ has diameter less than $(L_m - 2)d_{m - 1}$. But note that if $(L_m - 2)d_{m - 1} < L_{k - 1}$, then $b^{(k)}$ would have satisfied the diameter conditions diam$(b^{(k)}) < L_k$ in an earlier step. Hence all blocks $b^{(k)}$ satisfy a minimum volume condition:
\begin{equation}
|b^{(k)}| \ge L_{m - 1}, \text{where $m$ is the smallest integer such that} \; (L_m - 2)d_{m - 1} \ge L_{k - 1}. \label{(4.21)}
\end{equation}
Recall that $d_m \equiv \exp(L^{1/2}_{m + m_0})$ and $L_k \equiv (\frac{15}{8})^k$. Thus (\ref{(4.21)}) implies that $L_{m - 1 + m_0} \ge c_1 (k - 1)^2$, which means that
\begin{equation}
|b^{(k)}| \ge 1 + c_2 k^2 \ge 1 + c_3 (\log (|x - y| / 4 \vee 1))^2. \label{(4.22)}
\end{equation}
Here we make use of the fact that if $x,y \in \bar{b}^{(k)}$, $|x - y| \le 4 L_k$. (The bound $|\bar{b}^{(k)}| \le 4 L_k$ follows as in the discussion at the end of Subsection \ref{4.2.1}: the diameter of $b^{(k)}$ is less than $L_k$ in the metric $|\cdot|^{(j)}$; add a collar of width $L_k$ on each side and allow for a small expansion of distance due to blocks on smaller scales.) The key aspect of (\ref{(4.20)}) is that it establishes a minimum density of factors of $\varepsilon^s$ throughout the core volume of $b^{(k)}$. (This includes the extra volume coming from the minimum volume condition on $B^{(k)}$, established just below (\ref{dist}).) Thus we obtain (\ref{(4.23)}). 

Note that $R^{(k)}_{xy}$  is a sum of $Q^{(i)}_{xy}$ over $i$ such that $|x - y| \le 4L_i \le 4L_k$. Summing (\ref{(4.23)}) over $i \le k$, we obtain (\ref{(4.24)}). \qed

We have obtained a rate of decay uniform in $k$ for the probability that $x,y$ belong to the same small block. The decay is faster than any power of $|x - y|$. This is the rate that governs our estimates on averaged correlations. Statements regarding exponential decay of correlations with probability 1 depend on the low density of blocks at each scale, decreasing with $k$ so that $|x - y|^{(k)}$ remains comparable with $|x - y|$. The situation parallels that of \cite{Frohlich1983}, which established exponential decay with probability 1 in the one-body context.

\subsection{Perturbation Step and Proof of Inductive Bounds}\label{4.3}
Having established diluteness of the resonant blocks in Section \ref{4.2}, we focus here on deterministic estimates on graphs in the nonresonant region.
\subsubsection{Bounds on $A^{(k)}$}\label{4.3.1}
Here we perform the rotations that eliminate low-order interactions from the Hamiltonian. Then, after resumming long graphs, we will obtain the inductive bound (\ref{(4.2)}) for $k = j+1$.

As in Section 3.2, we write
\begin{equation}
J^{(j')} = J^{(j')\text{per}} + J^{(j')\text{res}}, \label{(4.25)}
\end{equation}
with
\begin{align}
 J^{(j')\text{per}}_{\sigma \tilde{\sigma}} &= \sum_{g_{j'}: \sigma \rightarrow \tilde{\sigma}, \, L_{k - 1} \le |g_{j'}| < L_k,\, g_{j'} \cap \calS_k = \varnothing, \,\sigma \ne \tilde{\sigma}} J^{(j')}_{\sigma \tilde{\sigma}} (g_{j'}), \label{(4.26)} \\
 A^{(k)}_{\sigma \tilde{\sigma}} &= \sum_{g_{j'}: \sigma \rightarrow \tilde{\sigma}} A^{(k)}_{\sigma \tilde{\sigma}}(g_{j'}) = \sum_{g_{j'}: \sigma \rightarrow \tilde{\sigma}} \frac{J^{(j')\text{per}} (g_{j'})}{E^{(j')}_{\sigma} - E^{(j')}_{\tilde{\sigma}}}. \label{(4.27)}
\end{align}

\textit{Long and short graphs; jump transitions.} 
We say a graph $g_{j'}$ is \textit{long} if $|g_{j'}| > \frac{8}{7} |I(g_{j'})|$. Otherwise, it is \textit{short}.
We will need to resum terms with long graphs, for given initial and final spin configurations $\sigma, \tilde{\sigma}$ and a given interval $I = I(g_{1'})$.
The data $\{\sigma, \tilde{\sigma}, I\}$ determine a \textit{jump transition}. 
Long graphs are extra small, by (\ref{(4.2)}), so for probability estimates we do not need to keep track of individual graphs, and we can take the supremum over the randomness. 
Let $g_{j''}$ denote either a short graph from $\sigma$ to $\tilde{\sigma}$ or a  jump transition taking $\sigma$ to $\tilde{\sigma}$ on an interval $I$.
The length of $g_{j''}$ is defined to be $|g_{j''}|=|I|\vee \tfrac{7}{8}L_j$.
The jump transition represents the collection of all long graphs from $\sigma$ to $\tilde{\sigma}$ with a given $I(g_{j'})$.
Then put
\begin{equation}
A^{(k)}_{\sigma \tilde{\sigma}}(g_{j''}) = 
\begin{cases}
A^{(k)}_{\sigma \tilde{\sigma}}(g_{j'}), \; \text{if} \; g_{j''} = g_{j'}, \text{a short graph};\\
\sum\limits_{\text{long} \, g_{j'}: \sigma \rightarrow \tilde{\sigma}} A^{(k)}_{\sigma \tilde{\sigma}}(g_{j'}),\; \text{if} \; g_{j''} \; \text{is long}.
\end{cases}
\label{(4.28)}
\end{equation}
With $\Omega^{(k)} = \exp(-A^{(k)})$, we define $H^{(k)} = \Omega^{(k)\tr} H^{(j')} \Omega^{(k)}$ and then as in (\ref{(3.13)}) we can write
\begin{equation}
H^{(k)} = H^{(j')}_0 + J^{(j')\res} + J^{(j)\lint} + J^{(k)}, \label{(4.29)}
\end{equation}
with
\begin{equation}
J^{(k)} = \sum\limits_{n = 1}^{\infty} \frac{n}{(n + 1)!} (\text{ad}\,A^{(k)})^n J^{(j')\text{per}} + \sum\limits_{n = 1}^{\infty} \frac{(\text{ad}\,A^{(k)})^n}{n!} J^{(j')\res}. \label{(4.30)}
\end{equation}

We may now prove the inductive bound (\ref{(4.2)}) for $k = j + 1$. 
\begin{proposition}\label{prop:A}
Let $\gamma$ be sufficiently small. Then
\begin{equation}
|A^{(k)}_{\sigma \tilde{\sigma}} (g_{j''})| \le
\begin{cases}
(\gamma / \varepsilon)^{|g_{j''}|} / (g_{j''}!)^{2/9}, \;\text{in general}; \\
\gamma^{|g_{j''}|}, \;\text{if} \; g_{j''} \; \text{is a jump step}. 
\end{cases}
\label{(4.29a)}
\end{equation}
\end{proposition}
\textit{Proof.} For short graphs, we have
\begin{equation}
|A^{(k)}_{\sigma \tilde{\sigma}} (g_{j'})| \le (\gamma / \varepsilon)^{|g_{j'}|} / (g_{j'}!)^{2/9},
\label{(4.31)}
\end{equation}
by the resonant condition (\ref{(4.8)}II). For long graphs, we bound numerator and denominator separately in (\ref{(4.27)}). The inductive bound (\ref{(4.6)}) applies to the numerator, and the resonant condition (\ref{(4.8)}I) bounds the denominator from below. After summing over the long graphs that contribute to $g_{j''}$, we have
\begin{equation}
A^{(k)}_{\sigma \tilde{\sigma}}(g_{j''}) \le (c \gamma / \varepsilon^2)^{\frac{8}{7}|I(g_{j''})| \vee L_{k - 1}}, \label{(4.32)}
\end{equation}
because all long graphs have $|g_{j'}| \ge \frac{8}{7}|I(g_{j'})| \vee L_{k - 1} = \frac{8}{7}|g_{j''}|$, see definition (\ref{(4.4)}). Recall that $\gamma = \varepsilon^{20}$, so $c^{8/7}\gamma^{1/7} \varepsilon^{-16/7} \le 1$, and we obtain
\begin{equation}
|A^{(k)}_{\sigma \tilde{\sigma}}(g_{j''})| \le \gamma^{|g_{j''}|}, \label{(4.33)}
\end{equation}
which completes the proof. (Since (\ref{(4.33)}) is a stronger estimate, we actually have $|A^{(k)}_{\sigma \tilde{\sigma}}(g_{j''})| \le (\gamma / \varepsilon)^{|g_{j''}|}/ (g_{j''}!)^{2/9}$ for all $g_{j''}$ -- recall that $g_{j''}!$ is defined without any factorials from jump steps.) \qed

\subsubsection{Bounds on $J^{(k)}$}\label{4.3.2}
In this subsection we prove the bound (\ref{(4.6)})   for $k = j + 1$, while at the same time giving details on how $J^{(k)}$ is expressed as a sum of graphs. By (\ref{(4.30)}), $J^{(k)}$ is a sum of terms involving one or more commutators of $J^{(j')}$ with $A^{(k)}$. The simplest situation is when there is no gap between $g_{j''}$ and $g_{j'}$ in $[A^{(k)}_{\sigma \tilde{\sigma}} (g_{j''}), J^{(j')}_{\tau \tilde{\tau}} (g_{j'})]$. Here we need to be careful about what we mean by a gap. For any block $\bar{b}^{(i)}$ involved in either graph, we use the fattened version $\bar{\bar{b}}^{(i)}$, defined as $\bar{b}^{(i)}$ plus a collar of width $\frac{15}{14} L_{i - 1}$. Then we specify that gaps do not include any sites in any of the $\bar{\bar{b}}^{(i)}$ involved in the graphs on either side. Obtaining decay in $\bar{\bar{b}}^{(i)}$ is problematical, because of the dependence of the block energies of $\bar{b}^{(i)}$ on variables in $\bar{\bar{b}}^{(i)}\backslash \bar{b}^{(i)}$. But we deal with the wider collars by contracting blocks $\bar{\bar{b}}^{(i)}$ to points when defining the metric $|\cd|^{(i)}$. With this definition in mind, consider the case with no gap. Then $g_{j''}$ and $g_{j'}$ are combined in the new graph $g_k$ for $J^{(k)}$, and we may bound the terms of the commutator separately, using (\ref{(4.2)}), (\ref{(4.6)}). This leads to an estimate
\begin{equation}
|J^{(k)}_{\sigma \tilde{\sigma}}(g_k)| \le \gamma (\gamma / \varepsilon)^{|g_k| - 1} / (g_k !)^{2/9}, \label{(4.34)}
\end{equation}
which matches up with (\ref{(4.6)}).

If there is a gap between $g_{j''}$ and the graph generated by $J^{(j')}$ or by previous commutators with $J^{(j')}$, then we need to exploit cancellation between terms to obtain decay in the gap. Let us consider the case of a single commutator $[A^{(k)}_{\sigma \tilde{\sigma}} (g_{j''}), J^{(j')}_{\tau \tilde{\tau}} (g_{j'})]$. The two terms $AJ$ and $JA$ differ in that energy denominators in $A$ are computed two ways, that is, before and after the transition $\tau \rightarrow \tilde{\tau}$. Likewise, the energy denominators in $J$ are computed before and after the transition $\sigma \rightarrow \tilde{\sigma}$. But let us focus on the effect the transition $\tau \rightarrow \tilde{\tau}$ has on $A$;
the effect that $A$ has on $J$'s denominators is similar.
 The energies in $A$'s denominators are $E^{(i')}_{\nu}$ for various spin configurations $\nu$ and scales $0 \le i \le j$. Differences $D^{(i')}_{\nu \tilde{\nu}} = E^{(i')}_{\nu} - E^{(i')}_{\tilde{\nu}}$ have a graphical expansion, see (\ref{(4.11)}). The expansion exhibits the non-local dependence on the spin configuration. For each denominator in $A^{(k)}_{\sigma \tilde{\sigma}} (g_{j''})$, we write
\begin{equation}
\frac{1}{D^{(i')}_{\nu \tilde{\nu}}(\tau)} - \frac{1}{D^{(i')}_{\nu \tilde{\nu}}(\tilde{\tau})} = \frac{D^{(i')}_{\nu \tilde{\nu}}(\tilde{\tau}) - D^{(i')}_{\nu \tilde{\nu}}(\tau)}{D^{(i')}_{\nu \tilde{\nu}}(\tau) D^{(i')}_{\nu \tilde{\nu}}(\tilde{\tau})}, \label{(4.35)}
\end{equation}
where the dependence on $\tau$ or $\tilde{\tau}$ is written explicitly. The commutator can be written as a sum of terms switching each denominator in turn from $\tau$ to $\tilde{\tau}$ using (\ref{(4.35)}). In the numerator we write
\begin{equation}
D^{(i')}_{\nu \tilde{\nu}}(\tilde{\tau}) - D^{(i')}_{\nu \tilde{\nu}}(\tau) = \sum\limits^{i - 1}_{m = 1} \sum_{g_{m'}} \delta_{\tau \tilde{\tau}} J ^{(m')}_{\sigma\sigma} (g_{m'}), \label{(4.36)}
\end{equation}
where $\delta_{\tau \tilde{\tau}}$ takes the difference between values at $\tau$ and $\tilde{\tau}$. As in the discussion following (\ref{(4.11)}), we repeat the process by applying (\ref{(4.35)}) to the denominators of $J ^{(m')}_{\sigma\sigma} (g_{m'})$. We are doing a discrete version of the chain rule of Subsection \ref{4.2.2} to probe the dependence on $\tau$. (Here, however, we use fattened blocks $\bar{\bar{b}}^{(i)}$, so we do not need to worry about dependence of block energies on $\tau$; we are investigating only their perturbative corrections.) The process can stop for any term whose chain of graphs crosses the gap between $g_{j''}$ and $g_{(j - 1)'}$. As in Subsection \ref{4.2.2}, each jump step of $g_{j''}$ is written as the sum of its constituent graphs, see (\ref{(4.28)}). This is necessary so that we may probe the dependence on $\tau$ everywhere it occurs. Each step of the process lowers the scale index on the denominators by one or more. At some point, the graphs must span the gap since otherwise the last $\delta J$ would vanish -- the energies $E^{(1')}$ depend only on $\tau$ one step away.

There is a limit to the range of dependence of energies on $\tau$. Energies $E^{(m')}$, $m\le j$ appear in $A^{(k)}_{\sigma \tilde{\sigma}} (g_{j''})$. They are diagonal entries $J^{(m-1)'}_{\sigma\sigma} (g_{(m-1)'})$, so their order is $< L_{m}$. This means the range is $< \frac{1}{2} L_{m}$, since $g_{(m-1)'}$ has to double back to undo any flips performed in its first half. But energies $E^{(\tilde{m}')}$ appear in these graphs for $\tilde{m}<m$, extending the range of dependence. The greatest possible total range is for a sequence $g_{(j - 1)'}, g_{(j - 2)'}, \ldots, g_{1'}$, leading to a maximum range of 
\begin{equation}\label{range}
\tfrac{1}{2}(L_j + L_{j - 1} + \ldots) = \tfrac{1}{2}L_j (1 + \tfrac{8}{15} + \ldots) \le \tfrac{15}{14} L_j.
\end{equation}
As mentioned earlier, this bound is important because it limits the number of spin configurations for which we need to control resonance probabilities. An exponential number of configurations is controlled by exponentially small resonance probabilities. This calculation validates the definition of $\bar{\bar{b}}^{(k)}$ as $\bar{b}^{(k)}$ plus a collar of width $\frac{15}{14}L_j$. (It shows that $\bar{\bar{b}}^{(k)}$ contains the region of dependence of the interactions $J^{(j')}_{\sigma\tilde{\sigma}} (\tilde{g}_{j'})$ for $\tilde{g}_{j'}$ contained in $\bar{b}^{(k)}$; these are the only ones involved when block rotations are performed -- see (\ref{(4.38)}).)

The double-back nature of the energy graphs implies that their graph length $|g_{i'}|$ is at least twice the length of the intervals spanned, $I(g_{i'})$. Hence they all become jump steps $g_{i''}$ with an improved rate of decay $\sim \gamma^{|g_{i''}|}$ as in (\ref{(4.2)}), instead of $(\gamma / \varepsilon^2)^{|g_{i'}|}$. There is an important caveat, however: the first doubled denominator from the chain rule may be greater in span than the size of the energy graphs generated. This would happen if the gap between $g_{j''}$ and $g_{(j - 1)'}$ is smaller than $L_j$. (Subsequent double denominators are internal to $g$ and they may be bounded using (\ref{(4.8)}I), leading to the $(\gamma / \varepsilon^2)^{|g|}$ bound.) See Figure \ref{gap}. 
\begin{footnotesize}
\begin{figure}[h]
\centering
\includegraphics[width=.9\textwidth]{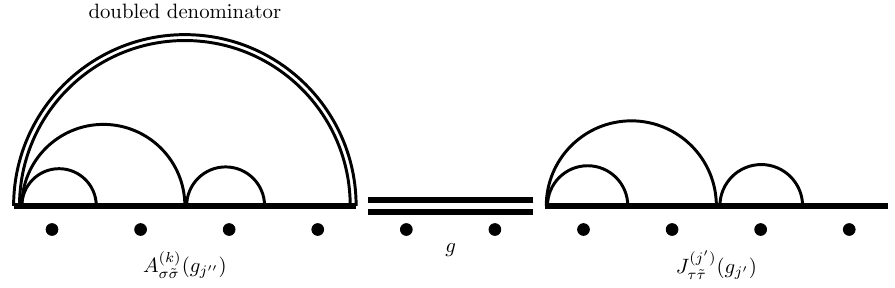}
\caption{Energy graph $g$ is double the length of the gap between $g_{j''}$ and $g_{j'}$. A denominator in $A^{(k)}_{\sigma \tilde{\sigma}}(g_{j''})$ is doubled because the commutator creates a difference, which is re-expressed using (\ref{(4.35)}). \label{gap}}
\end{figure} 
\end{footnotesize}
As a result, the first double denominator has to be treated probabilistically (Markov inequality) along with all the other non-jump step denominators in $A^{(k)}_{\sigma \tilde{\sigma}}(g_{j''})$. Otherwise the extra $\varepsilon^{-1}$ factors -- potentially as many as $|g_{j''}|$ -- would lead to non-uniformity of the decay estimates with $k$. Further details on this will be given below -- see the last paragraph of this section. 

Our immediate goal is to ensure that the expansion can be organized so that not too many extra denominators arise. To this end, consider what happens when 
multi-commutators from $(\text{ad}\,A^{(k)})^n$ act on $J^{(j')}$. We need to demonstrate that decay at rate $\gamma$ can be obtained for all gaps. But now that we are dealing with graphs with doubled denominators, it could happen that in a later step a doubled denominator gets differenced again as in (\ref{(4.35)}). We need to cap the multiplicity of denominators at 3, because unlimited powers would force the fractional moment exponent $s$ to zero. We allow a denominator to be differenced to close a gap to the left and to the right. 
See Appendix \ref{A} for a discussion of how we organize these gap-closing differences to achieve this, while at the same time maintaining manageable bounds on graphical sums.
The net result is a graphical expansion for $(\text{ad}\,A^{(k)})^n J^{(j')}$ as a sum of graphs $g_k$ involving the usual subgraphs $g_{j'}$ plus a collection of jump step graphs $g_{i''}$ with $i < j$ that connects them all together. We call them \textit{gap graphs}. 
Thus gap graphs consist of all the energy graphs from (\ref{(4.36)}) that were generated in the process of bridging the gaps. The doubled denominators themselves remain with the main portion of $g_k$, unless they were already in a jump step. The gap graphs can be summed up, and as for other jump steps, we obtain decay at rate $\gamma$ across gaps.
We obtain the required bound (\ref{(4.34)}), now with an understanding that $g_k$ includes the additional gap graphs as subgraphs. As with the other jump steps, gap graphs are ``spectator" graphs -- replaced with uniform bounds -- in the Markov inequality. 
This means that, like jump steps, they do not contribute factorials to (\ref{(4.3)}).
Note that the ``active" denominators do not depend on the ``spectator" parts of the graph as they see only the initial and final configuration of the jump.

The bound just proven for $J^{(k)}$ leads to the corresponding estimate (\ref{(4.6)}) for $J^{(k')}$ after the block rotations are performed. Note that by (\ref{(4.4)}), the minimum size of a graph $g_{j''}$ in an $A^{(k)}$ term is $\frac{7}{8}  \, L_{k - 1}$. The minimum size of a $J^{(j')}$ graph is $L_{k-1}$. Combining these, we obtain a minimum size of $\frac{15}{8} L_{k-1} = L_k$ for graphs $g_k$. This has been assumed throughout, so it needs to be verified as part of our inductive assumptions.

We return to the issue of the doubled or tripled denominators, and their effect on the Markov inequality. The main difference is that (\ref{(4.17)}) requires $s < \tfrac{1}{3}$ if we want $\mathbb{E} (\Delta E)^{-3s}$ to be finite. This is consistent with our choice $s = \tfrac{2}{7}$. 
Also, it is somewhat inconvenient having the two denominators $D_{\nu \tilde{\nu}}^{(\tau)}$ and $D_{\nu \tilde{\nu}}^{(\tilde{\tau})}$ in (\ref{(4.35)}). So we use a Schwartz or H\"{o}lder inequality to bound the expectation of doubled or tripled denominators by a geometric mean of expectations where multiple denominators are actually 2\textsuperscript{nd} or 3\textsuperscript{rd} powers. Thus it will be sufficient to prove bounds on expectations where the denominators have that structure. This is helpful because the Jacobian bound of Subsection \ref{4.2.2} can then be used to estimate the $-s$ moments of the denominators as before.

\subsection{Diagonalization and Conclusion of Proof}\label{4.4}

We reorganize terms as in Section \ref{3.3}:
\begin{equation}
J^{(j')\text{res}} + J^{(j)\text{lint}} + J^{(k)} = J^{(k)\text{ext}} + J^{(k)\text{sint}} + J^{(k)\text{lint}}. \label{(4.37)}
\end{equation}
Terms whose graph intersects $\calS_k$ and is contained in $\overline{\calS}_k$ are put in $J^{(k)\text{sint}}$ (small block terms) or $J^{(k)\text{lint}}$ (large block terms). Diagonal terms of order less than $L_k$ are included in $J^{(k)\text{sint}}$. This ensures that in the next step, all interactions will be of order at least $L_k$. 

Let $O^{(k)}$ be the matrix that diagonalizes small blocks, so that
\begin{equation}
H_0^{(k')} = O^{(k)\text{tr}} (H_0^{(j')} + J^{(k)\text{sint}}) O^{(k)} \label{(4.38)}
\end{equation}
is diagonal. 
(The diagonal entries of $H_0^{(k')}$ are the energies $E^{(k')}_\sigma$, which now include all effects up to order $L_k-1$.)
Then put
\begin{equation}
H^{(k')} =  O^{(k)\text{tr}} H^{(k)} O^{(k)} = H_0^{(k')} + J^{(k')} + J^{(k)\text{lint}}. \label{(4.39)}
\end{equation}
Here $J^{(k')}$ is the rotated version of $J^{(k)\text{ext}}$. It has a graphical expansion with bounds as in (\ref{(4.5)}), (\ref{(4.6)}), as proven in the last section. Graphs $g_{k'}$ include matrix elements of $O^{(k)}, O^{(k)\text{tr}}$, as appropriate.

Define the cumulative rotation
\begin{equation}
R^{(k')} = R^{(j')} \Omega^{(k)} O^{(k)}. \label{(4.40)}
\end{equation}
Then as in earlier steps we prove that
\begin{equation}
\mathbb{E} \; \text{Av}_{\alpha} \bigg| \sum_{\sigma \tilde{\sigma}} R_{\alpha \sigma}^{(k')\text{tr}} S_0^\mathrm{z} R_{\tilde{\sigma} \alpha}^{(k')} \bigg| = 1 + O(\varepsilon^s). \label{(4.41)}
\end{equation}
The probability that 0 is in a small block is less than $c \rho_1 \varepsilon^s$, by (\ref{(4.24)}). The rotation of $S^\mathrm{z}_0$ generates a graphical expansion much like the one for the rotation of $H$. The leading term is $\pm 1$ and the corrections are $O(\gamma / \varepsilon)$ -- see (\ref{(2.20)}) and the discussion after.

Let us consider the behavior of connected correlations $\langle \mathcal{O}_x ; \mathcal{O}_y \rangle_{\alpha}^{(k)}$. Cancellation of graphs forces graphs to span the distance from $i$ to $j$. Graphs do not penetrate large blocks, and there is no rotation in large blocks, therefore $\langle \mathcal{O}_x ; \mathcal{O}_y \rangle_{\alpha}^{(k)}$ vanishes if any large block intervenes between $i$ and $j$. Suppose that $|x - y|/ 8 \in [L_{m - 1}, L_m)$. Then as in the discussion following (\ref{(4.9)}), no more than half the distance from $x$ to $y$ could be covered by blocks $\bar{b}^{(j)}$ with $j < m$. The probability that a larger scale block covers part of the segment from $x$ to $y$ is bounded by $(c \rho_1 \varepsilon^s)^{1 + \tilde{c}_2 m^2}$, by (\ref{(4.23)}). Hence we have an estimate:
\begin{equation}
|\langle \mathcal{O}_x ; \mathcal{O}_y \rangle_{\alpha}^{(k)}| \le (c \gamma / \varepsilon)^{|x - y| / 2}, \; \text{with probability} \; 1 - (c \rho_1 \varepsilon^s)^{1 + c_3 (\log (|x - y|/ 8 \vee 1))^2}. \label{(4.42)}
\end{equation}
Of course, in step $k$ there are no blocks on scales $>k$, so we would actually have exponential decay for $\langle \mathcal{O}_x ; \mathcal{O}_y \rangle_{\alpha}^{(k)}$ with probability 1 for $|x - y| > 8 L_{k + 1}$. But (\ref{(4.42)}) gives a bound that is valid for all $x, y, k$, so it carries over to the limit $k \rightarrow \infty$. (The limit will be discussed below.) We obtain exponential decay except on a set whose probability decays rapidly with the distance. Averaged correlations are dominated by the probabilities of rare events (\textit{i.e.} blocks). Thus
\begin{equation}
\mathbb{E} \; \text{Av}_{\alpha} |\langle \mathcal{O}_x ; \mathcal{O}_y \rangle_{\alpha}^{(k)}| \le (c \rho_1 \varepsilon^s)^{1 + c_3 (\log (|x - y|/ 8 \vee 1))^2}, \label{(4.43)}
\end{equation}
which decays faster than any power of $|x - y|$, but not exponentially.

If we let the procedure run to $k = \infty$, off-diagonal matrix elements vanish in the limit. Then the eigenvalues of the starting Hamiltonian are given by the diagonal elements of $H^{\infty}_{0} \equiv \text{lim}_{k \rightarrow \infty} H_0^{(k')}$. The eigenfunctions are given by the columns of $R^{(\infty)} \equiv \text{lim}_{k \rightarrow \infty} R^{(k')}$. Block formation has to stop eventually in a finite volume $\Lambda$, by the Borel-Cantelli lemma, because by (\ref{(4.23)}) their probabilities are summable. After that, all the matrices involved converge rapidly. The bounds (\ref{(4.41)})-(\ref{(4.43)}) remain true in the limit, which completes the proof of Theorem \ref{th1.1}, with assumption \textbf{A1}($c_b$) replacing \textbf{LLA}($\nu,C$).

\subsection{Infinite Volume Limit and Local State-Labeling Operators}\label{4.5}

We show that eigenvalue differences and local expectations converge in the $\Lambda \rightarrow \infty$ limit, with probability one. The eigenstates are labeled by spin/metaspin configurations $\alpha$ for each finite volume $\Lambda$. Away from resonant blocks, there is a one-to-one correspondence between an ordinary spin configuration and the state label $\alpha$. As (\ref{(4.41)}) shows, $S^\mathrm{z}$ expectations are close to the labeling configuration values, so in that sense eigenstates resemble the $\gamma = 0$ eigenstates, which are concentrated on the labeling spin configuration. This is analogous to the concentration of eigenfunctions near individual sites in the one-body Anderson model at high disorder. Of course, resonant blocks interfere with this na\"{i}ve labeling scheme, because of mixing and entanglement of the unperturbed states in the block. In a block, labels are assigned when the diagonalization step is performed there. This involves a finite-dimensional matrix that has unique eigenvalues with probability one. (The discriminant cannot vanish on a set of positive measure without being identically zero, and as we explained earlier one can find parameter values where it is nonzero.) Therefore, except for a set of measure zero, one can label block states in order of increasing energy. We have been calling the block states ``metaspins" because like ordinary spin variables, they label the local state in the block. A block of size $n$ has $2^n$ metaspin values, hence they can be put into one-to-one correspondence with ordinary spin configurations in the block. Our constructions and estimates show that local expectations are determined up to errors of order $(\gamma /\varepsilon)^{\ell}$ by the spin/metaspin configuration out to a distance $\ell$ in the $k \rightarrow \infty$ limit.

The abovementioned properties are equivalent to the existence of an extensive set of quasi-local operators that commute with the Hamiltonian \cite{Serbyn2013,Huse2013,Ros2014}. We may construct such operators as follows. Working in the basis we have constructed, in which $H$ is diagonal, define an operator that assigns the spin/metaspin value to each eigenstate possessing that label. Such operators are diagonal in this basis, as is the Hamiltonian. So after returning to the original basis we obtain operators that commute with $H$ and that are quasi-local (because $R^{(\infty)}$ is given by a convergent product of local rotations).

Our procedure produces convergent expressions for the eigenvalues in a box $\Lambda$. However, if we wish to investigate their behavior in the limit $\Lambda \rightarrow \mathbb{Z}$, we should work with eigenvalue differences corresponding to states whose labeling configurations $\alpha, \beta$ differ only locally in a fixed region $\mathcal{R}$. As discussed in Subsection \ref{4.2.2}, eigenvalue differences have graphical expansions with exponential decay localized to $\mathcal{R}$ -- see (\ref{(4.6)}), (\ref{(4.11)}). Graphs generated in step $k$ depend on the random couplings only in a neighborhood of width $O(L_k)$ about the graph (dependence arises because of denominator energies). Expectations of observables localized in $\mathcal{R}$ likewise have local graphical expansions. These expansions can be used to demonstrate convergence of eigenvalue differences and expectations as $\Lambda$ increases to $\mathbb{Z}$ through a sequence of intervals $\Lambda_K \equiv [-K, K]$. 
Convergence of the infinite volume limit was demonstrated in the one-body context in \cite{Imbrie2014a} for eigenfunctions and eigenvalues. We outline a similar argument here for eigenvalue differences and expectations.

When investigating convergence as $\Lambda_K \rightarrow \mathbb{Z}$, it is convenient to use a $K$-independent construction of resonant blocks.
In each step of our procedure, a graph will be considered resonant if it is resonant for any value of $K$. Then, in addition to the usual graphical sums for estimating probabilities for resonances, there is a sum over values of $K$ that lead to distinct resonant conditions for a given graph.
Recall that we sum over every background spin configuration in a neighborhood $\calN$ of $\bar{g}$ so as to catch every possible resonance.
There are no more than diam($\calN$) possibilities for $K$, so this sum can easily be handled along with the background spin sum. With this setup, we maintain probability bounds 
such as (\ref{(4.23)}).

Compare the graphical expansions of an expectation or local energy difference associated with $\calR$ in two different boxes, $\Lambda_{K_1}$ and $\Lambda_{K_2}$, with $K_1 < K_2$. 
The difference involves graphs that extend from $\calR$ to $\Lambda_{K_1}^{\text{c}}$. (Jump steps need to be rewritten as sums of constituent graphs so as to isolate the ones extending to $\Lambda_{K_1}^{\text{c}}$.) Consider the event $\calE_K(\calR)$ in which there exists a path from $\calR$ to $\Lambda_{K}^{\text{c}}$ with length less than $\frac{1}{2}\text{dist}(\calR,\Lambda_{K}^{\text{c}})$,
in the metric where blocks $\bar{\bar{b}}^{(i)}$ are contracted to points for all $i\ge1$. By summing over the two paths from $\calR$ to $\Lambda_{K_1}^{\text{c}}$ and over configurations of blocks along each path, it should be clear that $P(\calE_K(\calR))$ decays exponentially like
$\gamma^{\kappa\text{dist}(\calR,\Lambda_K^{\text{c}})}$ for some 
$\kappa > 0$. By Borel-Cantelli, there is almost surely a $K_0$ such that $\calE_K(\calR)$ fails for all $K > K_0$. Bounds on a given graph
are governed by the distance it covers between blocks. Hence, as long as $K_1 > K_0$, we obtain bounds that decay exponentially in $\text{dist}(\calR,\Lambda_{K_1}^{\text{c}})$.
In this way, we obtain almost sure exponential convergence of local quantities in the $\Lambda \rightarrow \mathbb{Z}$ limit. 

\section{Level Statistics}\label{5}

At this point, we have proven many-body localization (MBL) under assumption $\textbf{A1}(c_b)$, which states that energy differences move to first order with the randomness -- see (\ref{(4.10)}). Here we use MBL as a shorthand for all of our conclusions, including
\begin{enumerate}[(i)]
  \item Existence of a labeling system for eigenstates by spin/metaspin configurations, with metaspins needed only on a dilute collection of resonant blocks.
  \item Bounds on the probability of resonant blocks, (\ref{(4.20)}), (\ref{(4.23)}), (\ref{(4.24)}), which establish their diluteness.
  \item Diagonalization of $H$ via a sequence of local rotations defined via convergent graphical expansions with bounds as in (\ref{(4.2)}).
  \item Bounds establishing closeness of expectations of local observables in any eigenstate to their na\"{i}ve $(\gamma = 0)$ values, when observables are not in resonant blocks. These lead to statements like (\ref{(4.41)}), which show that most states resemble the $\gamma = 0$ states locally.
\item Almost sure convergence of local energy differences and expectations of local observables as $\Lambda \rightarrow \mathbb{Z}$.
\item Exponential decay of connected correlations, except on a set of rapidly decaying probability, see (\ref{(4.42)}).
\item Faster-than-power-law decay of averaged connected correlations as in (\ref{(4.43)}).
\end{enumerate}

We would like to show how MBL can be obtained under weaker assumptions that (1)
allow for violation of the minimum level-spacing condition on a set of small probability, and (2) do not refer to properties of effective Hamiltonians. In effect, $\textbf{A1}(c_b)$ is a working hypothesis that we need in order to continue the induction, but the tools we have developed are flexible enough to prove $\textbf{A1}(c_b)$ with high probability as we go along. (If $\textbf{A1}(c_b)$ fails, we can define a new resonant block and check it again at a longer length scale.) Here is our fundamental assumption on level statistics. It depends on parameters $\nu, \varepsilon_0$:
\begin{assumptiona2}
(Unlikeliness of small eigenvalue differences) Consider the Hamiltonian $H$ in a box of size  $n$. Its eigenvalues satisfy
\begin{equation}
P\Big(\min_{\alpha \ne \beta} |E_{\alpha} - E_{\beta}| < \tilde{\varepsilon}^n\Big) \le \tilde{\varepsilon}^{\nu n}, \label{(5.1)}
\end{equation}
for all $\tilde{\varepsilon} \le \varepsilon_0$ and for all $n$. 
\end{assumptiona2}
\noindent
The exponential decay of probability with $n$ is actually not needed -- we could make do with probability decay similar to (\ref{(4.24)}), since that is what is used to control the diluteness of resonant blocks. But let us work with $\textbf{A2}(\nu, \varepsilon_0)$ for simplicity.

\begin{theorem}\label{th5.1}
Fix  $\nu > 0$ and $\varepsilon_0>0$. Let $\varepsilon = \gamma^{1/20}$ be sufficiently small. If $\textbf{A2}(\nu, \varepsilon_0)$ holds, then MBL holds as well.
\end{theorem}

Before proving this, let us show how $\textbf{A2}(\nu, \varepsilon_0)$ follows from the more standard statement on level statistics that appears on the introduction:

\begin{assumptionlla}(Limited level attraction) Consider the Hamiltonian $H$ in a box of size $n$. Its eigenvalues satisfy
\begin{equation}
P\Big(\min_{\alpha \ne \beta} |E_{\alpha} - E_{\beta}| < \delta\Big) \le \delta^{\nu} C^n, \label{(5.2)}
\end{equation}
for all $\delta > 0$ and all $n.$ 
\end{assumptionlla}

Clearly, we may take $\delta = \tilde{\varepsilon}^n$ in (\ref{(5.2)}) and then $\textbf{A2}(\nu',\varepsilon_0)$ holds for any $\nu' < \nu$, provided $\varepsilon_0$ is small enough (depending only on $C, \nu, \nu'$). Thus we have

\begin{corollary}\label{co5.2}
Let $\nu, C$ be fixed, and let $\gamma$ be sufficiently small. Assume \textbf{LLA}$(\nu, C)$. Then MBL holds.
\end{corollary}
As explained earlier, we can handle level statistics that are neutral (like Poisson, $\nu = 1$), or repulsive ($\nu>1$ as for GOE), and we can handle values of $\nu$ smaller than 1, which correspond to level attraction. Thus we obtain MBL, provided there is a bound uniform in $n$ on the level attraction exponent $\nu$. Note that, speaking broadly, level statistics are expected to be repulsive or neutral. However, we are not aware of any general result of that type.

\textit{Proof of Theorem \ref{th5.1}.} We need to verify $\textbf{A1}(c_b)$ on a set of sufficiently high probability. The idea is to compare the energy differences of $H_0^{(j - 1)'} + J^{(j)\text{sint}}$ that are associated with the block $\bar{b}^{(j)}$ with those of another Hamiltonian $\bar{\bar{H}}$ in the volume $\bar{\bar{b}}^{(j)}\cap\Lambda$. (Recall that $\bar{\bar{b}}^{(j)}$ is $\bar{b}^{(j)}$ plus an additional collar of width $\frac{15}{14} L_{j - 1}$, measured as usual in the metric $|\cd|^{(j - 1)}$.) This allows for the maximum range of dependence in the $\bar{b}^{(j)}$ eigenvalues (through energies $E_{\sigma}^{(j - 1)'}$ in $H^{(j - 1)'}$). Here $\bar{\bar{H}}$ is the original Hamiltonian in $\Lambda$, restricted to $\bar{\bar{b}}^{(j)}$, which means that spins outside of $\bar{\bar{b}}^{(j)}$ are fixed at $+1$ (as are the spins in $\Lambda^{\mathrm{c}}$).
As in our discussion on volume-dependence in Section \ref{4.5}, we may define resonant graphs and blocks by including every possible configuration of spins and of $\Lambda$ within the relevant range. Therefore, the set of resonant blocks $\bar{b}^{(j)}$ does not depend on $\Lambda$. 

Let us assume that in $\bar{\bar{b}}^{(j)}$, the eigenvalues of $\bar{\bar{H}}$ satisfy
\begin{equation}
\min_{\alpha \ne \beta} |E_{\alpha} - E_{\beta}| \ge \varepsilon^{s \varkappa n}, \label{(5.2a)}
\end{equation}
where $n$ is the number of sites in $\bar{b}^{(j)}$, and $\varkappa$ is a small constant to be chosen below. Note that the size of $\bar{\bar{b}}^{(j)}$ is no greater than some multiple $m$ of $n$, after allowing for the expansion of distance due to blocks at lower scales. So if we let $\tilde{\varepsilon} = \varepsilon^{s \varkappa /m}$ 
and take $\tilde{\varepsilon}  \le \varepsilon_0$, then $\textbf{A2} (\nu,\varepsilon_0)$ implies that (\ref{(5.2a)}) occurs with probability at least $1 - \tilde{\varepsilon}^{\nu m n} = 1 - \varepsilon^{s \nu \varkappa n}$. We will discuss below the case where (\ref{(5.2a)}) does not hold.

We may perform all our expansions on $\bar{\bar{H}}$ in $\bar{\bar{b}}^{(j)}\cap\Lambda$. We obtain block energies $\bar{\bar{E}}^{(j')}_{\alpha}$. These agree with the corresponding energies $E^{(j')}_{\alpha}$ obtained from the expansion in $\Lambda$, because as explained earlier, the range of dependence on $\sigma$ is less than
 the width of the collar $\bar{\bar{b}}^{(j)} \setminus \bar{b}^{(j)}$. Furthermore, all remaining terms in $\bar{\bar{H}}^{(j')}$(the Hamiltonian $\bar{\bar{H}}$ after the $j$\textsuperscript{th} step) are of order $L_{j}$, so by (\ref{(4.6)}) they are exponentially small -- in total no greater in norm than $(c \gamma / \varepsilon)^{n/4}$. (Note that separation conditions keep other blocks $\bar{b}^{(j)}$ or $\bar{B}^{(j')}$ out of $\bar{\bar{b}}^{(j)}$. The size of $\bar{b}^{(j)}$ is no greater than $4L_j$, from the collar of width $L_j-1$ about $b^{(j)}$, plus some additional expansion from smaller blocks.) Thus the minimum eigenvalue spacing in (\ref{(5.2a)}) transfers to $\bar{\bar{E}}^{(j')}_{\alpha}$ and hence to $E^{(j')}_{\alpha}$. (Since $\gamma = \varepsilon^{20}$, the corrections are much smaller than the minimum gap.) At the heart of our method is a way to extract local ``quasi-mode" energies for transitions approximated on a length scale $L$, with errors exponentially small in $L$. Rotations were performed for graphs connecting $\bar{b}^{(j)}$ to its complement, up to scale $L_j$, so any residual effects are exponentially small.

In order to obtain assumption $\textbf{A1}(c_b)$ we need to compare derivatives of $\bar{\bar{E}}^{(j')}_{\alpha}$ with those of $\bar{\bar{E}}_{\alpha}$. Since we are making a perturbation with exponentially small norm and derivatives, bounds on derivatives of $\bar{\bar{E}}_{\alpha}$ carry over to those of $E^{(j')}_{\alpha} = \bar{\bar{E}}^{(j')}_{\alpha}$ via second-order perturbation theory. There will be a sum over intermediate states (no more than exponential in $n$) and an energy denominator (bounded below by (\ref{(5.2a)})). Thus derivatives of $E^{(j')}_{\alpha}$ (or more precisely of differences $E^{(j')}_{\alpha} - E^{(j')}_{\beta}$) agree with those of $\bar{\bar{E}}_{\alpha}$ up to errors of order $(c\gamma / \varepsilon^{1 + s \varkappa})^n$.

Next we use (\ref{(5.2a)}) to prove a lower bound on the radial derivative of $\bar{\bar{E}}_{\alpha} - \bar{\bar{E}}_{\beta}$. Note that the radial variable $r$ appears as a multiplicative factor in $\bar{\bar{H}}$, since all couplings $h_i, J_i, \gamma_i$ are proportional to $r$. Therefore energies (and their differences) are strictly proportional to $r$. Let us fix the angular variables. Then there is some $r_0$ such that the minimum eigenvalue separation is $\varepsilon^{s \varkappa n}$. Any eigenvalue difference can be written as
\begin{equation}
E_{\alpha}(r) - E_{\beta}(r) \equiv D_{\alpha \beta}(r) = D_{\alpha \beta}(r_0) \frac{r}{r_0}. \label{(5.2b)}
\end{equation}
Therefore,
\begin{equation}
\frac{\partial}{\partial r} D_{\alpha \beta}(r) = \frac{D_{\alpha \beta}(r_0)}{r_0} \ge \frac{\varepsilon^{s \varkappa n}}{\sqrt{3mn}}, \label{(5.3)}
\end{equation}
where we use the radius of the integration domain $[-1, 1]^{mn}$ to bound $r_0$ from above. 
(A larger value of $r_0$ would mean (\ref{(5.2a)}) fails throughout the integration domain on the ray we are considering, in which case there is nothing to prove.)
Thus we obtain
\begin{equation}
\left|\frac{\partial}{\partial r}\left(E^{(j')}_{\alpha}(r) - E^{(j')}_{\beta}(r)\right)\right|^{-1} \le (c \varepsilon^{-s \varkappa})^n, \label{(5.4)}
\end{equation}
since as explained above, the derivatives of $E^{(j')}_{\alpha}$ agree with these of $E_{\alpha}$ up to terms much smaller than $\varepsilon^{s \varkappa n}$.

Note that (\ref{(5.4)}) compares with (\ref{(4.10)}) of assumption $\textbf{A1}(c_b)$, with $c_b = c \varepsilon^{-s \varkappa}$. Each time we do a denominator integral in proving an estimate like (\ref{(4.19)}), we pick up a factor of $c_b$, which through (\ref{(4.10)}) controls the rate of change of energy differences. So now that $c_b = c \varepsilon^{-s \varkappa}$, we need to absorb factors of $\varepsilon^{-s \varkappa}$ into (\ref{(4.19)}). 
As noted in Subsection \ref{4.2.3}, $\rho_1$ depends on $c_b$, but it grows no faster than a fixed power of $c_b$ (based on the fact that graph lengths are always at least some fixed multiple of the block sizes involved). In (\ref{(4.19)}), each factor $\rho_1$ is mated with a factor $\veps^{s/4}$, and so the additional factors 
of $\varepsilon^{- s \varkappa}$ can be handled with a reduction of the power from $s/4$ to $s/8$, for some small value of $\varkappa$. 
Thus our estimates work with only this minor modification.

It remains for us to discuss the case where (\ref{(5.2a)}) does not hold, an event whose probability is no greater than $\varepsilon^{s \nu \varkappa n}$. We may consider any block $\bar{\bar{b}}^{(j)}$ with a too-small level spacing as part of $\calS_k$, the next singular region. Its probability is exponentially small in the size of $\bar{\bar{b}}^{(j)}$, so we obtain a bound similar to (\ref{(4.20)}):
\begin{equation}
P^{(k)}_{xy} \le c \varepsilon^{s \nu \varkappa |x - y|}. \label{(5.4a)}
\end{equation}
The rate of decay is $\varepsilon^{s \nu \varkappa}$ instead of $\varepsilon^s$, but as long as $\varepsilon$ is chosen sufficiently small, the proof of diluteness of resonant blocks works as before. 
(Diluteness is the bound (\ref{(4.24)}) giving rapid falloff of the probability that $x,y$ belong to the same resonant block. 
It should be clear that the smaller power of $\gamma = \varepsilon^{20}$ just means that the value of $\kappa$ that can be achieved in Theorem \ref{th1.1} is a bit smaller than what we obtained by assuming $\textbf{A1}(c_b)$.)
The proof relies on bounds like (\ref{(4.20)}), which provide a probability factor $\varepsilon^s$ for each site of $b^{(k)}$. Then separation conditions tied to the volume control the sum over admissible $b^{(k)}$ and provide a minimum volume for a given diameter. All this would work if we defined the volume of a block $\bar{\bar{b}}^{(j)}$ of size $n$ as $O((\log n)^2)$, the same as the minimum volume achievable from sites of $\calS_k$ via the original construction. This makes it clear that we could make do with a weaker form of $\textbf{A2}(\nu,\varepsilon_0)$, replacing (\ref{(5.1)}) with
\begin{equation}
{P} \Big( \min_{\alpha \ne \beta} |E_{\alpha} - E_{\beta}| < \tilde{\varepsilon}^n \Big) \le \tilde{\varepsilon}^{\nu (1 + c_4 (\log n)^2)}, \label{(5.5)}
\end{equation}
for some constant $c_4$. (This could potentially be useful if, in the future, better methods are developed to control minimum level spacings.) The basic mechanism at play here is that by Borel-Cantelli, (\ref{(5.5)}) guarantees that there is some scale $k$ at which the minimum level spacing holds. When that happens, we obtain the needed variation of $E^{(k')}_{\alpha} - E^{(k')}_{\beta}$ with $r$, and the rest of the proof of MBL works as described in Chapter \ref{4}. This completes the proof of Theorem \ref{th5.1}. Corollary \ref{co5.2} then gives the full version of Theorem \ref{th1.1}. \qed

\appendix
\section{Extending Graphs Across Gaps}\label{A}
Here we describe how to organize the graphical expansions of Section \ref{4.3}  to exhibit decay across gaps between graphs while limiting the duplication of denominators. 

Consider the general expression for a term in $J^{(k)}$:
\begin{equation}\label{(A.1)}
\left(\text{ad}\,A^{(k)}(g_{j'',n})\right) \cdots \left(\text{ad}\,A^{(k)}(g_{j'',1})\right) J^{(j')}(g_{j',0}).
\end{equation}
Divide the $g_{j'',p}$ into non-overlapping groups. 
The actions of a commutator of a graph in one group with a graph in another group is simple, as the operators in question involve disjoint sets of spin indices. As indicated in (\ref{(4.35)}), the energy denominators are computed before and after a spin transition and the difference taken. 
We have the freedom to use either side of (\ref{(4.35)}) when working with these denominator differences. We use the right-hand side of (\ref{(4.35)}) and the graphical expansion (\ref{(4.36)}) only for a minimal set of differences, sufficient to close the gaps between groups. For the remaining differences, we bound each term on the left-hand side  of (\ref{(4.35)}) separately: $m$ differences result in $2^m$ terms; a factor of 2 per graph is easily controlled, as discussed in Subsection \ref{4.2.1}. 

We use the following algorithm to generate differences and select ones that need to be expanded as in (\ref{(4.36)}). We proceed through the sequence of graphs $g_{j'',1}, g_{j'',2},\ldots$ in (\ref{(A.1)}) and the associated commutators. If a graph $g_{j'',p}$ appears as the first representative of one of the groups, the associated commutator is written as a sum of terms involving the commutators with each of the previous graphs $g_{j'',p-1}, g_{j'',p-2},\ldots, g_{j'',1},g_{j',0}$, by Leibniz's rule for commutators. As $g_{j'',p}$ is the first in its group, there is no overlap between it and the previous graphs. Then, as was just explained, the commutator can be written as the sum of two differences. For example, one contains the effect of $A^{(k)}(g_{j'',p})$ on $J^{(j')}(g_{j',0})$ and the other contains the effect of $J^{(j')}(g_{j',0})$ on $A^{(k)}(g_{j'',p})$.
This may be indicated graphically with a line from one graph to the other and an arrow indicating the direction of the effect. (The arrow can go in either direction, because each graph's denominators are affected by changes in spin configuration induced by the other graph. However, the direction of the arrow is unimportant in what follows, as either direction will be sufficient to create a graphical connection across the gap between the graphs.) Note that the expanding set of lines ensures that each group is connected to its predecessors as soon as one of its graphs appears. Hence when all commutators have been performed, the graph of difference lines (the \textit{difference graph}) connects all of the groups. Many of the lines are redundant, however.

We define a minimal subgraph of the difference graph using a lace construction similar to the one introduced in \cite{Brydges1985}. Starting from the left-most group, we take the first line of the lace to be the one reaching as far as possible to the right. (Note that there can be no more than one line between two groups, by construction, as our algorithm specifies that each line have one endpoint in a \textit{new} group. Thus there is no ambiguity about which line reaches the farthest to the right.) The next line of the lace is taken as the one reaching the group farthest to the right from any of the groups spanned by the first line. (If two lines reach that group, we take the one originating from the group farthest to the right. This is unique because, as explained above, there cannot be two lines connecting the same two groups.)
We continue, always choosing the line reaching the group farthest to the right from the groups spanned by the expanding lace graph, and breaking ties by choosing the line originating from the group farthest to the right. 
See Figure \ref{lace}.
\begin{footnotesize}
\begin{figure}[h]
\centering
\includegraphics[width=.9\textwidth]{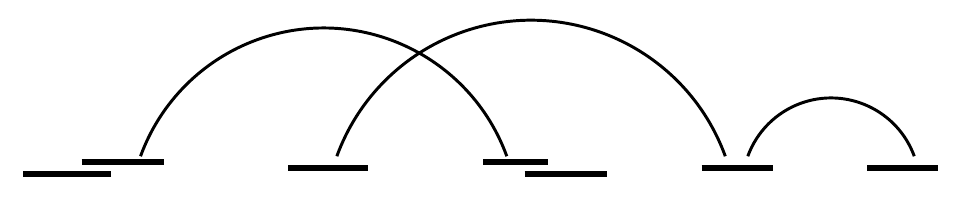}
\caption{A lace connecting five groups. Each arch represents a difference operation that provides a graphical connection and exponential decay across gaps between groups. \label{lace}}
\end{figure} 
\end{footnotesize}

All lines not in the lace are left as is: a sum of two terms, as in the left-hand side of (\ref{(4.35)}). They are not needed for generating decay across the gaps between graphs. The lines of the lace graph extend all the way from the left-most group to the right-most one. Therefore, by exhibiting decay between graphs connected by lace lines, we obtain decay across all gaps between groups.

The lace graph has the property that no more than two lines emanate from any group. There can be one line to the left and one to the right, but a second on either side is impossible, by the rules for constructing the lace graph. As a consequence, we see that no more than two differences will be applied to any graph. This is important because of the need to limit the extent of duplication of denominators.

The first difference creates a double denominator as in  (\ref{(4.35)}). Then  (\ref{(4.36)}) can be used to exhibit decay across the gap between the two graphs. The second difference creates a triple denominator as follows:
When two differences are applied, we have four denominators, which may be written as $d_{11}$, $d_{12}$, $d_{21}$, $d_{22}$, with the first index indicating which spin configuration is involved on the left, and the second index indicating the one on the right. Then we may use Leibniz's rule to write
\begin{align}\label{(A.2)}
&\left(\frac{1}{d_{22}} - \frac{1}{d_{21}}\right) - \left(\frac{1}{d_{12}} - \frac{1}{d_{11}}\right)
= \frac{d_{21} - d_{22}}{d_{22}d_{21}} - \frac{d_{11} - d_{12}}{d_{12}d_{11}}
\\&=\frac{d_{21} - d_{22}}{d_{22}d_{21}} - \frac{d_{21} - d_{22}}{d_{12}d_{21}}
+\frac{(d_{21}-d_{22})-(d_{11}-d_{12})}{d_{12}d_{21}}
+\frac{d_{11} - d_{12}}{d_{12}d_{21}} - \frac{d_{11} - d_{12}}{d_{12}d_{11}} \nonumber
\\&=\frac{(d_{21} - d_{22})(d_{21}-d_{22})}{d_{22}d_{21}d_{12} }
+\frac{(d_{21}-d_{22})-(d_{11}-d_{12})}{d_{12}d_{21}}
+\frac{(d_{11} - d_{21})(d_{11}-d_{12})}{d_{12}d_{21}d_{11} }.\nonumber
\end{align}
For each term in the final expression, we may obtain decay across the gaps to the left and to the right when denominator differences are expanded as in (\ref{(4.36)}). (If a graph fails to cross both gaps, the result is zero because it cannot feel both changes in spin configuration.) Note that  denominators can be tripled if differenced from both sides, but (as claimed in the main text) no  power higher than 3 occurs.

We need to provide estimates on the number of terms produced from applications of Leibniz's rule in the above algorithm. Recall that the range of dependence of denominators on the spin configuration is no greater than $\tfrac{15}{14}L_j$ -- see (\ref{range}). This becomes a bound on the number of groups that can be reached with a difference line originating from a particular 
$\text{ad}\,A^{(k)}(g_{j'',p})$ operation. When a difference operator hits a group, there is a sum over the graphs in the group; this may be controlled most simply by assigning a combinatoric factor 2 to each graph in the group. The number of lines incident on a group cannot exceed $\tfrac{15}{14}L_j$ (the maximum number of groups in range). Hence each graph receives no more than a combinatoric factor $2^{(15/14)L_j}$. There are also factors of 2 per graph that were mentioned above, coming from the representation of a commutator as a sum of two terms or as a sum of two differences. The product of all these combinatoric factors is bounded by $c^{L_j} \le c^{|g|}$ for each $g \in \{g_{j',0},g_{j'',1}, \ldots , g_{j'',n} \}$, in view of the minimum graph size in this step. Thus the counting factors are in line with ones already considered in Subsection \ref{4.2.1}, and are controlled by the exponential bounds on graphs as in 
(\ref{(4.2)}),(\ref{(4.6)}). As one often finds in this type of argument, dangerous counting factors from applications of Leibniz's rule are limited by geometrical considerations -- in this case by the range limitation and the cap of one on the number of commutators per group that are expanded with Leibniz's rule.

\section{Index of Definitions and Notations}\label{B}
Here we provide a list of important definitions and notations, along with locations where they are introduced in the text.

\begin{list}{}{%
\setlength{\topsep}{0pt}%
\setlength{\leftmargin}{0.2in}%
\setlength{\listparindent}{-0.1in}%
\setlength{\itemindent}{-.2in}%
\setlength{\parsep}{0pt}%
\setlength{\itemsep}{0pt}%
}%
\item
$\rho_0$, bound on probability densities: after (\ref{(1.2)}). 
\item
$\rho_1$, constant governing probability of an energy difference lying in a small interval: above (\ref{(3.3a)}), start of Subsection \ref{4.2.3}.
\item
Level spacing assumptions. \textbf{LLA}($\nu, C$): (\ref{(1.3)}); \textbf{A1}($c_b$): (\ref{(4.10)}); \textbf{A2}($\nu$,$\varepsilon_0$): (\ref{(5.1)}).
\item
$L_k =(15/8)^k$, length scales: start of Section \ref{1.3}.
\item
$\sigma^{(i)}$, spin configuration flipped at $i$: (\ref{(2.1)}).
\item
$E(\sigma)$, energy of spin configuration:  (\ref{(2.2)}).
\item
``$i$ is resonant''; $\varepsilon = \gamma^{1/20}$; $\calS_1$, resonant set;  $B^{(1)}$, its components: after (\ref{(2.3)}).
\item
$J^{(0)}$, $J^{(0)\text{per}}$, $J^{(0)\text{res}}$, interaction and its perturbative and resonant parts:  (\ref{(2.4)}).
\item
$A$, $A(i)$, generator of rotation: (\ref{(2.5)}),(\ref{(2.6)}).
\item
$\Omega$, the associated rotation; $H_0$, diagonal part of $H$: above (\ref{(2.7)}).
\item
$J^{(1)}$, new interaction:  (\ref{(2.7)}); $J^{(1)}_{\sigma\tilde{\sigma}}(g_1)$, the term associated with a particular graph:  (\ref{(2.9)}).
\item
$|g_1|$, first step graph: above  (\ref{(2.10)}).
\item
$d_m=\exp(L_{m+m_0}^{1/2})$, extended separation distances:  start of Section \ref{2.3}.
\item
$m_0$: chosen in Lemmas \ref{lem:4.1} and \ref{lem:4.2}.
\item
$b^{(1)}$, small block; $\bar{b}^{(1)}$, collared version; $\overline{\calS}_1$, collared version of $\calS_1$;  $\calS_{1'}$, region of large blocks; $\overline{\calS}_{1'}$, collared version; $\bar{B}^{(1')}$, $B^{(1')}$, large blocks;  above (\ref{(2.11)}). (Prime means large blocks only; bar means collar is included.)
\item
$J^{(1)\text{int}}$, interaction terms internal to blocks; $J^{(1)\text{sint}}$, $J^{(1)\text{lint}}$, internal to small, large blocks, respectively; $J^{(1)\text{ext}}$, non-internal terms: (\ref{(2.11)}).
\item
$O$, small block rotation; $\bar{\bar{b}}^{(1)}$, second collar block: after (\ref{(2.12)}).
\item
$H^{(1')}$, post-rotation effective Hamiltonian; $H_0^{(1')}$, its diagonal part; $J^{(1')}$, rotated interaction: (\ref{(2.13)})-(\ref{(2.15)}).
\item
$g_{1'}$, graph extended by rotation matrix elements: above (\ref{(2.16)}).
\item
$J^{(1')}(g_{1'})$, the associated interaction term: (\ref{(2.16a)}). 
\item
$A^{(2)\text{prov}}$, provisional rotation generator; $E_\sigma^{(1')}$, post-rotation energy:  (\ref{(3.1)}).
\item
$I(g_1)$, interval of $g_1$; $|I(g_1)|$, its size: after (\ref{(3.2)}).
\item
$B^{(2)}$, new blocks:  above (\ref{(3.3)}).
\item
$b^{(2)}$, new step 2 small blocks; $|b^{(2)}|$, its size; separation conditions; $\calS_{2'}$, new large block region; $\overline{\calS}_{2'}$, collared version;  $\bar{B}^{(2')}$, $B^{(2')}$, large blocks; $\calS_{2}$ is $\calS_{2'}$ plus small blocks $b^{(2)}$; $\overline{\calS}_2$, $\bar{b}^{(2)}$, collared versions: (\ref{(3.3)}) and after.
\item
$P_{ij}^{(2)}$, connectivity function for $B^{(2)}$ blocks; $|i-j|^{(1)}$, metric with $\bar{\bar{b}}^{(1)}$ blocks contracted to points; $s = \frac{2}{7}$: Proposition \ref{prop:3.1}.
\item
$Q_{ij}^{(2)}$, connectivity function for small blocks; $\bar{b}^{(2)}$:  above (\ref{(3.8)}).
\item
$J^{(1')\text{per}}$, $J^{(1')\text{res}}$, perturbative and resonant interactions for step 2: (\ref{(3.9)}).
\item
$A^{(2)}$, $A^{(2)}(g_{1'})$, generators of rotations: (\ref{(3.10)}).
\item
Long graphs, short graphs, jump transitions; $g_{1''}$,  graph with jump steps representing sums of long graphs; $|g_{1''}|$, its length: after (\ref{(3.10)}).
\item
$A^{(2)}(g_{1''})$, generator of rotations with long graphs resummed: (\ref{(3.11)}). 
\item
$\Omega^{(2)}$, the associated rotation: after (\ref{(3.11)}). 
\item
$H^{(2)}$, new Hamiltonian; $J^{(2)}$, new interaction: (\ref{(3.12)}),(\ref{(3.13)}).
\item
$g_2$, step 2 graph; $|g_2|$, its length; $J^{(2)}(g_2)$, the associated interaction term: (\ref{(3.15)}) and above. 
$g_2!$: after (\ref{(3.16)}).
\item
$O^{(2)}$, small block rotation matrix; $H^{(2')}$, new Hamiltonian; $H_0^{(2')}$, diagonal part; $E_\sigma^{(2')}$, its diagonal entries (energies); $J^{(2')}$, off-diagonal part: (\ref{(3.18)})-(\ref{(3.20)}) and after.
\item
$g_{2'}$, graph extended by rotation matrix elements: after (\ref{(3.21)}). 
\item
$J^{(2')}(g_{2'})$, the associated interaction term: (\ref{(3.21)}).
\item
$R^{(1')}$, $R^{(2')}$, cumulative rotations: (\ref{(3.22)}).
\item
Multigraphs; level $i$ subgraphs; $g^{\text{s}}_{j'}$, spatial graph; $g^{\text{d}}_{j'}$, denominator graph: above (\ref{(4.2)}).
\item
$g_{j''}!$, inductive definition of the factorial of $g_{j''}$; $|g_{i''}|$, length of $g_{j''}$;   (\ref{(4.3)})-(\ref{(4.4)}).
\item
$|x-y|^{(i)}$, metric with blocks $\bar{\bar{b}}^{(\tilde{i})}$, $\tilde{i}\le i$ contracted to points: after (\ref{(4.4)}).
\item
$\bar{g}_{j'}$, reduced graph for probability sums; $|\bar{g}_{j'}|$, its length; $\bar{g}_{j'}!$, its factorial: above (\ref{(4.7)}).
\item
$A^{(k)\text{prov}}$, provisional rotation generator; $\tilde{J}_{j'}$: (\ref{(4.7)}) and after.
\item
``$g_{j'}$ is resonant''; $I(g_{j'})$, interval of $g_{j'}$; $|I(g_{j'})|$, its length: (\ref{(4.8)}) and after.
\item
$B^{(k)}$, new block in step $k$; above (\ref{dist}).
\item
$b^{(k)}$, small block in step $k$; separation conditions; $|b^{(i)}|$, size of block: near (\ref{dist}).
\item
$\calS_{k'}$, new large block region; $B^{(k')}$, large block; collared versions $\overline{\calS}_{k'}$, $\bar{B}^{(k')}$, $\bar{b}^{(k)}$: after (\ref{dist}).
\item
Combinatoric factor: start of Subsection \ref{4.2.1}.
\item
$T_\alpha$, looping segment; $|T_\alpha|$, its weighted size: above (\ref{(4.18a)}). 
\item
Erased subgraphs: after (\ref{(4.18a)}). 
\item
$\bar{g}^{\text{e}}_{j'}!$, modified factorial eliminating those from erased sections: after (\ref{(4.19)}).
\item
$P_{ij}^{(k)}$, connectivity function for $B^{(k)}$ blocks: Proposition \ref{prop:4.2}.
\item
$Q_{ij}^{(k)}$, connectivity function for small blocks $\bar{b}^{(k)}$; $R_{ij}^{(k)}$, multiscale connectivity function for small blocks $\bar{b}^{(i)}, i \le k$: Proposition \ref{prop:4.3}.
\item
$J^{(j')\text{per}}$, $J^{(j')\text{res}}$, perturbative and resonant interactions for step $k$: (\ref{(4.25)}),(\ref{(4.26)}).
\item
$A^{(k)}(g_{k''})$, generator of rotations: (\ref{(4.27)}).
\item
Long graphs, short graphs, jump transitions; $g_{j''}$,  graph with jump steps representing sums of long graphs; $|g_{1''}|$, its length: after (\ref{(4.27)}).
\item
$A^{(k)}(g_{k''})$, generator of rotations with long graphs resummed: (\ref{(4.28)}).
\item
$\Omega^{(k)}$, the associated rotation: after (\ref{(4.28)}).
\item
$H^{(k)}$, new Hamiltonian; $J^{(k)}$, new interaction: (\ref{(4.29)}),(\ref{(4.30)}).
\item
Gap graphs: after (\ref{range}).
\item
$J^{(k)\text{ext}}$, $J^{(k)\text{sint}}$, $J^{(k)\text{lint}}$, interaction terms external to blocks, internal to small blocks, internal to large blocks: (\ref{(4.37)}).
\item
$O^{(k)}$, small block rotation matrix; $H^{(k')}$, new Hamiltonian; $H_0^{(k')}$, diagonal part; $E^{(k')}_\sigma$, diagonal entry; $J^{(k')}$, rotated interaction: (\ref{(4.38)}),(\ref{(4.39)}).
\item
$g_{k'}$, graph extended by rotation matrix elements: after (\ref{(4.39)}).
\item
$R^{(k')}$, cumulative rotation: (\ref{(4.40)}).
\end{list}
\section*{Acknowledgement}
The author would like to thank Tom Spencer for a collaboration over several years, during which time many of the ideas were developed. Thanks also to David Huse, who suggested this problem, and was helpful in numerous conversations. 
\begin{footnotesize}
\bibliographystyle{acm}

\end{footnotesize}
\end{document}